\documentclass[aps,prb,reprint,amsmath,amssymb]{revtex4-2}
\usepackage{graphicx}
\usepackage{bm}
\usepackage{color}
\usepackage{mathrsfs}

\begin{document}

\title{Projective-symmetry-group analysis of inelastic light scattering in Kitaev spin balls}

\author{Taku Kimura and Shoji Yamamoto${}^*$}

\affiliation{Department of Physics, Hokkaido University, Sapporo 060-0810, Japan}

\date{\today}

\begin{abstract}
Projective symmetry groups are applied to Raman observations of
the Kitaev quantum spin liquids in spherical lattice geometries
realized by Platonic and Archimedean polyhedra.
Parton single excitations in Kitaev spin polyhedra are characterized by
double-valued irreducible representations of their belonging projective symmetry groups,
whereas parton geminate excitations relevant to Raman scattering are decomposed into
single-valued irreducible representations of the corresponding point symmetry groups.
We combine a standard point-symmetry-group analysis of the Loudon-Fleury vertices
and an elaborate projective-symmetry-group analysis of itinerant spinons against
the ground gauge fields to reveal \textit{hidden selection rules} for Raman scattering
in $\mathbb{Z}_2$ spin liquids.
\end{abstract}

\maketitle

\section{Introduction}
\label{S:I}

   The Kitaev honeycomb model \cite{K2} sparked a brandnew interest in quantum spin liquids
(QSLs) \cite{S016502,Z025003,K451,M012002}.
It is exactly solvable to have a QSL ground state accompanied by $\mathbb{Z}_{2}$ gauge fields,
whose excitations are fractional, decomposing into itinerant ``spinons" and local gapped ``visons".
Jackeli and Khaliullin \cite{J017205} designed Mott insulators in the strong spin-orbit coupling
limit for the Kitaev model, leading to many candidate materials such as
$\mathrm{Na_{2}IrO_{3}}$ \cite{S064412},
$\alpha$-$\mathrm{Li_{2}IrO_{3}}$ \cite{Y127203},
$\mathrm{H_{3}LiIr_{2}O_{6}}$ \cite{T554},
and $\alpha$-$\mathrm{RuCl_{3}}$ \cite{P041112}.
The pure Kitaev model is hard to realize but often accompanied by not only usual Heisenberg
interactions, whether intralayer \cite{C027204,C097204} or interlayer
\cite{S115159,T094403,S155101,T174424}, but also off-diagonal exchanges referred to as
the $\Gamma$ term \cite{R077204,K013056,Y107201}.
Since fractional excitations remain possible in
such ``effective" Kitaev models
\cite{K451,M012002,S184411,Y174425,G075126,K134432,K187201,N912,P060408,P104427,P184429,R045117},
inelastic-neutron-scattering \cite{C127204,B733,B1055,D1079},
x-ray-absorption \cite{P041112}, and Raman-scattering \cite{S147201}
measurements have been performed on them in an attempt to diagnose QSLs.
Raman spectroscopy is particularly useful in detecting spinons separately from visons
\cite{K187201,N912,Y012003}.

   The Kitaev QSL is realizable with any lattice of coordination number three.
$\beta$-$\mathrm{Li_{2}IrO_{3}}$ \cite{T077202} and $\gamma$-$\mathrm{Li_{2}IrO_{3}}$ \cite{M4203},
consisting of
``hyperhoneycomb" \cite{M024426,N197205} and ``stripyhoneycomb" \cite{K205126} lattices,
respectively, are such candidates in three dimensions.
While they both exhibit gapless spinon excitations coming from nodal rings, the degeneracy of
the Fermi level strongly depends on the lattice geometry in general.
A normal Fermi surface is emergent in a ``hyperoctagon" lattice \cite{H235102,O085101},
whereas it reduces to what they call Weyl points in ``hypernonagon" \cite{O085101,K174409}
and ``hyperhexagon" \cite{O085101,S235146} lattices.
Spinon excitations may be gapped from the ground state \cite{O085101}.
Kitaev models in lower than two dimensions also attract much interest.
Kitaev honeycomb nanoribbons with both zigzag and armchair edges are discussed
in an attempt to optically distinguish between different topological phases \cite{S012046}
and investigated with particular interest in a bulk-edge correspondence \cite{T235434},
i.e. a possible relation between gapped states in the bulk and gapless states in the boundary.
A Kitaev spin ladder maps onto a one-dimensional $p$-wave superconductor in terms of Dirac
fermions to reveal the equivalence between spontaneous global $\mathbb{Z}_2$
symmetry breaking and emergent isolated Majorana modes \cite{D065028},
while that with inhomogeneous exchange interactions exhibits coexistent different topological
phases with Majorana end states inbetween \cite{P205412}.
\begin{figure}[b]
\centering
\includegraphics[width=86mm]{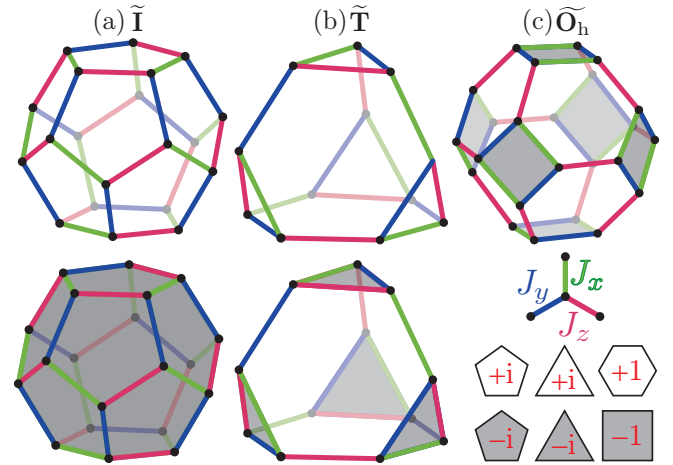}
\caption{(Color online)
         Kitaev spin balls consisting of dodecahedral (a), truncated tetrahedral (b), and
         truncated octahedral (c) lattices in their ground flux configurations.
         The ground state of the truncated octahedron is unique, whereas
         those of the dodecahedron and truncated tetrahedron are both degenerate \cite{Y247203}
         with their constituent pentagons arrangeable into either $\{W_p=+i;\,p=1,\cdots,12\}$
         or $\{W_p=-i;\,p=1,\cdots,12\}$ and triangles arrangeable into either
         $\{W_p=+i;\,p=1,\cdots,4\}$ or $\{W_p=-i;\,p=1,\cdots,4\}$.}
\label{F:SpinBalls}
\end{figure}

   In such circumstances, Mellado, Petrova, and Tchernyshyov (MPT) \cite{M041103} discuss
the Kitaev spin model in a spherical lattice geometry realized by Archimedean solids.
Analyzing the projective symmetry \cite{W165113,W174423} of the gauge-ground
Majorana fermionic Hamiltonian (cf. Appendix \ref{AS:prsym}) rather than the point symmetry of
the background lattice, they claim that a parton behaves like an electrically charged particle
in a radial (monopole) magnetic field within
the continuum---in the sense of a perfect sphere---approximation.
This parton has a \textit{half-odd-integral} orbital angular momentum due to the magnetic monopole
located at the center of the cluster.

   Motivated by the MPT theory, we present a symmetry argument of optical observations of
``Kitaev spin balls"---QSLs in a spherical lattice geometry realized by Platonic and Archimedean
polyhedra (cf. Fig. \ref{F:SpinBalls}).
Since Raman scattering within the Loudon-Fleury (LF) scheme \cite{F514,S1068,S365} is mediated by
spinons in pair, we make direct-product representations out of
irreducible representations of the corresponding projective symmetry group
and then decompose them into irreducible representations again.
In order to reveal how each spinon geminate excitation behaves under spatial inversion, which is
vitally important in the context of Raman scattering,
we go so far as to take \textit{gauged inversion}, if any, as well as gauged rotations,
into the projective symmetry.
\textit{Kitaev spin balls made only of $2l$-sided polygons} ($l\in\mathbb{N}$)
require such an elaborate formulation,
namely, making direct-product representations of the \textit{extended} binary polyhedral group,
i.e. the double cover of the \textit{full} icosahedral or octahedral group,
to obtain \textit{inversion-symmetry-definite} single-valued irreducible representations.

\section{Kitaev Models on Platonic and Archimedean Polyhedra}
\label{S:KMonPAP}

  The Kitaev Hamiltonian (Fig. \ref{F:SpinBalls}) reads
\begin{align}
   \mathscr{H}
 =-\sum_{\lambda=x,y,z}\sum_{<m,n>_{\lambda}}
   J_\lambda\sigma_m^\lambda\sigma_n^\lambda,
   \label{E:Hspin}
\end{align}
where $(\sigma_l^x,\sigma_l^y,\sigma_l^z)\,(l=1,\cdots,L)$ are the Pauli matrices and
$<m,n>_{\lambda}\,(\lambda=x,y,z)$ each run over a different set of $L/2$ nearest-neighbor
bonds between the $\lambda$ components.
We set this model in various polyhedral geometries, i.e. on dodecahedral, truncated-tetrahedral,
and truncated-octahedral lattices, whose point symmetry groups are given by
$\mathbf{I}_{\mathrm{h}}=\mathbf{I}\times\mathbf{C}_{\mathrm{i}}$,
$\mathbf{T}_{\mathrm{d}}=\mathbf{T}+IC_4\mathbf{T}$, and
$\mathbf{O}_{\mathrm{h}}=\mathbf{O}\times\mathbf{C}_{\mathrm{i}}
                        =\mathbf{T}_{\mathrm{d}}\times\mathbf{C}_{\mathrm{i}}$,
respectively.
$J_x$, $J_y$, and $J_z$ are all set to $J>0$ in the following.

   By representing the spin operators in terms of four Majorana fermions,
$\sigma_l^\lambda=i\eta_l^\lambda c_l$,
with anticommutation relations between them,
$\{\eta_m^\mu,\eta_n^\nu\}=2\delta_{mn}\delta_{\mu\nu}$,
$\{c_m,c_n\}=2\delta_{mn}$, and $\{\eta_m^\lambda,c_n\}=0$,
and then introducing bond operators,
$\hat{u}_{<m,n>_{\lambda}}\equiv i\eta_m^\lambda\eta_n^\lambda$,
the spin Hamiltonian (\ref{E:Hspin}) is rewritten into
\begin{align}
   \mathscr{H}
  =iJ\sum_{\lambda=x,y,z}\sum_{<m,n>_{\lambda}}
   \hat{u}_{<m,n>_{\lambda}}c_m c_n.
   \label{E:HMajorana}
\end{align}
Since $[\hat{u}_{<m,n>_\lambda},\mathscr{H}]=0$ and $\hat{u}_{<m,n>_\lambda}^2=1$,
$\hat{u}_{<m,n>_\lambda}$ reads a $\mathbb{Z}_2$ classical variable, $u_{<m,n>_\lambda}=\pm 1$.
Numbering the constituent polygons of a polyhedra, $p=1,\cdots,\frac{L}{2}+2$,
we define a flux operator \cite{K2,P134404} for each by multiplying its $N_p$ spin operators in
the anticlockwise manner viewed from the outside of the polyhedron,
\begin{align}
   \hat{W}_p
   &
   \equiv
   e^{i\hat{\varPhi}_p}
  =\prod_{<m,n>_{\lambda}\in \partial p}
   \sigma_m^\lambda\sigma_n^\lambda
   \nonumber \\
   &
  =(-i)^{N_p}
   \prod_{<m,n>_\lambda\in \partial p}
   \hat{u}_{<m,n>_\lambda}.
   \label{E:Wp}
\end{align}
$\hat{W}_p$ also commutes with (\ref{E:HMajorana}) and thus behaves as a classical variable,
$W_p=\pm 1$ or $\pm i$ according as $N_p$ is even or odd.
A $\mathrm{U(1)}$ gauge flux, $W_p\equiv e^{i\varPhi_p}\,(-\pi<\varPhi_p\leq\pi)$, pierces
the constituent polygon $p$.
Every Kitaev spin ball consists of $\frac{L}{2}+2$ gauged polygons with their flux
variables satisfying $\prod_{p=1}^{\frac{L}{2}+2}W_p=1$.
The Hilbert space of the spin Hamiltonian (\ref{E:Hspin}) is block-diagonal with respect to flux
configurations $\{W_p\}$, consisting of $2^{\frac{L}{2}+1}$ blocks of dimension
$2^{\frac{L}{2}-1}\times 2^{\frac{L}{2}-1}$,
while that of the augmented Majorana Hamiltonian (\ref{E:HMajorana}) is block-diagonal with respect
to bond configurations $\{u_{<m,n>_{\lambda}}\}$ as well as $\{W_p\}$, consisting of
$2^{\frac{3L}{2}}$ blocks of dimension $2^{\frac{L}{2}}\times 2^{\frac{L}{2}}$.
Four Majorana fermions at each site have $2^{2L}$ degrees of freedom, containing
``unphysical states" \cite{Y217202,P165414} to be projected out by the operator
\cite{Y217202,P165414,Z014403,U220404}
\begin{align}
   \mathscr{P}
  =\prod_{l=1}^{L}
   \frac{1}{2}(1+\eta_l^x\eta_l^y\eta_l^z c_l).
\end{align}

   Once a set of the $3L/2$ gauge fields $\{u_{<m,n>_\lambda}\}$ is given,
we have a Majorana quadratic Hamiltonian to be solved,
\begin{align}
   &
   \mathscr{H}
  =\frac{i}{2}\sum_{m=1}^L\sum_{n=1}^L
   \mathcal{H}_{mn}
   c_m c_n;
   \nonumber \\
   &
   \mathcal{H}_{mn}
 =-\mathcal{H}_{nm}
  \equiv
   Ju_{<m,n>_{\lambda}}.
   \label{E:GaugeFixedMajoranaHamiltonian}
\end{align}
The real skew-symmetric matrix $\mathcal{H}$ can be block-diagonalized by a real orthogonal matrix
$\bm{\Psi}$,
\begin{align}
   &
   \mathscr{H}
  =\frac{i}{2}
   {}^{\mathrm{t}}\!\bm{c}\bm{\Psi}
   {}^{\mathrm{t}}\!\bm{\Psi}\mathcal{H}\bm{\Psi}
   {}^{\mathrm{t}}\!\bm{\Psi}\bm{c}
  =\frac{i}{2}
   {}^{\mathrm{t}}\!\tilde{\bm{c}}\mathcal{E}\tilde{\bm{c}}
  =i\sum_{k=1}^{L/2}
   \frac{\varepsilon_k}{2}
   \tilde{c}_{2k-1}\tilde{c}_{2k};
   \nonumber \\
   &
   \bm{c}
  \equiv
   \left[
    \begin{array}{c}
     c_1 \\
     \vdots \\
     c_L \\
    \end{array}
   \right]
  =\left[
    \begin{array}{ccc}
     \psi_{1,1} & \cdots & \psi_{1,L} \\
     \vdots     & \ddots & \vdots     \\
     \psi_{L,1} & \cdots & \psi_{L,L} \\
    \end{array}
   \right]
  \left[
    \begin{array}{c}
     \tilde{c}_1 \\
     \vdots \\
     \tilde{c}_L \\
    \end{array}
   \right]
  \equiv
   \bm{\Psi}\tilde{\bm{c}},
   \nonumber \\
   &
   \tilde{\bm{c}}
  ={}^{\mathrm{t}}\!\bm{\Psi}
   \bm{c},\ 
   \mathcal{E}
   \equiv
   \frac{1}{2}
   \left[
    \begin{array}{ccccc}
     0 & \varepsilon_1 & & & \\
    -\varepsilon_1 & 0 & & & \\
    & & \ddots & & \\
    & & & \!\!\!0 & \varepsilon_{\frac{L}{2}} \\
    & & & \!\!\!-\varepsilon_{\frac{L}{2}} & 0
    \end{array}
   \right].
   \label{E:AlternatingMatrix}
\end{align}
We recomplexify Majorana fermions,
\begin{align}
   &
   \tilde{c}_{2k-1}
  =\alpha_k^\dagger+\alpha_k,\ 
   \tilde{c}_{2k}
  =i\bigl(\alpha_k^\dagger-\alpha_k\bigr),
   \nonumber \\
   &
   c_l
  =\sum_{k=1}^{L/2}
   (\psi_{l,2k-1}\tilde{c}_{2k-1}
   +\psi_{l,2k}  \tilde{c}_{2k}  )
   \nonumber \\
   &\ \ 
  =\sum_{k=1}^{L/2}
   \bigl[
    (\psi_{l,2k-1}+i\psi_{l,2k})\alpha_k^\dagger
   +(\psi_{l,2k-1}-i\psi_{l,2k})\alpha_k
   \bigr],
   \nonumber \\
   &
   \alpha_k
  =\frac{1}{2}
   (\tilde{c}_{2k-1}+i\tilde{c}_{2k})
   \color{black}
  =\frac{1}{2}\sum_{l=1}^L
   (\psi_{l,2k-1}+i\psi_{l,2k})c_l,
   \nonumber \\
   &
   \alpha_k^\dagger
  =\frac{1}{2}
   (\tilde{c}_{2k-1}-i\tilde{c}_{2k})
   \color{black}
  =\frac{1}{2}\sum_{l=1}^L
   (\psi_{l,2k-1}-i\psi_{l,2k})c_l,
   \label{E:c=>alpha}
\end{align}
so as to obtain a diagonal Hamiltonian,
\begin{align}
   \!\!
   \mathscr{H}
  =\sum_{k=1}^{L/2}
   \frac{\varepsilon_k}{2}
   \bigl(
      \alpha_k^\dagger\alpha_k
     -\alpha_{k}\alpha_k^\dagger
   \bigr)
  =\sum_{k=1}^{L/2}\varepsilon_k
   \left(
    \alpha_k^\dagger\alpha_k-\frac{1}{2}
   \right),
   \label{E:MajoranaHdiag}
\end{align}
with nonnegative eigenvalues $\varepsilon_k\ge 0$.
Note that all sets of the gauge fields 
$\{u_{<m,n>_\lambda};\,<m,n>_x,<m,n>_y,<m,n>_z=1,\cdots,\frac{L}{2}\}$
yielding the same flux configuration $\{W_p;\,p=1,\cdots,\frac{L}{2}+2\}$
give the same set of eigenvalues $\{\varepsilon_k;\,k=1,\cdots,\frac{L}{2}\}$.
$\mathscr{P}$ can be expressed in terms of the bond variables $u_{<m,n>_\lambda}$, mixing
coefficients $\psi_{l,l'}$, and quasiparticle occupation operators $\alpha_k^\dagger\alpha_k$
to act on quasiparticle (spinon) states labeled background gauge fields $\{u_{<m,n>_\lambda}\}$.
Physical (unphysical) spinon states in the ground (lowest-energy) gauge sector consist of
even (odd) numbers of emergent spinons $\alpha_k^\dagger\alpha_k$.
All the $2^{\frac{3L}{2}}$ gauge sectors each contain $2^{\frac{L}{2}-1}$ physical and
$2^{\frac{L}{2}-1}$ unphysical states, each consisting of either only even or only odd numbers
of spinons.

   The ground flux configurations of Kitaev spin balls (Fig. \ref{F:SpinBalls}) are such that
$W_p$ of every constituent $N_p$-sided polygon is $+1$, $-1$, or either of $+i$ and $-i$
according as $N_p$ is $4l+2$, $4l$, or $2l+1$ with $l\in\mathbb{N}$ \cite{M041103,P134404}.
With the time-reversal-invariant Hamiltonian, the ground state is at least doubly degenerate
unless all $N_p$'s are even \cite{Y247203}.
Considering that the eigenspectrum of (\ref{E:HMajorana}) depends on $\{u_{<m,n>_{\lambda}}\}$
only through $\{W_p\}$ and $W_p$'s each commute with (\ref{E:Hspin}) as well as
(\ref{E:HMajorana}), we describe the ground state, unless otherwise noted, as the spinon vacuum
against a ground flux configuration
\begin{align}
   |\{n_k\}\rangle_0\otimes|\{W_p\}\rangle_0\equiv|0\rangle,
   \label{E:|0>}
\end{align}
where we denote the $\kappa$th spinon state against the $q$th flux configuration by
$|\{n_k\}\rangle_\kappa\otimes|\{W_p\}\rangle_q$
$(\kappa=0,\cdots,2^{\frac{L}{2}-1}-1;\,q=0,\cdots,2^{\frac{L}{2}+1}-1)$,
allowing it to run over physical states only.

\section{Projective Symmetry Groups for Gauge-Ground Kitaev Polyhedra}
\label{S:PSGforGGKP}
\subsection{Single- and double-valued irreducible representations}

   Characterizing Raman scattering mediated by Majorana spinons emergent in the gauge-ground
Kitaev truncated octahedron in terms of its projective symmetry group is essentially twofold:
first we go further than MPT \cite{M041103} in obtaining a projective symmetry group for
single Majorana eigenmodes, i.e., construct the double cover of the $\mathrm{O}(3)$ superset of
a pure rotation group, and then analyze direct-product representations made of its double-valued
irreducible representations.
Let us denote the point symmetry group of a Kitaev spin ball and its arbitrary group element by
$\mathbf{P}$ and $P$, respectively, and the $\mathbb{Z}_{2}$-gauge extension of $\mathbf{P}$ and
resultant gauged point symmetry operations by $\widetilde{\mathbf{P}}$ and $\widetilde{P}$,
respectively.
Regular and semiregular polyhedral lattices of our interest have the same coordination number
three and their point symmetry groups are either the cubic ($\mathbf{T}_{\mathrm{d}}$,
$\mathbf{O}_{\mathrm{h}}$) or icosahedral ($\mathbf{I}_{\mathrm{h}}$) groups.
Therefore, $\mathbf{P}\subset\mathrm{O}(3)$ in general.
$P\in\mathbf{P}$ generally changes the ground gauge fields of the Majorana Hamiltonian.
We demonstrate in detail gauged point symmetry operations on gauge-ground Kitaev polyhedra
as well as pure point symmetry operations on their background lattices in Appendix \ref{AS:prsym}.
Any two bond configurations yielding the same set of fluxes can be converted to each other
by local gauge operations.
Every rotation $R\in\mathbf{R}$ leaves any flux configuration $\{W_p;\,p=1,\cdots,\frac{L}{2}+2\}$
unchanged, whereas inversion $I\in\mathbf{P}$ and every reflection $\sigma\in\mathbf{P}$ reverse
the signs of all imaginary $W_p$'s peculiar to polygons of odd $N_p$.
Only if the group action $P$ leaves the flux configuration $\{W_p\}$ unchanged, there exist a pair
of gauge transformations $\pm\varLambda(P)$ to recover the initial ground gauge fields,
$\pm\varLambda(P)P\{u_{<m,n>_{\lambda}}\}=\{u_{<m,n>_{\lambda}}\}$.
We denote a couple of gauged point symmetry operations $\pm\varLambda(P)P$ unifiedly as
$\widetilde{P}$ and distinguishably by $\overline{P}$ and $\underline{P}$.
The symmetry groups of the gauge-ground Kitaev dodecahedron and truncated tetrahedron are
the $\mathbb{Z}_2$-gauge extensions of $\mathrm{SO}(3)$ subgroups of their full point symmetry
groups, $\widetilde{\mathbf{I}}$ and $\widetilde{\mathbf{T}}$, respectively, whereas that of
the gauge-ground Kitaev truncated octahedron is the $\mathbb{Z}_2$-gauge extension of its
full point symmetry group, $\widetilde{\mathbf{O}_{\textrm{h}}}$.
While gauged rotations $\widetilde{R}$ with $R\in\mathbf{O}$
and gauged inversions $\widetilde{I}$ with $I\in\mathbf{C}_{\textrm{i}}$
are all symmetry operations of the gauge-ground Kitaev truncated octahedron,
they are not necessarily commutable
because every gauge transformation $\varLambda(P)$ is obedient to the preceding
point symmetry operation $P$.
All the
$g^{\widetilde{\mathbf{O}}}\times g^{\widetilde{\mathbf{C}_{\textrm{i}}}}
+g^{\widetilde{\mathbf{C}_{\textrm{i}}}}\times g^{\widetilde{\mathbf{O}}}
=384$
products between the $g^{\widetilde{\mathbf{O}}}$ elements of $\widetilde{\mathbf{O}}$ and
the $g^{\widetilde{\mathbf{C}_{\textrm{i}}}}$ elements of $\widetilde{\mathbf{C}_{\textrm{i}}}$
are indeed symmetry operations of the gauge-ground Kitaev truncated octahedron,
but they quadruply count the $g^{\widetilde{\mathbf{O}_{\textrm{h}}}}=96$ elements of
$\widetilde{\mathbf{O}_{\textrm{h}}}
=\widetilde{\mathbf{O}}+\overline{I}\widetilde{\mathbf{O}}$.
Note further that the symmetry group of the gauge-ground Kitaev truncated octahedron
is different from that of half-integral spins in an octahedral environment,
$\widetilde{\mathbf{O}}\times\mathbf{C}_{\textrm{i}}$ (cf. Appendix \ref{AS:ChTirrepsP&tildeP}),
where $\widetilde{\mathbf{O}}\subset\mathrm{SU}(2)$, being a double covering group for
the pure rotation group $\mathbf{O}\subset\mathrm{SO}(3)$, commutes with
$\mathbf{C}_{\textrm{i}}$ because inversion has no effect on any angular momentum \cite{D2008}.
\begin{figure*}
\centering
\includegraphics[width=176mm]{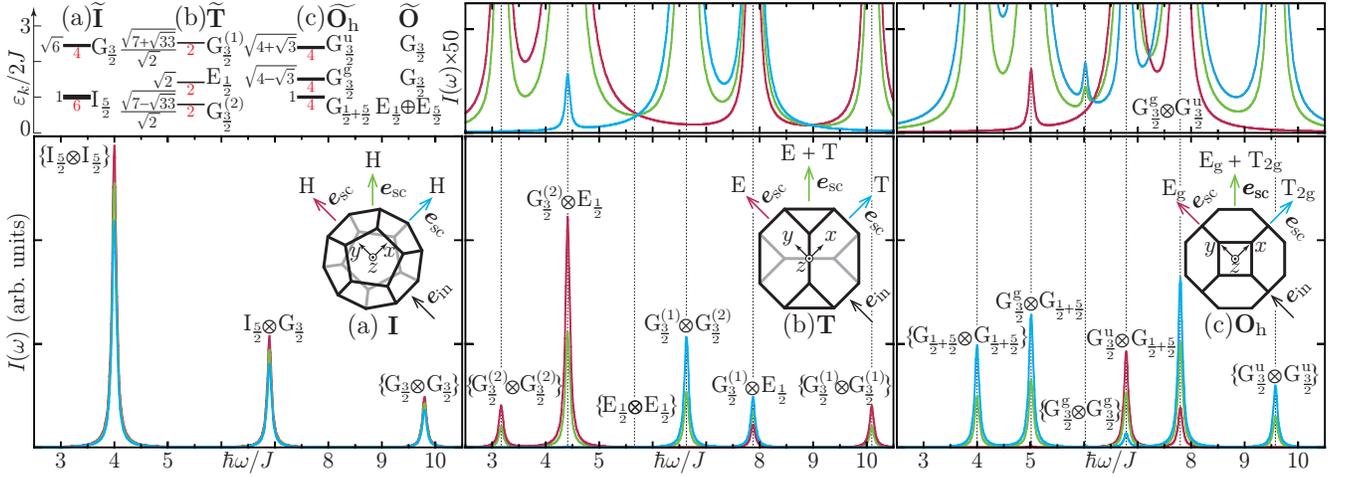}
\caption{(Color online)
         Spinon excitation energies $\varepsilon_k$ and Raman intensities $I(\omega)$ of
         Kitaev spin balls consisting of
         dodecahedral (a), truncated-tetrahedral (b), and truncated-octahedral (c) lattices
         in their ground flux configurations,
         where $\delta$-function peaks are Lorentzian-broadened by $0.05J$
         \cite{Z014403}.
         The eigenenergy, multiplicity, and irreducible representation are specified beside each
         eigenlevel.
         For the incident polarization $(\frac{\pi}{2},\frac{\pi}{2})$, we observe various
         scattered polarizations $(\frac{\pi}{2},\frac{l\pi}{4})\,(l=0,1,2)$, each
         consisting of peaks attributable to direct-product representations of the projective
         symmetry groups
         $\widetilde{\mathbf{I}}$, $\widetilde{\mathbf{T}}$, and
         $\widetilde{\mathbf{O}_{\mathrm{h}}}$
         ($\widetilde{\varXi}_i\otimes\widetilde{\varXi}_j$ in Table \ref{T:DPRintoIrrep})
         on one hand and
         containing one or more irreducible representations of the point symmetry groups
         $\mathbf{I}$, $\mathbf{T}$, and $\mathbf{O}_{\mathrm{h}}$
         ($\bigoplus_k\varXi_k$ in Table \ref{T:DPRintoIrrep})
         on the other hand.}
\label{F:Ek&I(w)}
\end{figure*}

   We are now in a position to construct the double group
$\widetilde{\mathbf{O}_{\mathrm{h}}}$.
The $48$ elements of $\mathbf{O}_{\mathrm{h}}$ divide into $10$ classes:
$\{E\}$,
$\{6C_4\}$,
$\{3C_2\}$,
$\{6C_2'\}$,
$\{8C_3\}$,
$\{I\}$,
$\{6IC_4\}$,
$\{3IC_2\}$,
$\{6IC_2'\}$,
$\{8IC_3\}$;
while the $96$ elements of $\widetilde{\mathbf{O}_{\mathrm{h}}}$ divide into $13$ classes:
$\{\overline{E}\}$, $\{\underline{E}\}$,
$\{6\overline{C_4},6\underline{C_4}\}$,
$\{3\overline{C_2},3\underline{C_2}\}$,
$\{6\overline{C_2'},6\underline{C_2'}\}$,
$\{8\overline{C_3}\}$, $\{8\underline{C_3}\}$,
$\{\overline{I},\underline{I}\}$,
$\{6\overline{IC_4},6\underline{IC_4}\}$,
$\{3\overline{IC_2},3\underline{IC_2}\}$,
$\{6\overline{IC_2'},6\underline{IC_2'}\}$,
$\{8\overline{IC_3}\}$, $\{8\underline{IC_3}\}$.
The $q$th class $\mathcal{C}_q$ ($q=1,\cdots,n_{\mathcal{C}}^{\widetilde{\mathbf{P}}}$) of
$\widetilde{\mathbf{P}}$ is generally obtained by gauging point symmetry operations
of the same type to yield $h_q$ elements in such ways that
$\{h_q\overline{P}_q\}$,
$\{h_q\underline{P}_q\}$, or
$\{\frac{h_q}{2}\overline{P}_q,\frac{h_q}{2}\underline{P}_q\}$.
Let us denote the $i$th irreducible representation of $\mathbf{P}$ ($\widetilde{\mathbf{P}}$)
by $\varXi_i$ ($\widetilde{\varXi}_i$) and its dimensionality by
$d_{\varXi_i}^{\mathbf{P}}$ ($d_{\widetilde{\varXi}_i}^{\widetilde{\mathbf{P}}}$).
Having in mind that all the single-valued irreducible representations of
$\mathbf{O}_{\mathrm{h}}$ remain unchanged in
$\widetilde{\mathbf{O}_{\mathrm{h}}}$,
$\widetilde{\varXi}_i=\varXi_i\,(i=1,\cdots,10)$,
we compare
\begin{align}
   \sum_{i=1}
       ^{n_{\mathcal{C}}^{\mathbf{O}_{\mathrm{h}}}\equiv 10}
   \left(d_{\varXi_i}^{\mathbf{O}_{\mathrm{h}}}\right)^2
  =g^{\mathbf{O}_{\mathrm{h}}},\ 
   \sum_{i=1}
       ^{n_{\mathcal{C}}^{\widetilde{\mathbf{O}_{\mathrm{h}}}}\equiv 13}
   \left(d_{\widetilde{\varXi}_i}^{\widetilde{\mathbf{O}_{\mathrm{h}}}}\right)^2
  =g^{\widetilde{\mathbf{O}_{\mathrm{h}}}}
\end{align}
to reveal that the three double-valued irreducible representations of
$\widetilde{\mathbf{O}_{\mathrm{h}}}$ have the same dimensionality,
$d_{\widetilde{\varXi}_i}^{\widetilde{\mathbf{O}_{\mathrm{h}}}}=4\,(i=11,12,13)$.
Since their characters satisfy
$\chi_{\widetilde{\varXi}_i}^{\widetilde{\mathbf{O}_{\mathrm{h}}}}(\overline{P})
=-\chi_{\widetilde{\varXi}_i}^{\widetilde{\mathbf{O}_{\mathrm{h}}}}(\underline{P})$,
we readily find that
$\chi_{\widetilde{\varXi}_i}^{\widetilde{\mathbf{O}_{\mathrm{h}}}}(\overline{E})
=-\chi_{\widetilde{\varXi}_i}^{\widetilde{\mathbf{O}_{\mathrm{h}}}}(\underline{E})
=4$ and
$\chi_{\widetilde{\varXi}_i}^{\widetilde{\mathbf{O}_{\mathrm{h}}}}(\widetilde{P})
=0\,(P=C_4,C_2, C_2',I,IC_4,IC_2,IC_2')$,
while the rest
$\chi_{\widetilde{\varXi}_i}^{\widetilde{\mathbf{O}_{\mathrm{h}}}}(\widetilde{P})\,(P=C_3,IC_3)$
are obtainable through the first orthogonality relation
(cf. Appendix \ref{AS:ChTirrepsP&tildeP})
\begin{align}
   \sum_{\widetilde{P}\in\widetilde{\mathbf{O}_{\mathrm{h}}}}
   \chi_{\widetilde{\varXi}_i}^{\widetilde{\mathbf{O}_{\mathrm{h}}}}(\widetilde{P})^*
   \chi_{\widetilde{\varXi}_j}^{\widetilde{\mathbf{O}_{\mathrm{h}}}}(\widetilde{P})
  =g^{\widetilde{\mathbf{O}_{\mathrm{h}}}}
   \delta_{ij}.
\end{align}
We name the thus-obtained double-valued irreducible representations
$\mathrm{G}_{\frac{3}{2}}^{\mathrm{g}}$,
$\mathrm{G}_{\frac{3}{2}}^{\mathrm{u}}$, and
$\mathrm{G}_{\frac{1}{2}+\frac{5}{2}}$
so that they signify the gerade- or ungerade-like response to a gauged point symmetry operation
as well as suggest the compatibility relations between $\widetilde{\mathbf{O}_{\mathrm{h}}}$ and
its subgroup
$\widetilde{\mathbf{O}}$,
$\mathrm{G}_{\frac{1}{2}+\frac{5}{2}}
 \downarrow\widetilde{\mathbf{O}}
=\mathrm{E}_{\frac{1}{2}}\oplus\mathrm{E}_{\frac{5}{2}}$
and
$\mathrm{G}_{\frac{3}{2}}^{\mathrm{g}}\downarrow\widetilde{\mathbf{O}}
=\mathrm{G}_{\frac{3}{2}}^{\mathrm{u}}\downarrow\widetilde{\mathbf{O}}
=\mathrm{G}_{\frac{3}{2}}$,
i.e.,
\begin{align}
   &
   \chi_{\mathrm{G}_{\frac{1}{2}+\frac{5}{2}}}^{\widetilde{\mathbf{O}_{\mathrm{h}}}}
   (\widetilde{P})
  =\chi_{\mathrm{E}_{\frac{1}{2}}}^{\widetilde{\mathbf{O}}}(\widetilde{P})
  +\chi_{\mathrm{E}_{\frac{5}{2}}}^{\widetilde{\mathbf{O}}}(\widetilde{P});
   \allowdisplaybreaks
   \label{E:tildeOh&tildeOcomptibility1} \\
   &
   \chi_{\mathrm{G}_{\frac{3}{2}}^{\mathrm{g}}}^{\widetilde{\mathbf{O}_{\mathrm{h}}}}
   (\widetilde{P})
  =\chi_{\mathrm{G}_{\frac{3}{2}}}^{\widetilde{\mathbf{O}}}(\widetilde{P});\ 
   \chi_{\mathrm{G}_{\frac{3}{2}}^{\mathrm{g}}}^{\widetilde{\mathbf{O}_{\mathrm{h}}}}
   (\widetilde{IC_3})
  =\sqrt{3}
   \chi_{\mathrm{G}_{\frac{3}{2}}^{\mathrm{g}}}^{\widetilde{\mathbf{O}_{\mathrm{h}}}}
   (\widetilde{C_3}),
   \allowdisplaybreaks
   \label{E:tildeOh&tildeOcomptibility2} \\
   &
   \chi_{\mathrm{G}_{\frac{3}{2}}^{\mathrm{u}}}^{\widetilde{\mathbf{O}_{\mathrm{h}}}}
   (\widetilde{P})
  =\chi_{\mathrm{G}_{\frac{3}{2}}}^{\widetilde{\mathbf{O}}}(\widetilde{P});\ 
   \chi_{\mathrm{G}_{\frac{3}{2}}^{\mathrm{u}}}^{\widetilde{\mathbf{O}_{\mathrm{h}}}}
   (\widetilde{IC_3})
 =-\sqrt{3}
   \chi_{\mathrm{G}_{\frac{3}{2}}^{\mathrm{u}}}^{\widetilde{\mathbf{O}_{\mathrm{h}}}}
   (\widetilde{C_3}).
   \label{E:tildeOh&tildeOcomptibility3}
\end{align}
The Majorana spinon spectrum of the gauge-ground $\widetilde{\mathbf{O}_{\mathrm{h}}}$
Kitaev polyhedron thus consists of three quadruplets
$3\times 4=L/2$
[see Fig. \ref{F:Ek&I(w)} together with Eq. (\ref{E:MajoranaHdiag})].
If we employ $\widetilde{\mathbf{O}}$ \cite{M041103} in this context, we have two doublets, 
$\mathrm{E}_{\frac{1}{2}}$ and $\mathrm{E}_{\frac{5}{2}}$, instead of the quadruplet
$\mathrm{G}_{\frac{1}{2}+\frac{5}{2}}$, and they look \textit{accidentally} degenerate with
each other.
Only the full symmetry group $\widetilde{\mathbf{O}_{\mathrm{h}}}$ can reveal the necessary
quadruplet.
All the $L/2$ Majorana spinon eigenmodes of the gauge-ground
$\widetilde{\mathbf{I}}$ and $\widetilde{\mathbf{T}}$ Kitaev polyhedra are also describable with
double-valued irreducible representations of their projective symmetry groups
[see Fig. \ref{F:Ek&I(w)} together with Eq. (\ref{E:MajoranaHdiag})].
The former consist of a sextuplet of $\textrm{I}_{\frac{5}{2}}$ and a quadruplet of
$\textrm{G}_{\frac{3}{2}}$, while the latter consist of three doublets of
$\textrm{G}_{\frac{3}{2}}^{(1)}$, $\textrm{G}_{\frac{3}{2}}^{(2)}$, and
$\textrm{E}_{\frac{1}{2}}$, where the $4$-dimensional real irreducible representation
$\textrm{G}_{\frac{3}{2}}$ splits into the $2$-dimensional complex ones
$\textrm{G}_{\frac{3}{2}}^{(1)}$ and $\textrm{G}_{\frac{3}{2}}^{(2)}$
due to the pure imaginary Hamiltonian (\ref{E:MajoranaHdiag}).
Irreducible representations of the double groups $\widetilde{\mathbf{I}}$,
$\widetilde{\mathbf{T}}$, $\widetilde{\mathbf{O}}$, and $\widetilde{\mathbf{O}_{\mathrm{h}}}$
are analyzed in further detail and listed with their characters in
Appendix \ref{AS:ChTirrepsP&tildeP}.

\subsection{Direct-product representations}

   Direct-product representations of a nonabelian group are not necessarily irreducible even though
they are made of irreducible representations.
Those of a projective symmetry group $\widetilde{\mathbf{P}}$ are generally decomposed into
irreducible representations of $\widetilde{\mathbf{P}}$,
\begin{align}
   &
   \widetilde{\varXi}_{i}\otimes\widetilde{\varXi}_j
  =\frac{1}{g^{\widetilde{\mathbf{P}}}}
   \mathop{\bigoplus}_{k=1}^{n_{\mathcal{C}}^{\widetilde{\mathbf{P}}}}\widetilde{\varXi}_k
   \sum_{q=1}^{n_{\mathcal{C}}^{\widetilde{\mathbf{P}}}}
   h_q
   \chi_{\widetilde{\varXi}_k}^{\widetilde{\mathbf{P}}}(\mathcal{C}_{q})^*
   \chi_{\widetilde{\varXi}_i\otimes\widetilde{\varXi}_j}^{\widetilde{\mathbf{P}}}
   (\mathcal{C}_{q});
   \nonumber \\
   &
   \chi_{\widetilde{\varXi}_i\otimes\widetilde{\varXi}_j}^{\widetilde{\mathbf{P}}}(\widetilde{P})
  =\chi_{\widetilde{\varXi}_i}^{\widetilde{\mathbf{P}}}(\widetilde{P})
   \chi_{\widetilde{\varXi}_j}^{\widetilde{\mathbf{P}}}(\widetilde{P}).
\end{align}
Two-spinon-mediated Raman scatterings in a Kitaev QSL are generally labeled with
direct-product representations of its projective symmetry group
$\widetilde{\mathbf{P}}$, $\widetilde{\varXi}_i\otimes\widetilde{\varXi}_j\,
(i,j=n_{\mathcal{C}}^{\mathbf{P}}+1,\cdots,n_{\mathcal{C}}^{\widetilde{\mathbf{P}}})$,
each decomposable into single-valued irreducible representations of the corresponding
point symmetry group $\mathbf{P}$,
\begin{align}
   \widetilde{\varXi}_{i}\otimes\widetilde{\varXi}_j
  =\mathop{\bigoplus}_{k=1}^{n_{\mathcal{C}}^{\mathbf{P}}}\varXi_k
   \sum_{q=1}^{n_{\mathcal{C}}^{\widetilde{\mathbf{P}}}}
      \frac{h_q}{g^{\widetilde{\mathbf{P}}}}
   \chi_{\varXi_k}^{\mathbf{P}}(\mathcal{C}_{q})^*
   \chi_{\widetilde{\varXi}_i\otimes\widetilde{\varXi}_j}^{\widetilde{\mathbf{P}}}
   (\mathcal{C}_{q}),
   \label{E:DPR}
\end{align}
having in mind that
\begin{align}
   \chi_{\widetilde{\varXi}_i\otimes\widetilde{\varXi}_j}^{\widetilde{\mathbf{P}}}(\overline{P})
  =\chi_{\widetilde{\varXi}_i\otimes\widetilde{\varXi}_j}^{\widetilde{\mathbf{P}}}(\underline{P})\ 
   (i,j=n_{\mathcal{C}}^{\mathbf{P}}+1,\cdots,n_{\mathcal{C}}^{\widetilde{\mathbf{P}}}).
\end{align}
Direct-product representations made of the two same irreducible representations further
decompose into symmetric and antisymmetric direct-product representations,
\begin{align}
   &
   \widetilde{\varXi}_i\otimes\widetilde{\varXi}_i
  =[ \widetilde{\varXi}_i\otimes\widetilde{\varXi}_i ]
  \oplus
   \{\widetilde{\varXi}_i\otimes\widetilde{\varXi}_i\}
  \equiv\mathop{\bigoplus}_{\sigma=\pm}
   (\widetilde{\varXi}_i\otimes\widetilde{\varXi}_i)_\sigma,
   \nonumber \\[-2mm]
   &
   (\widetilde{\varXi}_i\otimes\widetilde{\varXi}_i)_\pm
  =\mathop{\bigoplus}_{k=1}^{n_{\mathcal{C}}^{\mathbf{P}}}(\varXi_k)_\pm
   \sum_{q=1}^{n_{\mathcal{C}}^{\widetilde{\mathbf{P}}}}
      \frac{h_{q}}{g^{\widetilde{\mathbf{P}}}}
   \chi_{\varXi_{k}}^{\mathbf{P}}(\mathcal{C}_{q})^*
   \chi_{(\widetilde{\varXi}_{i}\otimes\widetilde{\varXi}_{i})_\pm}^{\widetilde{\mathbf{P}}}
   (\mathcal{C}_{q});
   \allowdisplaybreaks
   \nonumber \\
   &
   \chi_{(\widetilde{\varXi}_i\otimes\widetilde{\varXi}_i)_\pm}^{\widetilde{\mathbf{P}}}
   (\widetilde{P})
  =\frac{1}{2}
   \left[
    \chi_{\widetilde{\varXi}_i}^{\widetilde{\mathbf{P}}}(\widetilde{P})^2
   \pm
    \chi_{\widetilde{\varXi}_i}^{\widetilde{\mathbf{P}}}(\widetilde{P}^2)
   \right].
   \label{E:symDPR}
\end{align}
Spinon-geminate-excitation-relevant direct-product representations of the double groups
$\widetilde{\mathbf{I}}$, $\widetilde{\mathbf{T}}$, $\widetilde{\mathbf{O}}$,
and $\widetilde{\mathbf{O}_{\mathrm{h}}}$ are listed with their containing
single-valued irreducible representations of the corresponding point symmetry groups in
Table \ref{T:DPRintoIrrep} and with further details, including their characters, in
Appendix \ref{AS:ChTDPrepstildeP}.
\begin{table}
\centering
\caption{Spinon-geminate-excitation-relevant direct-product representations
         made of double-valued irreducible representations
         $\widetilde{\varXi}_i\otimes\widetilde{\varXi}_j$ and
         their decompositions into single-valued irreducible representations
         $\widetilde{\varXi}_k$,
         which are doubly or singly underlined when they are relevant to
         inelastic (Raman) or elastic (Rayleigh) scatterings,
         for various double groups $\widetilde{\mathbf{P}}$.
         Note that $\widetilde{\varXi}_k$ of $\widetilde{\mathbf{P}}$ is nothing but
         $\varXi_k$ of $\mathbf{P}$.}
\begin{tabular}{crcll}
\hline \hline
$\widetilde{\mathbf{P}}$               & \multicolumn{3}{c}
                                         {$\widetilde{\varXi}_i\otimes\widetilde{\varXi}_j$}
                                       & $\bigoplus_k\widetilde{\varXi}_k
                                         =\bigoplus_k\varXi_k$ \hfill \\
\hline
\raisebox{-5mm}[0mm][0mm]{$\widetilde{\mathbf{I}}$}
                                       & $\{\mathrm{I}_{\frac{5}{2}}$
                                       & \hspace{-2.7mm} $\otimes$ \hspace{-2.4mm}
                                       & $\mathrm{I}_{\frac{5}{2}}\}$
                                       & $\{\underline{\mathrm{A}}\}\!\oplus\!
                                          \{\mathrm{G}\}\!\oplus\!
                                          2\{\underline{\underline{\mathrm{H}}}\}$
                                       \\
                                       & $\mathrm{I}_{\frac{5}{2}}$
                                       & \hspace{-2.7mm} $\otimes$ \hspace{-2.4mm}
                                       & $\mathrm{G}_{\frac{3}{2}}$
                                       & $\mathrm{T}_{1}\!\oplus\!
                                          \mathrm{T}_{2}\!\oplus\!
                                         2\mathrm{G}\!\oplus\!
                                          2\underline{\underline{\mathrm{H}}}$   \\
                                       & $\{\mathrm{G}_{\frac{3}{2}}$
                                       & \hspace{-2.7mm} $\otimes$ \hspace{-2.4mm} 
                                       & $\mathrm{G}_{\frac{3}{2}}\}$
                                       & $\{\underline{\mathrm{A}}\}\!\oplus\!
                                          \{\underline{\underline{\mathrm{H}}}\}$   \\[1.5mm]
\hline
\raisebox{-14.5mm}[0mm][0mm]{$\widetilde{\mathbf{T}}$}              
                                       & $\{\mathrm{G}_{\frac{3}{2}}^{(2)}$
                                       & \hspace{-2.7mm} $\otimes$ \hspace{-2.4mm}
                                       & $\mathrm{G}_{\frac{3}{2}}^{(2)}\}$
                                       & $\{\underline{\underline{\mathrm{E}^{(1)}}}\}$   \\
                                       & $\mathrm{G}_{\frac{3}{2}}^{(2)}$
                                       & \hspace{-2.7mm} $\otimes$ \hspace{-2.4mm}
                                       & $\mathrm{E}_{\frac{1}{2}}$
                                       & $\underline{\underline{\mathrm{E}^{(2)}}}\!\oplus\!
                                          \underline{\underline{\mathrm{T}}}$   \\
                                       & $\{\mathrm{E}_{\frac{1}{2}}$
                                       & \hspace{-2.7mm} $\otimes$ \hspace{-2.4mm}
                                       & $\mathrm{E}_{\frac{1}{2}}\}$
                                       & $\{\underline{\mathrm{A}}\}$   \\
                                       & $\mathrm{G}_{\frac{3}{2}}^{(1)}$
                                       & \hspace{-2.7mm} $\otimes$ \hspace{-2.4mm}
                                       & $\mathrm{G}_{\frac{3}{2}}^{(2)}$
                                       & $\underline{\mathrm{A}}\!\oplus\!
                                          \underline{\underline{\mathrm{T}}}$   \\
                                       & $\mathrm{G}_{\frac{3}{2}}^{(1)}$
                                       & \hspace{-2.7mm} $\otimes$ \hspace{-2.4mm}
                                       & $\mathrm{E}_{\frac{1}{2}}$
                                       & $\underline{\underline{\mathrm{E}^{(1)}}}\!\oplus\!
                                          \underline{\underline{\mathrm{T}}}$   \\
                                       & $\{\mathrm{G}_{\frac{3}{2}}^{(1)}$
                                       & \hspace{-2.7mm} $\otimes$ \hspace{-2.4mm}
                                       & $\mathrm{G}_{\frac{3}{2}}^{(1)}\}$
                                       & $\{\underline{\underline{\mathrm{E}^{(2)}}}\}$  \\[2mm]
\hline
\raisebox{-13mm}[0mm][0mm]{$\widetilde{\mathbf{O}}$}
                                       & $\{\mathrm{E}_{\frac{1}{2}}$
                                       & \hspace{-2.7mm} $\otimes$ \hspace{-2.4mm}
                                       & $\mathrm{E}_{\frac{1}{2}}\}$
                                       & $\{\underline{\mathrm{A_{1}}}\}$   \\
                                       & $\mathrm{E}_{\frac{1}{2}}$
                                       & \hspace{-2.7mm} $\otimes$ \hspace{-2.4mm}
                                       & $\mathrm{E}_{\frac{5}{2}}$
                                       & $\mathrm{A_{2}}\!\oplus\!
                                          \underline{\underline{\mathrm{T_{2}}}}$   \\
                                       & $\{\mathrm{E}_{\frac{5}{2}}$
                                       & \hspace{-2.7mm} $\otimes$ \hspace{-2.4mm}
                                       & $\mathrm{E}_{\frac{5}{2}}\}$
                                       & $\{\underline{\mathrm{A_{1}}}\}$   \\
                                       & $\mathrm{G}_{\frac{3}{2}}$
                                       & \hspace{-2.7mm} $\otimes$ \hspace{-2.4mm}
                                       & $\mathrm{E}_{\frac{1}{2}}$
                                       & $\underline{\underline{\mathrm{E}}}\!\oplus\!
                                          \mathrm{T_{1}}\!\oplus\!
                                          \underline{\underline{\mathrm{T_{2}}}}$\\
                                       & $\mathrm{G}_{\frac{3}{2}}$
                                       & \hspace{-2.7mm} $\otimes$ \hspace{-2.4mm}
                                       & $\mathrm{E}_{\frac{5}{2}}$
                                       & $\underline{\underline{\mathrm{E}}}\!\oplus\!
                                          \mathrm{T_{1}}\!\oplus\!
                                          \underline{\underline{\mathrm{T_{2}}}}$   \\
                                       & $\mathrm{G}_{\frac{3}{2}}$
                                       & \hspace{-2.7mm} $\otimes$ \hspace{-2.4mm}
                                       & $\mathrm{G}_{\frac{3}{2}}$
                                       & $\{\underline{\mathrm{A_{1}}}\}\!\oplus\!
                                          [\mathrm{A_{2}}]\!\oplus\!
                                          \{\underline{\underline{\mathrm{E}}}\}\!\oplus\!
                                         2[\mathrm{T_{1}}]\!\oplus\!
                                          [\underline{\underline{\mathrm{T_{2}}}}]\!\oplus\!
                                          \{\underline{\underline{\mathrm{T_{2}}}}\}\!$   \\[1.5mm]
\hline
\raisebox{-14mm}[0mm][0mm]{$\widetilde{\mathbf{O}_{\mathrm{h}}}$} 
                                       & $\{\mathrm{G}_{\frac{1}{2}+\frac{5}{2}}$
                                       & \hspace{-2.7mm} $\otimes$ \hspace{-2.4mm}
                                       & $\mathrm{G}_{\frac{1}{2}+\frac{5}{2}}\}$
                                       & $\{\underline{\mathrm{A_{1g}}}\}\!\oplus\!
                                          \{\mathrm{A_{1u}}\}\!\oplus\!
                                          \{\mathrm{A_{2u}}\}\!\oplus\!
                                          \{\underline{\underline{\mathrm{T_{2g}}}}\}$   \\
                                       & $\mathrm{G}_{\frac{3}{2}}^{\mathrm{g}}$
                                       & \hspace{-2.7mm} $\otimes$ \hspace{-2.4mm}
                                       & $\mathrm{G}_{\frac{1}{2}+\frac{5}{2}}$
                                       & $\underline{\underline{\mathrm{E_{g}}}}\!\oplus\!
                                          \mathrm{E_{u}}\!\oplus\!
                                          \mathrm{T_{1g}}\!\oplus\!
                                          \mathrm{T_{1u}}\!\oplus\!
                                          \underline{\underline{\mathrm{T_{2g}}}}\!\oplus\!
                                          \mathrm{T_{2u}}$   \\
                                       & $\{\mathrm{G}_{\frac{3}{2}}^{\mathrm{g}}$
                                       & \hspace{-2.7mm} $\otimes$ \hspace{-2.4mm}
                                       & $\mathrm{G}_{\frac{3}{2}}^{\mathrm{g}}\}$
                                       & $\{\underline{\mathrm{A_{1g}}}\}\!\oplus\!
                                          \{\mathrm{E_{u}}\}\!\oplus\!
                                          \{\underline{\underline{\mathrm{T_{2g}}}}\}$   \\
                                       & $\mathrm{G}_{\frac{3}{2}}^{\mathrm{u}}$
                                       & \hspace{-2.7mm} $\otimes$ \hspace{-2.4mm}
                                       & $\mathrm{G}_{\frac{1}{2}+\frac{5}{2}}$
                                       & $\underline{\underline{\mathrm{E_{g}}}}\!\oplus\!
                                          \mathrm{E_{u}}\!\oplus\!
                                          \mathrm{T_{1g}}\!\oplus\!
                                          \mathrm{T_{1u}}\!\oplus\!
                                          \underline{\underline{\mathrm{T_{2g}}}}\!\oplus\!
                                          \mathrm{T_{2u}}$   \\
                                       & $\mathrm{G}_{\frac{3}{2}}^{\mathrm{g}}$
                                       & \hspace{-2.7mm} $\otimes$ \hspace{-2.4mm}
                                       & $\mathrm{G}_{\frac{3}{2}}^{\mathrm{u}}$
                                       & $\mathrm{A_{1u}}\!\oplus\!
                                          \mathrm{A_{2u}}\!\oplus\!
                                          \underline{\underline{\mathrm{E_{g}}}}\!\oplus\!
                                          \mathrm{T_{1g}}\!\oplus\!
                                          \mathrm{T_{1u}}\!\oplus\!
                                          \underline{\underline{\mathrm{T_{2g}}}}\!\oplus\!
                                          \mathrm{T_{2u}}\!\!\!$   \\
                                       & $\{\mathrm{G}_{\frac{3}{2}}^{\mathrm{u}}$
                                       & \hspace{-2.7mm} $\otimes$ \hspace{-2.4mm}
                                       & $\mathrm{G}_{\frac{3}{2}}^{\mathrm{u}}\}$
                                       & $\{\underline{\mathrm{A_{1g}}}\}\!\oplus\!
                                          \{\mathrm{E_{u}}\}\!\oplus\!
                                          \{\underline{\underline{\mathrm{T_{2g}}}}\}$   \\[2mm]
\hline\hline
\end{tabular}
\label{T:DPRintoIrrep}
\end{table}

\section{Raman Intensity Profiles}
\label{S:RIP}
\subsection{Point-symmetry argument}

   Within the LF theory \cite{F514,S1068,S365}, the Raman scattering intensity at absolute zero
reads \cite{K187201,P094439}
\begin{align}
   I(\omega)
   &
  =\frac{1}{2\pi\hbar L}
   \int_{-\infty}^\infty
   \langle 0|
    e^{\frac{{i}\mathscr{H}t}{\hbar}}
    \mathscr{R}
    e^{-\frac{{i}\mathscr{H}t}{\hbar}}
    \mathscr{R}
   |0\rangle
   e^{i\omega t}dt;
   \allowdisplaybreaks
   \nonumber \\
   \mathscr{R}
   &
  \equiv
  -J\sum_{\lambda=x,y,z}\sum_{<m,n>_{\lambda}}
   (\bm{e}_{\mathrm{in}}\cdot\bm{d}_{mn})
   (\bm{e}_{\mathrm{sc}}\cdot\bm{d}_{mn})
   \sigma_m^\lambda\sigma_n^\lambda
   \allowdisplaybreaks
   \nonumber \\
   &
 =iJ\sum_{\lambda=x,y,z}\sum_{<m,n>_{\lambda}}
   (\bm{e}_{\mathrm{in}}\cdot\bm{d}_{mn})
   (\bm{e}_{\mathrm{sc}}\cdot\bm{d}_{mn})
   \allowdisplaybreaks
   \nonumber \\
   &\quad
   \times
   \hat{u}_{<m,n>_{\lambda}}c_m c_n,
   \label{E:I(w)LF}
\end{align}
where
$\bm{e}_{\mathrm{in}}
\equiv
(\sin\vartheta_{\mathrm{in}}\cos\varphi_{\mathrm{in}},
 \sin\vartheta_{\mathrm{in}}\sin\varphi_{\mathrm{in}},
 \cos\vartheta_{\mathrm{in}})$ and
$\bm{e}_{\mathrm{sc}}
\equiv
(\sin\vartheta_{\mathrm{sc}}\cos\varphi_{\mathrm{sc}},
 \sin\vartheta_{\mathrm{sc}}\sin\varphi_{\mathrm{sc}},
 \cos\vartheta_{\mathrm{sc}})$
are the polarization vectors of incident and scattered lights, respectively,
while $\bm{d}_{mn}\equiv\bm{r}_m-\bm{r}_n$ are
the lattice vectors with $\bm{r}_m$ and $\bm{r}_n$ being the positions of neighboring sites.
When the ground state belongs to the double group $\widetilde{\mathbf{P}}$
\cite{SGof|0>},
it is useful to write the Raman operator (\ref{E:I(w)LF}) as \cite{D175}
\begin{align}
   \mathscr{R}
   {
  ={\sum_i}'
   \sum_{\mu=1}^{d_{\widetilde{\varXi}_i}^{\widetilde{\mathbf{P}}}}
   E_{\widetilde{\varXi}_i:\mu}^{\widetilde{\mathbf{P}}}
   \mathcal{R}_{\widetilde{\varXi}_i:\mu}^{\widetilde{\mathbf{P}}}
   }
  =\mathop{{\sum_i}'}
   \sum_{\mu=1}^{d_{\varXi_i}^{\mathbf{P}}}
   E_{\varXi_i:\mu}^{\mathbf{P}}
   \mathcal{R}_{\varXi_i:\mu}^{\mathbf{P}},
   \label{E:Rirrep}
\end{align}
where $\sum_i'$ runs over the LF-active irreducible representations $\widetilde{\varXi}_i$ of
$\widetilde{\mathbf{P}}$,
which are necessarily \textit{real} and \textit{single-valued} and
therefore equal to the irreducible representations $\varXi_i$ of the corresponding point
symmetry group $\mathbf{P}$,
and
$E_{\widetilde{\varXi}_i:\mu}^{\widetilde{\mathbf{P}}}$
($E_{\varXi_i:\mu}^{\mathbf{P}}$) and
$\mathcal{R}_{\widetilde{\varXi}_i:\mu}^{\widetilde{\mathbf{P}}}$
($\mathcal{R}_{\varXi_i:\mu}^{\mathbf{P}}$)
are the $\mu$th polarization-vector basis function and LF vertex for
$\widetilde{\varXi}_i$ ($\varXi_i$), respectively,
both of which are explicitly given in Appendix \ref{AS:RamanPD}.
Within the LF formulation, the nonvanishing vertices read
$\mathcal{R}_{\mathrm{A}:\mu}^{\mathbf{I}}$,
$\mathcal{R}_{\mathrm{H}:\mu}^{\mathbf{I}}$
for the dodecahedron,
$\mathcal{R}_{\mathrm{A}:\mu}^{\mathbf{T}}$,
$\mathcal{R}_{\mathrm{E}:\mu}^{\mathbf{T}}$,
$\mathcal{R}_{\mathrm{T}:\mu}^{\mathbf{T}}$
for the truncated tetrahedron,
$\mathcal{R}_{\mathrm{A}_{1\mathrm{g}}:\mu}^{\mathbf{O}_{\mathrm{h}}}$,
$\mathcal{R}_{\mathrm{E}_{\mathrm{g}}:\mu}^{\mathbf{O}_{\mathrm{h}}}$,
$\mathcal{R}_{\mathrm{T}_{2\mathrm{g}}:\mu}^{\mathbf{O}_{\mathrm{h}}}$
for the truncated octahedron, and, for reference,
$\mathcal{R}_{\mathrm{A}_1:\mu}^{\mathbf{C}_{6\mathrm{v}}}$,
$\mathcal{R}_{\mathrm{E}_2:\mu}^{\mathbf{C}_{6\mathrm{v}}}$
in two-dimensional lattices of triangular geometry \cite{P094439,C172406,K024414,P174412}.
In the spherical lattice geometry realized by Platonic and Archimedean polyhedra,
all the vertices of the identity representation, such as 
$\mathcal{R}_{\mathrm{A}:\mu}^{\mathbf{I}}$,
$\mathcal{R}_{\mathrm{A}:\mu}^{\mathbf{T}}$, and
$\mathcal{R}_{\mathrm{A}_{1\mathrm{g}}:\mu}^{\mathbf{O}_{\mathrm{h}}}$,
commute with the corresponding Hamiltonians and therefore reduce to Rayleigh scattering.
This is the case with $\mathcal{R}_{\mathrm{A}_1:\mu}^{\mathbf{C}_{6\mathrm{v}}}$ as well.

   Since the ground state (\ref{E:|0>}) is invariant under every symmetry operation of
$\mathbf{P}$, every expectation value between Raman vertices of different symmetry species for it
goes to zero \cite{D175,K024414,P094439},
\begin{align}
   &
   \frac{1}{2\pi\hbar L}
   \int_{-\infty}^\infty
   \langle 0|
    e^{\frac{{i}\mathscr{H}t}{\hbar}}
    \mathcal{R}_{\varXi_i:\mu}^{\mathbf{P}}
    e^{-\frac{{i}\mathscr{H}t}{\hbar}}
    \mathcal{R}_{\varXi_j:\nu}^{\mathbf{P}}
   |0\rangle
   e^{i\omega t}dt
   \nonumber \\
   &\ 
  =\frac{\delta_{ij}\delta_{\mu\nu}}{2\pi\hbar L}
   \int_{-\infty}^\infty
   \langle 0|
    e^{\frac{{i}\mathscr{H}t}{\hbar}}
    \mathcal{R}_{\varXi_i:\mu}^{\mathbf{P}}
    e^{-\frac{{i}\mathscr{H}t}{\hbar}}
    \mathcal{R}_{\varXi_i:\mu}^{\mathbf{P}}
   |0\rangle
   e^{i\omega t}dt
   \nonumber \\
   &\ 
  \equiv\delta_{ij}\delta_{\mu\nu}I_{\varXi_i:\mu}^{\mathbf{P}}(\omega),
   \label{E:RiRjOrthogonality}
\end{align}
and $I_{\varXi_i:\mu}^{\mathbf{P}}(\omega)\,(\mu=1,\cdots,d_{\varXi_{i}}^{\mathbf{P}})$
no longer depend on $\mu$ \cite{K187201,R045117,P094439,C172406}.
While the Raman spectra of gauge-ground Kitaev polyhedra are analyzable with direct-product
representations of their projective symmetry groups $\widetilde{\mathbf{P}}$,
they can be classified by irreducible representations of the corresponding point symmetry
groups $\mathbf{P}$.
Substituting the irreducible decomposition of the Raman operator $\mathscr{R}$ (\ref{E:Rirrep})
into the LF expression of the Raman intensity (\ref{E:I(w)LF}) and taking account of
the spectral degeneracy within each multidimensional irreducible representation
(cf. Appendix \ref{AS:RamanPD}), we have
\begin{align}
   &
   I(\omega)
  ={\sum_i}'
   \sum_{\mu=1}^{d_{\varXi_{i}}^{\mathbf{P}}}
   \left(E_{\varXi_{i}:\mu}^{\mathbf{P}}\right)^2
   I_{\varXi_{i}:\mu}^{\mathbf{P}}(\omega)
   \allowdisplaybreaks
   \nonumber \\
   &\qquad
  ={\sum_i}'
   I_{\varXi_{i}:1}^{\mathbf{P}}(\omega)
   \sum_{\mu=1}^{d_{\varXi_{i}}^{\mathbf{P}}}
   \left(E_{\varXi_{i}:\mu}^{\mathbf{P}}\right)^2.
   \label{E:I(w)irrep}
\end{align}
Having in mind that
$[\mathcal{R}_{\varXi_i:\mu}^{\mathbf{P}},\hat{W}_p]=0$,
${}_q\langle\{W_p\}|\{W_p\}\rangle_{q'}=\delta_{qq'}$, and
$c_l|0\rangle=\sum_{k=1}^{L/2}(\psi_{l,2k-1}+i\psi_{l,2k})\alpha_k^\dagger|0\rangle$,
the LF vertex $\mathcal{R}_{\varXi_i:\mu}^{\mathbf{P}}$ evokes two spinons without any vison
(for more details refer to Appendix \ref{AS:RamanPD}),
\begin{align}
   I_{\varXi_i:\mu}^{\mathbf{P}}(\omega)
   &
  =\int_{-\infty}^\infty
   \frac{dt\,e^{i\omega t}}{2\pi\hbar L}
   \sum_{q=0}^{2^{\frac{L}{2}+1}-1}\sum_{\kappa=0}^{2^{\frac{L}{2}-1}-1}
   {}_0\langle\{W_p\}|\otimes{}_0\langle\{n_k\}|
   \allowdisplaybreaks
   \nonumber \\
   &\quad
  \times
   e^{\frac{{i}\mathscr{H}t}{\hbar}}
   \mathcal{R}_{\varXi_i:\mu}^{\mathbf{P}}
   e^{-\frac{{i}\mathscr{H}t}{\hbar}}
   |\{n_k\}\rangle_\kappa\otimes|\{W_p\}\rangle_q
   \allowdisplaybreaks
   \nonumber \\
   &\quad
  \times
   {}_q\langle\{W_p\}|\otimes{}_\kappa\langle\{n_k\}|
   \mathcal{R}_{\varXi_i:\mu}^{\mathbf{P}}
   |\{n_k\}\rangle_0\otimes|\{W_p\}\rangle_0
   \allowdisplaybreaks
   \nonumber \\
   &
  =\frac{1}{L}
   \sum_{1=k<k'=\frac{L}{2}}
   \left|
     \langle 0|\alpha_{k}\alpha_{k'}
       \mathcal{R}_{\varXi_i:\mu}^{\mathbf{P}}
    |0\rangle
   \right|^{2}
   \allowdisplaybreaks
   \nonumber \\
   &\quad
  \times
   \delta(\hbar\omega-\varepsilon_{k}-\varepsilon_{k'}).
   \label{E:I(w)LFirrep}
\end{align}
\begin{figure}

\centering
\includegraphics[width=86mm]{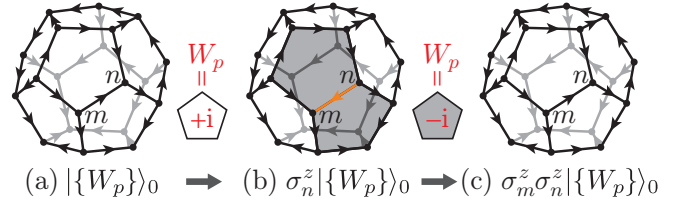}
\caption{(Color online)
         Actions of the spin operators
         $\sigma_n^z=ic_n\eta_n^z$ (b) and
         $\sigma_m^z\sigma_n^z=-i\hat{u}_{<m,n>_z}c_m c_n$ (c)
         on the gauge-ground Kitaev dodecahedron $|\{W_p\}\rangle_0$ (a)
         in the context of calculating the dynamic structure factor (\ref{E:Sqw}) and
         Raman scattering intensity (\ref{E:I(w)LFirrep}).}
\label{F:Sqw&I(w)}
\end{figure}

   We may be reminded that the above is not the case with any single spin operator.
Unlike the Raman response, visons (Fig. \ref{F:Sqw&I(w)}) as well as spinons are involved in
the dynamic spin response \cite{K207203,K115127}
\begin{align}
   &
   S^{\lambda\lambda}(\bm{q},\omega)
  =\frac{1}{2\pi\hbar L}
   \int_{-\infty}^\infty
   \sum_{m,n=1}^L
   e^{-i\bm{q}\cdot(\bm{r}_m-\bm{r}_n)}
   \nonumber \\
   &\qquad\qquad\quad\times
   \langle 0|
    e^{\frac{{i}\mathscr{H}t}{\hbar}}
    \sigma_m^\lambda
    e^{-\frac{{i}\mathscr{H}t}{\hbar}}
    \sigma_n^\lambda
   |0\rangle
   e^{i\omega t}dt
   \nonumber \\
   &\ 
  =\int_{-\infty}^\infty
   \frac{dt\,e^{i\omega t}}{2\pi\hbar L}
   \sum_{m,n=1}^L
   \sum_{q=0}^{2^{\frac{L}{2}+1}-1}\sum_{\kappa=0}^{2^{\frac{L}{2}-1}-1}
   e^{-i\bm{q}\cdot(\bm{r}_m-\bm{r}_n)}
   \nonumber \\
   &\ \times
   {}_0\langle\{W_p\}|\otimes{}_0\langle\{n_k\}|
   \allowdisplaybreaks
   e^{\frac{{i}\mathscr{H}t}{\hbar}}
   \sigma_m^\lambda
   e^{-\frac{{i}\mathscr{H}t}{\hbar}}
   |\{n'_k\}\rangle_\kappa\otimes|\{W_p\}\rangle_q
   \allowdisplaybreaks
   \nonumber \\
   &\ 
  \times
   {}_q\langle\{W_p\}|\otimes{}_\kappa\langle\{n'_k\}|
   \sigma_n^\lambda
   |\{n_k\}\rangle_0\otimes|\{W_p\}\rangle_0.
   \label{E:Sqw}
\end{align}
Indeed
${}_0\langle\{n_k\}|\alpha_k^\dagger\alpha_k|\{n_k\}\rangle_0=0$ $(k=1,\cdots,\frac{L}{2})$,
but
the spinon operator $\alpha_k$ and therefore vacuum state $|\{n_k\}\rangle_0$ depend on
the background flux configuration $|\{W_p\}\rangle_q$.
We denote those against an excited flux configuration $|\{W_p\}\rangle_{q\neq 0}$ by
$\alpha'_k$ and $|\{n'_k\}\rangle_0$ distinguishably from $\alpha_k$ and $|\{n_k\}\rangle_0$
against $|\{W_p\}\rangle_0$ in Eq. (\ref{E:Sqw}).
Since
spinons in an excited flux sector reads a linear combination of spinons in the ground flux sector,
$\alpha'_{k'}=\sum_{k=1}^{L/2}(\chi_{k',k}\alpha_k+\upsilon_{k',k}\alpha_k^\dagger)$
$(k'=1,\cdots,\frac{L}{2})$,
and their vacuum $|\{n'_k\}\rangle_0$ reads a linear combination of
the ground-flux-sector spinon vacuum and/or excited states, i.e.
either a linear combination of
$|\{n_k\}\rangle_0$,
$\alpha_{k_1}^\dagger\alpha_{k_2}^\dagger|\{n_k\}\rangle_0$,
$\alpha_{k_1}^\dagger\alpha_{k_2}^\dagger\alpha_{k_3}^\dagger\alpha_{k_4}^\dagger
 |\{n_k\}\rangle_0$,
$\cdots$
or that of
$\alpha_{k_1}^\dagger|\{n_k\}\rangle_0$,
$\alpha_{k_1}^\dagger\alpha_{k_2}^\dagger\alpha_{k_3}^\dagger|\{n_k\}\rangle_0$,
$\cdots$,
we can exactly calculate the dynamic structure factor (\ref{E:Sqw}) as well \cite{K115127}.
In higher dimensions, Eq. (\ref{E:Sqw}) is hard to calculate for sufficiently large systems,
with excited flux configurations $|\{W_p\}\rangle_{q\neq 0}$ being no longer invariant
under the primitive translation, but we can employ a Dyson equation instead
to accomplish the thermodynamic-limit calculation \cite{K207203,K115127}.

   Figure \ref{F:Ek&I(w)} shows the polarized Raman spectra of gauge-ground Kitaev spin balls
with light polarization vectors varying within the $xy$ plane.
The polarization dependence of the intensity is very weak in the dodecahedron
but significant and individual in the truncated tetrahedron and octahedron.
The former observations are similar to the case with the honeycomb Kitaev QSL \cite{K187201}.
For polarization vectors in the $xy$ plain,
$\vartheta_{\mathrm{in}}=\vartheta_{\mathrm{sc}}=\frac{\pi}{2}$ with varying
$\varphi_{\mathrm{in}}$ and $\varphi_{\mathrm{sc}}$, we have
\begin{align}
   &
   \sum_{\mu=1}^2
   \left(E_{\mathrm{E}_{2}:\mu}^{\mathbf{C}_{6\mathrm{v}}}\right)^2
  =\frac{1}{2};
   \allowdisplaybreaks
   \nonumber \\
   &
   \sum_{\mu=1}^5
   \left(E_{\mathrm{H}:\mu}^{\mathbf{I}}\right)^2
  =\frac{\cos^2(\varphi_{\mathrm{in}}-\varphi_{\mathrm{sc}})}{6}+\frac{1}{2};
   \allowdisplaybreaks
   \nonumber \\
   &
   \sum_{\mu=1}^2
   \left(E_{\mathrm{E}:\mu}^{\mathbf{T}}\right)^2
  =\sum_{\mu=1}^2
   \left(E_{\mathrm{E_{g}}:\mu}^{\mathbf{O}_{\mathrm{h}}}\right)^2
   \nonumber \\
   &\qquad\qquad\quad
  =\frac{\cos^{2}(\varphi_{\mathrm{in}}-\varphi_{\mathrm{sc}})}{6}
  +\frac{\cos^{2}(\varphi_{\mathrm{in}}+\varphi_{\mathrm{sc}})}{2},
   \allowdisplaybreaks
   \nonumber \\
   &
   \sum_{\mu=1}^3
   \left(E_{\mathrm{T}:\mu}^{\mathbf{T}}\right)^2
  =\sum_{\mu=1}^3
   \left(E_{\mathrm{T_{2g}}:\mu}^{\mathbf{O}_{\mathrm{h}}}\right)^2
  =\frac{\sin^{2}(\varphi_{\mathrm{in}}+\varphi_{\mathrm{sc}})}{2};
   \label{E:PolVecBasisFunSum}
\end{align}
hence the perfect depolarization of Raman scattering in a honeycomb QSL.
While the $\widetilde{\mathbf{I}}$ gauged dodecahedron also has one and only Raman-active
multidimensional irreducible representation and all the three relevant direct-product
representations of $\widetilde{\mathbf{I}}$ contain this
$\mathrm{H}$ mode, the sum of its five basis functions no longer reduces to a constant,
resulting in similar shapes peaked at the three fixed frequencies
$\hbar\omega/2J=2,\,1+\sqrt{6},\,2\sqrt{6}$
but different weights varying as Eq. (\ref{E:PolVecBasisFunSum})
of the polarized spectra.
The $\widetilde{\mathbf{T}}$ and $\widetilde{\mathbf{O}_{\mathrm{h}}}$ gauged polyhedra each have
two Raman-active modes to yield spectra peaking and weighing differently according to
the light polarization.
Such observations are also the case with $\widetilde{\mathbf{D}_{2\mathrm{h}}}$ harmonic
honeycomb Kitaev QSLs in three dimensions \cite{O085101,P094439}.
Full details of the polarized Raman intensity profiles of all the gauged polyhedra in question
are given in Appendix \ref{AS:RamanPD}.

\subsection{Projective-symmetry argument}

   The $\widetilde{\mathbf{T}}$ and $\widetilde{\mathbf{O}_{\mathrm{h}}}$ gauged polyhedra each
have three spinon modes to yield geminate excitations of $3+_3\!\mathrm{C}_2$ types.
There are $6$ pair-spinon-resonant frequencies in them each.
In the case of $\widetilde{\mathbf{T}}$,
one of them, $\{\mathrm{E}_{\frac{1}{2}}\otimes\mathrm{E}_{\frac{1}{2}}\}$
($\hbar\omega/2J=2\sqrt{2}$), is a Rayleigh channel,
while all the rest contain the Raman-active
$\mathrm{E}$ (detectable with $\varphi_{\mathrm{in}}\pm\varphi_{\mathrm{sc}}\neq\frac{\pi}{2}$)
and/or
$\mathrm{T}$ (detectable with $\varphi_{\mathrm{in}}+\varphi_{\mathrm{sc}}\neq 0,\pi$)
modes, where the two-dimensional real irreducible representation
$\mathrm{E}\equiv\mathrm{E}^{(1)}\oplus\mathrm{E}^{(2)}$
splits into two one-dimensional complex ones, $\mathrm{E}^{(1)}$ and $\mathrm{E}^{(2)}$,
bringing nonvanishing Raman intensities at all the six frequencies but
$\hbar\omega/2J=2\sqrt{2}$.
In the case of $\widetilde{\mathbf{O}_{\mathrm{h}}}$, all the direct-product representations
contain the Raman-active $\mathrm{T}_{2\mathrm{g}}$ mode
(detectable with $\varphi_{\mathrm{in}}+\varphi_{\mathrm{sc}}\neq 0,\pi$),
bringing nonvanishing Raman intensities at all the six frequencies.
On the other hand, only the three direct-product representations
$\mathrm{G}_{\frac{3}{2}}^{\mathrm{g}}\otimes\mathrm{G}_{\frac{1}{2}+\frac{5}{2}}$
($\hbar\omega/2J=1+\sqrt{4-\sqrt{3}}$),
$\mathrm{G}_{\frac{3}{2}}^{\mathrm{u}}\otimes\mathrm{G}_{\frac{1}{2}+\frac{5}{2}}$
($\hbar\omega/2J=1+\sqrt{4+\sqrt{3}}$), and
$\mathrm{G}_{\frac{3}{2}}^{\mathrm{g}}\otimes\mathrm{G}_{\frac{3}{2}}^{\mathrm{u}}$
($\hbar\omega/2J=\sqrt{4-\sqrt{3}}+\sqrt{4+\sqrt{3}}$)
contain another Raman-active mode $\mathrm{E}_{\mathrm{g}}$
(detectable with $\varphi_{\mathrm{in}}\pm\varphi_{\mathrm{sc}}\neq\frac{\pi}{2}$).
In this context, we should pay special attention to the geminate excitations labeled
$\{\mathrm{G}_{\frac{3}{2}}^{\mathrm{g}}\otimes\mathrm{G}_{\frac{3}{2}}^{\mathrm{g}}\}$
($\hbar\omega/2J=2\sqrt{4-\sqrt{3}}$) and
$\{\mathrm{G}_{\frac{3}{2}}^{\mathrm{u}}\otimes\mathrm{G}_{\frac{3}{2}}^{\mathrm{u}}\}$
($\hbar\omega/2J=2\sqrt{4+\sqrt{3}}$).
If we describe this gauged polyhedron in terms of $\widetilde{\mathbf{O}}$, rather than
$\widetilde{\mathbf{O}_{\mathrm{h}}}$, these two direct-product representations degenerate into
$\{\mathrm{G}_{\frac{3}{2}}\otimes\mathrm{G}_{\frac{3}{2}}\}
=\{\underline{\mathrm{A_{1}}}\}
 \oplus
 \{\underline{\underline{\mathrm{E}}}\}
 \oplus
 \{\underline{\underline{\mathrm{T_{2}}}}\}$
(see Table \ref{T:DPRintoIrrep})
to cause misunderstanding as if outgoing photons of $\varphi_{\mathrm{sc}}=\varphi_{\mathrm{in}}$
brought nonvanishing Raman intensities at the two frequencies
$\hbar\omega/2J=2\sqrt{4\mp\sqrt{3}}$ as well.
Under the pertinent $\widetilde{\mathbf{O}_{\mathrm{h}}}$ description,
the Raman intensities at the two frequencies $\hbar\omega/2J=2\sqrt{4\mp\sqrt{3}}$ in
the gauged truncated octahedron purely belongs to the $\mathrm{T}_{2\mathrm{g}}$ symmetry
species, because they are mediated by spinon geminate excitations belonging to
the direct-product representations
$\{\mathrm{G}_{\frac{3}{2}}^{\mathrm{g}}\otimes\mathrm{G}_{\frac{3}{2}}^{\mathrm{g}}\}$
and
$\{\mathrm{G}_{\frac{3}{2}}^{\mathrm{u}}\otimes\mathrm{G}_{\frac{3}{2}}^{\mathrm{u}}\}$,
both of which decompose into
$\{\underline{\mathrm{A_{1g}}}\}
 \oplus
 \{\mathrm{E_{u}}\}
 \oplus
 \{\underline{\underline{\mathrm{T_{2g}}}}\}$,
i.e.,
the Raman-active $\mathrm{T}_{2\mathrm{g}}$,
LF-Raman-inactive $\mathrm{A_{1g}}$, and
Raman-inactive $\mathrm{E_{u}}$ (instead of Raman-active $\mathrm{E_{g}}$)
symmetry species (see Table \ref{T:DPRintoIrrep}).

   In an attempt to describe partons in Kitaev truncated octahedron,
MPT \cite{M041103} restrict their symmetry argument to gauged rotations
$\widetilde{\mathbf{R}}\subset\mathrm{SU}(2)\cong\mathrm{Spin}(3)$,
i.e. double covers of pure rotation groups $\mathbf{R}\subset\mathrm{SO}(3)$,
because they employ projective symmetry groups with the aim to characterize an itinerant parton
as a charged particle in quantized orbital motion, and therefore need the isomorphism
$\mathrm{SU(2)}/\mathbb{Z}_{2}\cong\mathrm{SO}(3)$.
For partons emergent in a gauged truncated octahedron, they consider gauging the subgroup
$\mathbf{O}$ of the full octahedral group $\mathbf{O}_{\mathrm{h}}$.
On the other hand, in order to describe spinon geminate, rather than single, excitations
in the context of Raman scattering, we construct and have to construct the double cover of
$\mathbf{O}_{\mathrm{h}}\subset\mathrm{O}(3)$ \cite{T377} instead of that of
$\mathbf{O}\subset\mathrm{SO}(3)$.
It is not until we analyze the projective symmetry of Majorana spinons to the fullest extent that
we can correctly understand Raman scattering in a time-reversal-invariant gauged polyhedron.

\section{Summary and Future Aspect}
\label{S:SFA}

   Our approach to Raman observations of QSLs is feasible regardless of whatever geometry.
Kitaev nanoribbons \cite{S012046,T235434}, for instance, are describable with gauged space
groups, $\mathbf{L}\wedge\widetilde{\mathbf{P}}$,
where $\mathbf{L}$ is a one-dimensional translation group \cite{Y125124}.
Their eigenspectra are no longer discrete but consist of continuous bands.
Intraband and interband spinon geminate excitations are distinguished and identified by
light polarizations and direct-product representations of
$\mathbf{L}\wedge\widetilde{\mathbf{P}}$ \cite{Y}.

   Another extension of our approach is going beyond the LF vertices \cite{P060408,P104427}.
In the $\widetilde{\mathbf{T}}$ Kitaev spin ball, 
the direct-product representation $\{\mathrm{E}_{\frac{1}{2}}\otimes\mathrm{E}_{\frac{1}{2}}\}$
is Raman-inactive within the LF scheme (Table \ref{T:DPRintoIrrep}),
but an $\mathrm{E}_{\frac{1}{2}}$ multiple direct-product representation may become Raman active
in higher-order scatterings to visualize the Majorana spinon spectrum in a wider range.
Optical observation of partons in QSLs will be even more attractive with the language of
projective symmetry.

\acknowledgments

This work was supported by
the Ministry of Education, Culture, Sports, Science, and Technology of Japan.

\begin{widetext}

\begin{appendix}
\renewcommand{\appendixname}{APPENDIX}
\section{PROJECTIVE SYMMETRY OPERATIONS ON GAUGE-GROUND KITAEV POLYHEDRA}
\label{AS:prsym}

   Dodecahedral, truncated-tetrahedral, and truncated-octahedral lattices
belong to the point symmetry groups
$\mathbf{I}_{\mathrm{h}}$, $\mathbf{T}_{\mathrm{d}}$, and $\mathbf{O}_{\mathrm{h}}$, respectively.
We illustrate their symmetry operations with Fig. \ref{AF:ptSG}.
When we consider Kitaev models on these lattices, their free Majorana fermionic Hamiltonians with
given gauge fields are not generally invariant under the point group actions of their belonging
lattices.
Let us find gauged point symmetry operations of the ground gauge sectors of these Hamiltonians.
We illustrate symmetry operations of gauge-ground polyhedra with Fig. \ref{AF:prSG}.
Every gauge-ground Kitaev spin ball is such that
all $W_p$'s of $N_p=0$ mod $4$ are $-1$, all $W_p$'s of $N_p=2$ mod $4$ are $+1$, and
all $W_p$'s of odd $N_p$ are either of $+i$ and $-i$ \cite{M041103}.
Since the Kitaev spin Hamiltonian is time reversal invariant, its ground state is at least
doubly degenerate unless all $N_p$'s are even \cite{Y247203}.
\begin{figure}[b]
\centering
\includegraphics[width=140mm]{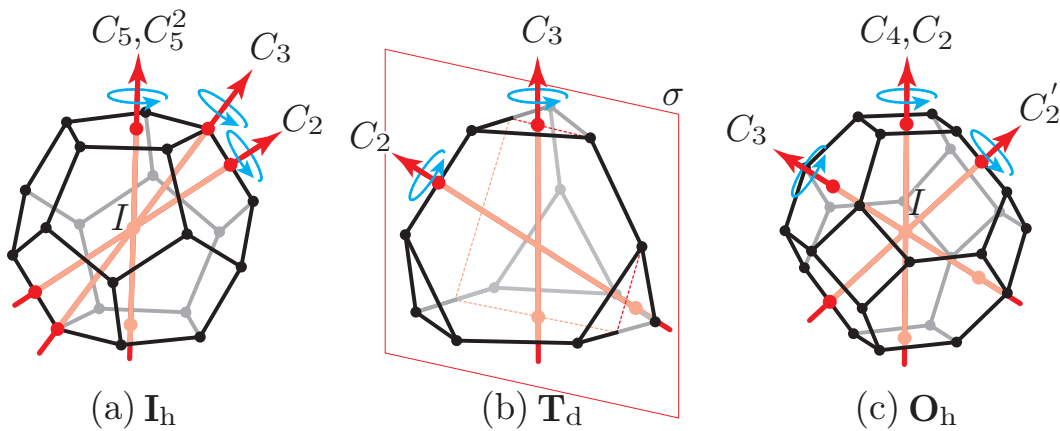}
\caption{(Color online)
         Point symmetry operations on
         dodecahedral (a), truncated-tetrahedral (b), and truncated-octahedral (c) lattices
         belonging to the full
         icosahedral ($\mathbf{I}_{\mathrm{h}}$),
         tetrahedral ($\mathbf{T}_{\mathrm{d}}$), and
         octahedral ($\mathbf{O}_{\mathrm{h}}$) groups, respectively.}
\label{AF:ptSG}
\end{figure}
\begin{figure}
\centering
\includegraphics[width=172mm]{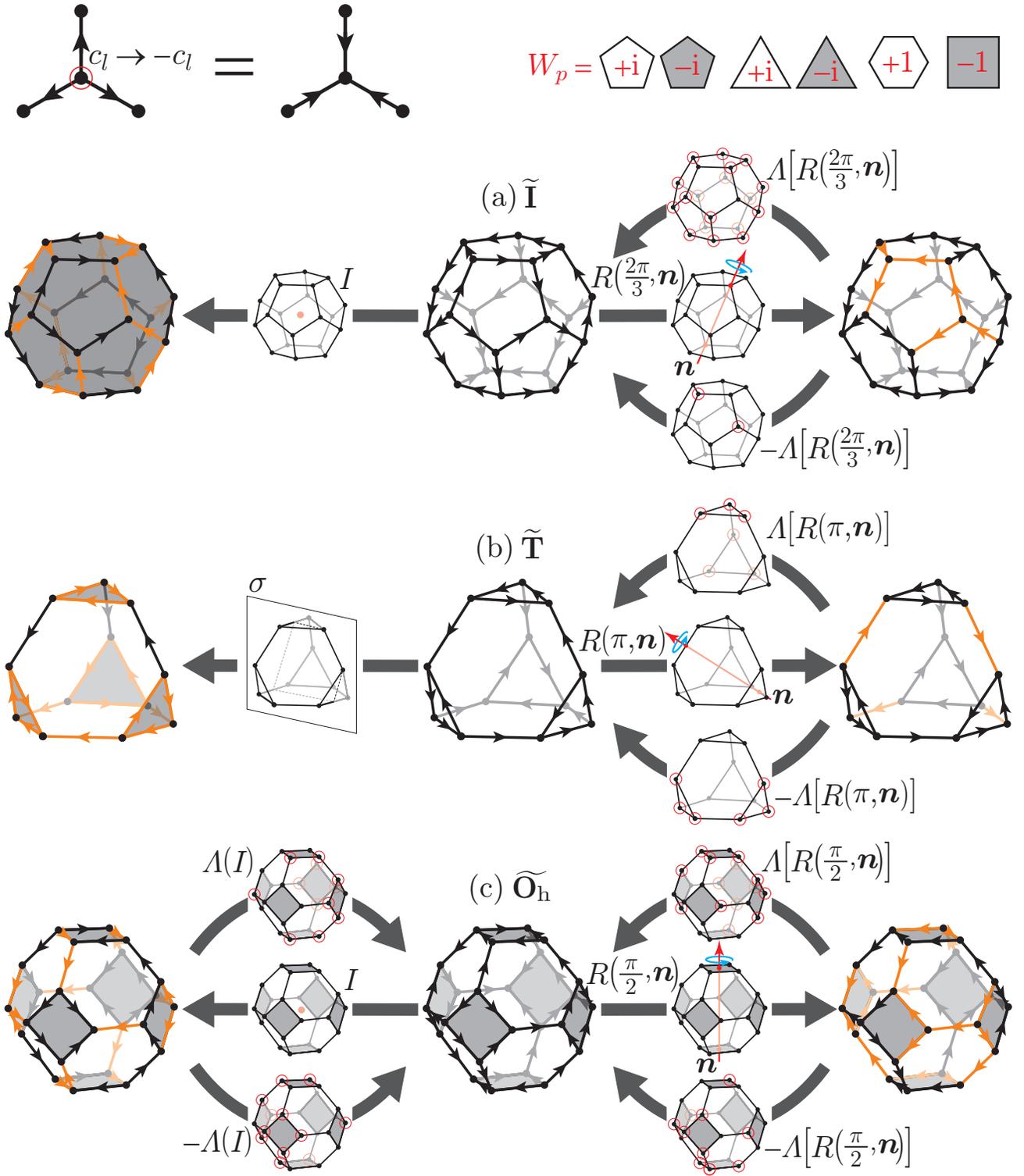}
\caption{(Color online)
         Gauged rotations, (gauged) inversion, and mirror operations of gauge-ground Kitaev
         spin balls consisting of
         dodecahedral (a), truncated-tetrahedral (b), and truncated-octahedral (c) lattices,
         whose symmetry groups read $\widetilde{\mathbf{I}}$, $\widetilde{\mathbf{T}}$, and
         $\widetilde{\mathbf{O}_{\mathrm{h}}}$, respectively.
         Inversion $I\in\mathbf{I}_{\mathrm{h}}$ of the gauged dodecahedron and
         mirror operations $\sigma\in\mathbf{T}_{\mathrm{d}}$ of the gauged truncated tetrahedron
         can be followed by no such gauge operation as to recover the initial bond
         configuration.}
\label{AF:prSG}
\end{figure}

   Figure \ref{AF:prSG}(a) illustrates a gauged rotation of the gauge-ground Kitaev dodecahedron.
Suppose we rotate it by $\frac{2\pi}{3}$ about one of the threefold axes $\bm{n}$,
which we shall denote by $R(\frac{2\pi}{3},\bm{n})$, and then gauge some Majorana fermions as
$c_l\to -c_l$, or equivalently, change the signs of their relevant bonds as
$u_{<l,l'>_\lambda}\to -u_{<l,l'>_\lambda}\ (\lambda=x,\,y,\,z)$, so as to recover the initial
bond configuration.
When a rotational symmetry operation $R(\varphi,\bm{n})$ $(0\leq\varphi<2\pi)$ is performed,
there exist two such local gauge operations, which we shall denote by
$\pm\varLambda\left[R(\varphi,\bm{n})\right]$ with a reminiscence of the double-valued nature of
rotation operators acting on half-integral spin states.
In the example of Fig. \ref{AF:prSG}(a),
$-\varLambda\left[R(\frac{2\pi}{3},\bm{n})\right]$ acts on two sites, while
$+\varLambda\left[R(\frac{2\pi}{3},\bm{n})\right]$ on all the rest,
where we make site assignment to $\pm\varLambda\left[R(\varphi,\bm{n})\right]$ in accordance with
$\mathrm{SU}(2)$ rotations.
How many and which sites to operate depend not only on the rotation axis $\bm{n}$ and
angle $\varphi$ but also the initial bond configuration.
We have $2^{\frac{L}{2}+1}$ flux configurations $\{W_p\}$ including the ground two,
each available from a set of $2^{\frac{3L}{2}}/2^{\frac{L}{2}+1}=2^{L-1}$ different
bond configurations $\{u_{<m,n>_{\lambda}}\}$.
We denote a couple of these serial transformations as
$+\varLambda\left[R(\frac{2\pi}{3},\bm{n})\right]
  R(\frac{2\pi}{3},\bm{n})
 \equiv
  \overline{R(\frac{2\pi}{3},\bm{n})}
$
and
$-\varLambda\left[R(\frac{2\pi}{3},\bm{n})\right]
  R(\frac{2\pi}{3},\bm{n})
 \equiv
  \underline{R(\frac{2\pi}{3},\bm{n})}
$.
Note that
$\left[\overline{R(\frac{2\pi}{3},\bm{n})}\right]^3\{u_{<m,n>_{\lambda}}\}
=-\{u_{<m,n>_{\lambda}}\}$,
while
$\left[\underline{R(\frac{2\pi}{3},\bm{n})}\right]^3\{u_{<m,n>_{\lambda}}\}
= \{u_{<m,n>_{\lambda}}\}$.

   Figure \ref{AF:prSG}(a) illustrates inversion of the gauge-ground Kitaev dodecahedron as well,
resulting in all $W_p$'s being reversed,
$\{W_p=+i;\,p=1,\cdots,12\}\to\{W_p=-i;\,p=1,\cdots,12\}$.
The constituent pentagons each initially have a flux of $\frac{\pi}{2}$ and all their fluxes
$\varPhi_p$ are reversed into $-\varPhi_p$ by inversion.
The flux variables $W_p\equiv e^{i\varPhi_p}$ are also all reversed.
Any local gauge transformation $c_l\to -c_l$ results in reversing the signs of bonds in pair
in the three surrounding polygons and therefore causes no change in their $W_p$'s.
We find that the symmetry group of the gauge-ground Kitaev dodecahedron is not the double cover
of the full point symmetry group, $\widetilde{\mathbf{I}_{\mathrm{h}}}$, but that of
an $\mathrm{SO}(3)$ subgroup, $\widetilde{\mathbf{I}}$.
This is the case with the gauge-ground Kitaev truncated tetrahedron as well
[Fig. \ref{AF:prSG}(b)].
Since a mirror operation $\sigma\in\mathbf{T}_{\mathrm{d}}$ reverses $W_p$'s of
its four constituent triangles, its symmetry group is not $\widetilde{\mathbf{T}_{\mathrm{d}}}$
but $\widetilde{\mathbf{T}}$.
On the other hand, inversion causes no change in $W_p$'s of the gauge-ground truncated octahedron
[Fig. \ref{AF:prSG}(c)].
This is because the truncated octahedron consists only of $2l$-sided polygons ($l\in\mathbb{N}$),
whose fluxes are either $0$ or $\pi$.
Even though inversion reverses such fluxes as $\varPhi_p\to -\varPhi_p$,
the corresponding flux variables $W_p\equiv e^{i\varPhi_p}$ remain unchanged.
Any two bond configurations $\{u_{<m,n>_\lambda}\}$ yielding the same set of fluxes $\{W_p\}$
can be converted to each other by local gauge operations.
Inversion of the gauge-ground truncated octahedron can be followed by two such local gauge
operations as to recover the initial bond configuration, 
which we shall denote by $\pm\varLambda(I)$, each to act on different halves of the lattice sites.
We generally denote a couple of gauged point symmetry operations $\pm\varLambda(P)P$ unifiedly as
$\widetilde{P}$ and distinguishably by $\overline{P}$ and $\underline{P}$.

\section{IRREDUCIBLE REPRESENTATIONS OF DOUBLE GROUPS FOR GAUGE-GROUND KITAEV POLYHEDRA}
\label{AS:ChTirrepsP&tildeP}

   We denote the orders of a point symmetry group $\mathbf{P}$ and its double covering group
$\widetilde{\mathbf{P}}$ by $g^{\mathbf{P}}$ and $g^{\widetilde{\mathbf{P}}}$, respectively.
Suppose the double cover $\widetilde{\mathbf{P}}$ to be the $\mathbb{Z}_2$-gauge extension
of $\mathbf{P}\subset\mathrm{O}(3)$.
Two group elements
$\widetilde{P}_{1}\in\widetilde{\mathbf{P}}$ and
$\widetilde{P}_{2}\in\widetilde{\mathbf{P}}$
are conjugate when we find such an element $\widetilde{P}\in\widetilde{\mathbf{P}}$ as to
satisfy
\begin{align}
   \widetilde{P}_{2}
  =\widetilde{P}\widetilde{P}_{1}\widetilde{P}^{-1}.
\end{align}
Every set of conjugate elements forms a class.
The classes of the double groups of our interest read
\begin{align*}
   \widetilde{\mathbf{I}}:
   &
   \{\overline{E}\},\{\underline{E}\},
   \{12\overline{C_{5}}\},\{12\underline{C_{5}}\},
   \{12\overline{C_{5}^{2}}\},\{12\underline{C_{5}^{2}}\},
   \{20\overline{C_{3}}\},\{20\underline{C_{3}}\},
   \{15\overline{C_{2}},15\underline{C_{2}}\};
   \\
   \widetilde{\mathbf{T}}:
   &
   \{\overline{E}\},\{\underline{E}\},
   \{3\overline{C_{2}},3\underline{C_{2}}\},
   \{4\overline{C_{3}}\},\{4\underline{C_{3}}\},
   \{4\overline{C_{3}^{2}}\},\{4\underline{C_{3}^{2}}\};
   \\
   \widetilde{\mathbf{O}}:
   &
   \{\overline{E}\},\{\underline{E}\},
   \{6\overline{C_{4}}\},\{6\underline{C_{4}}\},
   \{3\overline{C_{2}},3\underline{C_{2}}\},
   \{6\overline{C_{2}'},6\underline{C_{2}'}\},
   \{8\overline{C_{3}}\},\{8\underline{C_{3}}\};
   \\
   \widetilde{\mathbf{O}_{\mathrm{h}}}:
   &
   \begin{aligned}
   {}
   &
   \{\overline{E}\},\{\underline{E}\},
   \{6\overline{C_{4}},6\underline{C_{4}}\},
   \{3\overline{C_{2}},3\underline{C_{2}}\},
   \{6\overline{C_{2}'},6\underline{C_{2}'}\},
   \{8\overline{C_{3}}\},\{8\underline{C_{3}}\},
   \\[-1mm]
   &
   \{\overline{I},\underline{I}\},
   \{6\overline{IC_{4}},6\underline{IC_{4}}\},
   \{3\overline{IC_{2}},3\underline{IC_{2}}\},
   \{6\overline{IC_{2}'},6\underline{IC_{2}'}\},
   \{8\overline{IC_{3}}\},\{8\underline{IC_{3}}\}.
   \end{aligned}
\end{align*}
Supposing the $q$th class $\mathcal{C}_q$ ($q=1,\cdots,n_{\mathcal{C}}^{\widetilde{\mathbf{P}}}$)
of $\widetilde{\mathbf{P}}$ to consist of $h_q$ elements, it reads
$\{h_q\overline{P}_q\}$,
$\{h_q\underline{P}_q\}$, or
$\{\frac{h_q}{2}\overline{P}_q,\frac{h_q}{2}\underline{P}_q\}$.

   The number of (complex) irreducible representations equals how many classes are in the group.
Since all the single-valued (complex) irreducible representations of $\mathbf{P}$, amounting to
$n_{\mathcal{C}}^{\mathbf{P}}$, remain unchanged in $\widetilde{\mathbf{P}}$, we find
$n_{\mathcal{C}}^{\widetilde{\mathbf{P}}}-n_{\mathcal{C}}^{\mathbf{P}}$
double-valued (complex) irreducible representations in $\widetilde{\mathbf{P}}$.
When we denote the $i$th (complex) irreducible representation of
$\mathbf{P}$ ($\widetilde{\mathbf{P}}$) by $\varXi_i$ ($\widetilde{\varXi}_i$) and
its dimensionality by
$d_{\varXi_i}^{\mathbf{P}}$ ($d_{\widetilde{\varXi}_i}^{\widetilde{\mathbf{P}}}$), we have
\begin{alignat}{2}
   &
   \sum_{i=1}
       ^{n_{\mathcal{C}}^{\mathbf{I}}\equiv 5}
   \left(d_{\varXi_i}^{\mathbf{I}}\right)^2
  =g^{\mathbf{I}}
  =60,
   &&\ 
   \sum_{i=1}
       ^{n_{\mathcal{C}}^{\widetilde{\mathbf{I}}}\equiv 9}
   \left(d_{\widetilde{\varXi}_i}^{\widetilde{\mathbf{I}}}\right)^2
  =g^{\mathbf{I}}
  +\sum_{i=n_{\mathcal{C}}^{\mathbf{I}}+1}
       ^{n_{\mathcal{C}}^{\widetilde{\mathbf{I}}}\equiv 9}
   \left(d_{\widetilde{\varXi}_i}^{\widetilde{\mathbf{I}}}\right)^2
  =g^{\widetilde{\mathbf{I}}}
  =120;
   \\
   &
   \sum_{i=1}
       ^{n_{\mathcal{C}}^{\mathbf{T}}\equiv 4}
   \left(d_{\varXi_i}^{\mathbf{T}}\right)^2
  =g^{\mathbf{T}}
  =12,
   &&\ 
   \sum_{i=1}
       ^{n_{\mathcal{C}}^{\widetilde{\mathbf{T}}}\equiv 7}
   \left(d_{\widetilde{\varXi}_i}^{\widetilde{\mathbf{T}}}\right)^2
  =g^{\mathbf{T}}
  +\sum_{i=n_{\mathcal{C}}^{\mathbf{T}}+1}
       ^{n_{\mathcal{C}}^{\widetilde{\mathbf{T}}}\equiv 7}
   \left(d_{\widetilde{\varXi}_i}^{\widetilde{\mathbf{T}}}\right)^2
  =g^{\widetilde{\mathbf{T}}}
  =24;
   \\
   &
   \sum_{i=1}
       ^{n_{\mathcal{C}}^{\mathbf{O}}\equiv 5}
   \left(d_{\varXi_i}^{\mathbf{O}}\right)^2
  =g^{\mathbf{O}}
  =24,
   &&\ 
   \sum_{i=1}
       ^{n_{\mathcal{C}}^{\widetilde{\mathbf{O}}}\equiv 8}
   \left(d_{\widetilde{\varXi}_i}^{\widetilde{\mathbf{O}}}\right)^2
  =g^{\mathbf{O}}
  +\sum_{i=n_{\mathcal{C}}^{\mathbf{O}}+1}
       ^{n_{\mathcal{C}}^{\widetilde{\mathbf{O}}}\equiv 8}
   \left(d_{\widetilde{\varXi}_i}^{\widetilde{\mathbf{O}}}\right)^2
  =g^{\widetilde{\mathbf{O}}}
  =48;
   \\
   &
   \sum_{i=1}
       ^{n_{\mathcal{C}}^{\mathbf{O}_{\mathrm{h}}}\equiv 10}
   \left(d_{\varXi_i}^{\mathbf{O}_{\mathrm{h}}}\right)^2
  =g^{\mathbf{O}_{\mathrm{h}}}
  =48,
   &&\ 
   \sum_{i=1}
       ^{n_{\mathcal{C}}^{\widetilde{\mathbf{O}_{\mathrm{h}}}}\equiv 13}
   \left(d_{\widetilde{\varXi}_i}^{\widetilde{\mathbf{O}_{\mathrm{h}}}}\right)^2
  =g^{\mathbf{O}_{\mathrm{h}}}
  +\sum_{i=n_{\mathcal{C}}^{\mathbf{O}_{\mathrm{h}}}+1}
       ^{n_{\mathcal{C}}^{\widetilde{\mathbf{O}_{\mathrm{h}}}}\equiv 13}
   \left(d_{\widetilde{\varXi}_i}^{\widetilde{\mathbf{O}_{\mathrm{h}}}}\right)^2
  =g^{\widetilde{\mathbf{O}_{\mathrm{h}}}}
  =96
\end{alignat}
in an attempt to determine the dimensionalities of the double-valued (complex) irreducible
representations $d_{\widetilde{\varXi}_i}^{\widetilde{\mathbf{P}}}\,
 (i=n_{\mathcal{C}}^{\mathbf{P}}+1,\cdots,n_{\mathcal{C}}^{\widetilde{\mathbf{P}}})$.
The characters of $\widetilde{\varXi}_i$ are such that
\begin{alignat}{2}
   \chi_{\widetilde{\varXi}_{i}}^{\widetilde{\mathbf{P}}}(\overline{P})
   &
  =\ \ \chi_{\widetilde{\varXi}_{i}}^{\widetilde{\mathbf{P}}}(\underline{P})
   &&\ 
   (i=1,\cdots,n_{\mathcal{C}}^{\mathbf{P}}),
   \label{AE:ChiSingle}
   \\
   \chi_{\widetilde{\varXi}_{i}}^{\widetilde{\mathbf{P}}}(\overline{P})
   &
  =-\chi_{\widetilde{\varXi}_{i}}^{\widetilde{\mathbf{P}}}(\underline{P})
   &&\ 
   (i=n_{\mathcal{C}}^{\mathbf{P}}+1,\cdots,n_{\mathcal{C}}^{\widetilde{\mathbf{P}}}).
   \label{AE:ChiDouble}
\end{alignat}
When $\overline{P}$ and $\underline{P}$ belong to the same class, i.e.,
$
   \chi_{\widetilde{\varXi}_{i}}^{\widetilde{\mathbf{P}}}(\overline{P})
  =\chi_{\widetilde{\varXi}_{i}}^{\widetilde{\mathbf{P}}}(\underline{P})
$,
we immediately find
\begin{align}
   \chi_{\widetilde{\varXi}_{i}}^{\widetilde{\mathbf{P}}}(\overline{P})
   &
  =\chi_{\widetilde{\varXi}_{i}}^{\widetilde{\mathbf{P}}}(\underline{P})
  =0\ 
   (i=n_{\mathcal{C}}^{\mathbf{P}}+1,\cdots,n_{\mathcal{C}}^{\widetilde{\mathbf{P}}}).
   \label{AE:ZeroClass}
\end{align}
The character orthogonality theorems of the first and second kinds read \cite{D2008}
\begin{align}
   &
   \sum_{q=1}^{n_{\mathcal{C}}^{\widetilde{\mathbf{P}}}}
   h_q
   \chi_{\widetilde{\varXi}_i}^{\widetilde{\mathbf{P}}}(\mathcal{C}_q)^*
   \chi_{\widetilde{\varXi}_j}^{\widetilde{\mathbf{P}}}(\mathcal{C}_q)
  =g^{\widetilde{\mathbf{P}}}\delta_{ij},
   \label{AE:Orthogonal1}
   \\
   &
   \sum_{i=1}^{n_{\mathcal{C}}^{\widetilde{\mathbf{P}}}}
   \chi_{\widetilde{\varXi}_i}^{\widetilde{\mathbf{P}}}(\mathcal{C}_q)^*
   \chi_{\widetilde{\varXi}_i}^{\widetilde{\mathbf{P}}}(\mathcal{C}_r)
  =\frac{g^{\widetilde{\mathbf{P}}}}{h_q}\delta_{qr}.
   \label{AE:Orthogonal2}
\end{align}
When we denote the $h_q$ elements of $\mathcal{C}_q$ distinguishably as
$\{\widetilde{P}_q^{(1)},\cdots,\widetilde{P}_q^{(h_q)}\}$,
we can define structure constants as
\vspace{-3mm}
\begin{align}
   \sum_{i=1}^{h_q}\widetilde{P}_q^{(i)}
   \sum_{j=1}^{h_r}\widetilde{P}_r^{(j)}
  =\sum_{s=1}^{n_{\mathcal{C}}^{\widetilde{\mathbf{P}}}}
   c_{qr:s}
   \sum_{k=1}^{h_k}\widetilde{P}_s^{(k)}
\end{align}
to have another relation,
\vspace{-3mm}
\begin{align}
   h_q h_r
   \chi_{\widetilde{\varXi}_i}^{\widetilde{\mathbf{P}}}(\mathcal{C}_q)
   \chi_{\widetilde{\varXi}_i}^{\widetilde{\mathbf{P}}}(\mathcal{C}_r)
  =d_{\widetilde{\varXi}_i}^{\widetilde{\mathbf{P}}}
   \sum_{s=1}^{n_{\mathcal{C}}^{\widetilde{\mathbf{P}}}}
   h_s c_{qr:s}
   \chi_{\widetilde{\varXi}_i}^{\widetilde{\mathbf{P}}}(\mathcal{C}_s).
   \label{AE:Classmulti}
\end{align}
With Eqs. (\ref{AE:ZeroClass}), (\ref{AE:Orthogonal1}), (\ref{AE:Orthogonal2}), and
(\ref{AE:Classmulti}) in mind,
we can obtain characters of both single- and double-valued (complex) irreducible
representations of any double group $\widetilde{\mathbf{P}}$, which are listed in
Tables \ref{AT:characterItilde}--\ref{AT:characterOhtilde} with particular emphasis on
the relation between $\widetilde{\mathbf{P}}$ and $\mathbf{P}$.

\begin{table}[htb]
\caption{Irreducible representations of the double group $\widetilde{\mathbf{I}}$
         and their characters.}
\begin{tabular}{cccccccccccccccc}
\cline{3-16}\\[-3.8mm]\cline{3-16}
                         &
                         & \rule{0mm}{7.75mm}
                         & \raisebox{2mm}[0mm][0mm]{$\{\overline{{E}}\}$}
                         & \raisebox{2mm}[0mm][0mm]{$\{\underline{{E}}\}$}
                         & \,\,\,
                         & \raisebox{2mm}[0mm][0mm]{$\{12\overline{{C}_{5}}\}$}
                         & \raisebox{2mm}[0mm][0mm]{$\{12\underline{{C}_{5}}\}$}
                         & \,\,\,
                         & \raisebox{2mm}[0mm][0mm]{$\{12\overline{{C}_{5}^{2}}\}$}
                         & \raisebox{2mm}[0mm][0mm]{$\{12\underline{{C}_{5}^{2}}\}$}
                         & \,\,\,
                         & \raisebox{2mm}[0mm][0mm]{$\{20\overline{{C}_{3}}\}$}
                         & \raisebox{2mm}[0mm][0mm]{$\{20\underline{{C}_{3}}\}$}
                         & \,\,\,
                         & \shortstack{
                           $\{15\overline{{C}_{2}},$ \\
                           $\,\,\,\,\,15\underline{{C}_{2}}\,\}$}
\\
\end{tabular}
\centering
\label{AT:characterItilde}
\end{table}
\vspace{-3mm}
\begin{table}[htb]
\caption{Irreducible representations of the double group $\widetilde{\mathbf{T}}$
         and their characters.}
\begin{tabular}{ccclcccccccccccccccccccccc}
\cline{3-21}
\\[-3.8mm]
\cline{3-21}
     & \rule{0mm}{7.75mm}               &
     &
     & \multicolumn{2}{c}{\raisebox{2mm}[0mm][0mm]{$\{\overline{{E}}\}$}}
     & \multicolumn{2}{c}{\raisebox{2mm}[0mm][0mm]{$\{\underline{{E}}\}$}}
     & 
     & \multicolumn{2}{c}{\shortstack{
       $\{3\overline{{C}_{2}},$ \\
       $\,\,\,\,\,\,\,3\underline{{C}_{2}}\,\}$
       }}
     & 
     & \multicolumn{2}{l}{\raisebox{2mm}[0mm][0mm]{\,\,\,$\{4\overline{{C}_{3}}\}$}}
     & \multicolumn{2}{l}{\raisebox{2mm}[0mm][0mm]{\,\,\,$\{4\underline{{C}_{3}}\}$}}
     &
     & \multicolumn{2}{l}{\raisebox{2mm}[0mm][0mm]{\,\,\,$\{4\overline{{C}_{3}^{2}}\}$}}
     & \multicolumn{2}{l}{\raisebox{2mm}[0mm][0mm]{\,\,\,$\{4\underline{{C}_{3}^{2}}\}$}}
\\[0.5mm]
\cline{3-21}
\raisebox
{-15mm}
[0mm][0mm]
{\smash{
 $\widetilde{\mathbf{T}}
  \!
  \left\{
  \rule{0mm}{19mm}
  \right.$}}
\hspace{-4mm}
                         & \raisebox
                           {-6.5mm}
                           [0mm][0mm]
                           {\smash{
                            $\mathbf{T}\!
                             \left\{
                             \rule{0mm}{10mm}
                             \right.$}}
                           \hspace{-3mm}
                         & $ \mathrm{A} $\vphantom{{\Large S}}
                         &
                         & \multicolumn{4}{c}{\hspace{-4mm}$ 1 $}
                         &
                         & \multicolumn{2}{c}{\,\,$ 1 $}
                         & \,\,\qquad\quad
                         & \multicolumn{4}{c}{\hspace{-11mm}$ 1 $}
                         &
                         & \multicolumn{4}{c}{\hspace{-10mm}\,\,\,\,$ 1 $}
\\
\end{tabular}
\centering
\label{AT:characterTtilde}
\end{table}
\vspace{-5mm}
\begin{table}[htb]
\caption{Irreducible representations of the double group $\widetilde{\mathbf{O}}$
         and their characters.}
\begin{tabular}{ccccccccccccccccc}
\cline{3-15}
\\[-3.8mm]
\cline{3-15}
                         & \rule{0mm}{8mm}
                         &
                         & \raisebox{2.2mm}[0mm][0mm]{$\{\overline{{E}}\}$}\hspace{0.0mm}
                         & \hspace{0.0mm}\raisebox{2.2mm}[0mm][0mm]{$\{\underline{{E}}\}$}
                         & \,\,\,\quad
                         & \raisebox{2.2mm}[0mm][0mm]{$\{6\overline{{C}_{4}}\}$}\hspace{0.0mm}
                         & \hspace{0.0mm}\raisebox{2.2mm}[0mm][0mm]{$\{6\underline{{C}_{4}}\}$}
                         & \,\quad
                         & \shortstack{
                           $\{3\overline{{C}_{2}},$ \\
                           $\,\,\,\,\,\,\,3\underline{{C}_{2}}\,\}$}
                         & \,\quad
                         & \shortstack{
                           $\{6\overline{{C}_{2}'},$ \\
                           $\,\,\,\,\,\,\,6\underline{{C}_{2}'}\,\}$}
                         & \,\quad
                         & \raisebox{2.2mm}[0mm][0mm]{$\{8\overline{{C}_{3}}\}$}\hspace{0.0mm}
                         & \hspace{0.0mm}\raisebox{2.2mm}[0mm][0mm]{$\{8\underline{{C}_{3}}\}$}
\\
\end{tabular}
\centering
\label{AT:characterOtilde}
\end{table}

\begin{turnpage}
\begin{table}[htb]
\caption{Irreducible representations of
         the double group $\widetilde{\mathbf{O}_{\mathrm{h}}}$ and their characters.
         Those of the direct-product group $\widetilde{\mathbf{O}}\times\mathbf{C}_{\textrm{i}}$
         are also presented.}
{\begin{tabular}{ccccccccccccccccccccccccccccccccccccccccc}
\cline{3-32}
\\[-3.8mm]
\cline{3-32}
    &                            & \rule{0mm}{8.3mm}
    & \raisebox{2.2mm}[0mm][0mm]{$\{\overline{{E}}\}$}\hspace{0.0mm}
    & \hspace{0.0mm}\raisebox{2.2mm}[0mm][0mm]{$\{\underline{{E}}\}$}
    & \,\,
    & \multicolumn{2}{c}{
      \shortstack{
      $\{6\overline{{C}_{4}},$ \\
      $\,\,\,\,\,6\underline{{C}_{4}}\,\}$}
      \raisebox{2.2mm}[0mm][0mm]
      } 
    & \,\,
    & \multicolumn{2}{c}{
      \shortstack{
      $\{3\overline{{C}_{2}},\,\,$ \\
      $\,\,\,3\underline{{C}_{2}}\,\}$}
      }
    & \,\,
    & \multicolumn{2}{c}{
      \!\!\!\!\!
      \shortstack{
      $\{6\overline{{C}_{2}'},$ \\
      $\,\,\,\,\,6\underline{{C}_{2}'}\,\}$}
      \!\!\!\!\!
      }
    & \,\,
    & \raisebox{2.2mm}[0mm][0mm]{$\{8\overline{{C}_{3}}\}$}\hspace{0.0mm}
    & \hspace{0.0mm}\raisebox{2.2mm}[0mm][0mm]{$\{8\underline{{C}_{3}}\}$}
    & \,\,
    & \multicolumn{2}{c}{
      \shortstack{
      $\{\overline{{I}},$ \\
      $\,\,\,\,\,\underline{{I}}\,\}$}
      }
    & \,\,
    & \multicolumn{2}{c}{
      \shortstack{
      $\{6\overline{{IC}_{4}},$ \\
      $\,\,\,\,\,6\underline{{IC}_{4}}\,\}$}
      \raisebox{2.2mm}[0mm][0mm]
      }
    & \,\,
    & \multicolumn{2}{c}{
      \shortstack{
      $\{3\overline{{IC}_{2}},$ \\
      $\,\,\,\,\,3\underline{{IC}_{2}}\,\}$}
      }
    & \,\,
    & \multicolumn{2}{c}{
      \shortstack{
      $\{6\overline{{IC}_{2}'},$ \\
      $\,\,\,\,\,6\underline{{IC}_{2}'}\,\}$}
      }
    & \,\,
    & \raisebox{2.2mm}[0mm][0mm]{$\{8\overline{{IC}_{3}}\}$}\hspace{0.0mm}
    & \hspace{0.0mm}\raisebox{2.2mm}[0mm][0mm]{$\{8\underline{{IC}_{3}}\}$}
    &
\\
\end{tabular}
}
\label{AT:characterOhtilde}
\centering
\end{table}
\end{turnpage}
\clearpage

\section{DIRECT-PRODUCT REPRESENTATIONS OF DOUBLE GROUPS FOR GAUGE-GROUND KITAEV POLYHEDRA}
\label{AS:ChTDPrepstildeP}

   Since Raman scattering within the LF scheme \cite{F514,S1068,S365,K187201}
is caused by spinons in pair, we make direct-product representations out of double-valued
irreducible representations of double covers $\widetilde{\mathbf{P}}$ of the corresponding
point symmetry groups $\mathbf{P}\subset\mathrm{O}(3)$.
Direct-product representations of a nonabelian group are not necessarily irreducible even though
the constituent representations are irreducible.
We take interest in spinon-geminate-excitation-relevant direct-product representations
$\widetilde{\varXi}_i\otimes\widetilde{\varXi}_j\,
(i,j=n_{\mathcal{C}}^{\mathbf{P}}+1,\cdots,n_{\mathcal{C}}^{\widetilde{\mathbf{P}}})$
of $\widetilde{\mathbf{P}}$, which are decomposed into single-valued irreducible
representations of the corresponding point symmetry group $\mathbf{P}$,
\begin{align}
   \widetilde{\varXi}_{i}\otimes\widetilde{\varXi}_j
  =\mathop{\bigoplus}_{k=1}^{n_{\mathcal{C}}^{\widetilde{\mathbf{P}}}}\widetilde{\varXi}_k
   \sum_{q=1}^{n_{\mathcal{C}}^{\widetilde{\mathbf{P}}}}
      \frac{h_q}{g^{\widetilde{\mathbf{P}}}}
   \chi_{\widetilde{\varXi}_k}^{\widetilde{\mathbf{P}}}(\mathcal{C}_{q})^*
   \chi_{\widetilde{\varXi}_i\otimes\widetilde{\varXi}_j}^{\widetilde{\mathbf{P}}}
   (\mathcal{C}_{q})
  =\mathop{\bigoplus}_{k=1}^{n_{\mathcal{C}}^{\mathbf{P}}}\varXi_k
   \sum_{q=1}^{n_{\mathcal{C}}^{\widetilde{\mathbf{P}}}}
      \frac{h_q}{g^{\widetilde{\mathbf{P}}}}
   \chi_{\varXi_k}^{\mathbf{P}}(\mathcal{C}_{q})^*
   \chi_{\widetilde{\varXi}_i\otimes\widetilde{\varXi}_j}^{\widetilde{\mathbf{P}}}
   (\mathcal{C}_{q}),
   \label{AE:DPR}
\end{align}
having in mind that
\begin{align}
   \chi_{\widetilde{\varXi}_i\otimes\widetilde{\varXi}_j}^{\widetilde{\mathbf{P}}}(\overline{P})
  =\chi_{\widetilde{\varXi}_i}^{\widetilde{\mathbf{P}}}(\overline{P})
   \chi_{\widetilde{\varXi}_j}^{\widetilde{\mathbf{P}}}(\overline{P})
  =\chi_{\widetilde{\varXi}_i}^{\widetilde{\mathbf{P}}}(\underline{P})
   \chi_{\widetilde{\varXi}_j}^{\widetilde{\mathbf{P}}}(\underline{P})
  =\chi_{\widetilde{\varXi}_i\otimes\widetilde{\varXi}_j}^{\widetilde{\mathbf{P}}}(\underline{P})\ 
   (i,j=n_{\mathcal{C}}^{\mathbf{P}}+1,\cdots,n_{\mathcal{C}}^{\widetilde{\mathbf{P}}}).
   \label{AE:character(DPRij)}
\end{align}
Direct-product representations made of the two same irreducible representations consist of
symmetric (bosonic) and antisymmetric (fermionic) parts,
\begin{align}
   \widetilde{\varXi}_{i}\otimes\widetilde{\varXi}_{i}
  =[\widetilde{\varXi}_{i}\otimes\widetilde{\varXi}_{i}]
   \oplus
   \{\widetilde{\varXi}_{i}\otimes\widetilde{\varXi}_{i}\},
   \label{AE:DPRii=symDPRii+antisymDPRii}
\end{align}
which are decomposed into symmetric and antisymmetric single-valued irreducible representations
of the corresponding point symmetry group $\mathbf{P}$, respectively,
\begin{align}
   &
   [\widetilde{\varXi}_i\otimes\widetilde{\varXi}_i]
  =\mathop{\bigoplus}_{k=1}^{n_{\mathcal{C}}^{\mathbf{P}}}[\varXi_k]
   \sum_{q=1}^{n_{\mathcal{C}}^{\widetilde{\mathbf{P}}}}
      \frac{h_{q}}{g^{\widetilde{\mathbf{P}}}}
   \chi_{\varXi_{k}}^{\mathbf{P}}(\mathcal{C}_{q})^*
   \chi_{[\widetilde{\varXi}_{i}\otimes\widetilde{\varXi}_{i}]}^{\widetilde{\mathbf{P}}}
   (\mathcal{C}_{q}),
   \label{AE:symDPRii}
   \\
   &
   \{\widetilde{\varXi}_i\otimes\widetilde{\varXi}_i\}
  =\mathop{\bigoplus}_{k=1}^{n_{\mathcal{C}}^{\mathbf{P}}}\{\varXi_k\}
   \sum_{q=1}^{n_{\mathcal{C}}^{\widetilde{\mathbf{P}}}}
      \frac{h_{q}}{g^{\widetilde{\mathbf{P}}}}
   \chi_{\varXi_{k}}^{\mathbf{P}}(\mathcal{C}_{q})^*
   \chi_{\{\widetilde{\varXi}_{i}\otimes\widetilde{\varXi}_{i}\}}^{\widetilde{\mathbf{P}}}
   (\mathcal{C}_{q}).
   \label{AE:antisymDPRii}
\end{align}
Note that characters of symmetric and antisymmetric direct-product representations are given by
\begin{align}
   \chi_{[\widetilde{\varXi}_{i}\otimes\widetilde{\varXi}_{i}]}^{\widetilde{\mathbf{P}}}
   (\widetilde{P})
   &
  =\frac{1}{2}
   \left[
      \chi_{\widetilde{\varXi}_{i}}^{\widetilde{\mathbf{P}}}(\widetilde{P})^{2}
     +\chi_{\widetilde{\varXi}_{i}}^{\widetilde{\mathbf{P}}}(\widetilde{P}^{2})
   \right],
   \label{AE:character(symDPRii)}
   \\
   \chi_{\{\varXi_{i}\otimes\varXi_{i}\}}^{\widetilde{\mathbf{P}}}(\widetilde{P})
   &
  =\frac{1}{2}
   \left[
     \chi_{\widetilde{\varXi}_{i}}^{\widetilde{\mathbf{P}}}(\widetilde{P})^{2}
    -\chi_{\widetilde{\varXi}_{i}}^{\widetilde{\mathbf{P}}}(\widetilde{P}^{2})
   \right].
   \label{AE:character(antisymDPRii)}
\end{align}

   We can obtain characters of any direct-product representation using
Eqs. (\ref{AE:character(symDPRii)}) and (\ref{AE:character(antisymDPRii)}) as well as
(\ref{AE:character(DPRij)}),
which of our interest are listed in
Tables \ref{AT:characterDPRItilde}--\ref{AT:characterDPROhtilde}.
Direct-product representations for geminate excitations of different Majorana spinon eigenmodes
are not necessarily made of different irreducible representations but may be made of the same ones.
Those made of different irreducible representations can be decomposed into irreducible
representations by Eq. (\ref{AE:DPR}), while those made of the same ones by
Eqs. (\ref{AE:symDPRii}) and (\ref{AE:antisymDPRii}).
Direct-product representations for geminate excitations of degenerate Majorana spinon eigenmodes
are also the latter case.
The thus-obtained decompositions into irreducible representations are all listed in
Table \ref{AT:DPRintoIrrep}.
\begin{table}[htb]
\caption{Direct-product representations made of double-valued irreducible representations of
         the double group $\widetilde{\mathbf{I}}$ and their characters.}\vspace{2mm}
\begin{tabular}{rclccccccccccccccccc}
\hline\hline
\raisebox{2mm}[0mm][0mm]{$\widetilde{\varXi}_{i}$} & \hspace{-2.5mm}
\raisebox{2mm}[0mm][0mm]{$\otimes$}                & \hspace{-2.5mm}
\raisebox{2mm}[0mm][0mm]{$\widetilde{\varXi}_{j}$}
                         & \raisebox{2mm}[0mm][0mm]{$\{\overline{E}\}$}\hspace{-0.5mm}
                         & \hspace{-0.5mm}\raisebox{2mm}[0mm][0mm]{$\{\underline{E}\}$}
                         &
                         & \raisebox{2mm}[0mm][0mm]{$\{12\overline{C_{5}}\}$}\hspace{-0.5mm}
                         & \hspace{-0.5mm}\raisebox{2mm}[0mm][0mm]{$\{12\underline{C_{5}}\}$}
                         &
                         & \raisebox{2mm}[0mm][0mm]{$\{12\overline{C_{5}^{2}}\}$}\hspace{-0.5mm}
                         & \hspace{-0.5mm}\raisebox{2mm}[0mm][0mm]{$\{12\underline{C_{5}^{2}}\}$}
                         &
                         & \raisebox{2mm}[0mm][0mm]{$\{20\overline{C_{3}}\}$}\hspace{-0.5mm}
                         & \hspace{-0.5mm}\raisebox{2mm}[0mm][0mm]{$\{20\underline{C_{3}}\}$}
                         &
                         & \rule{0mm}{7.8mm}
                           \shortstack{
                           $\{15\overline{{C}_{2}},$ \\
                           $\,\,\,\,\,15\underline{{C}_{2}}\,\}$}
\\[0.6mm]
\hline
$[\mathrm{I}_{\frac{5}{2}}$    & \hspace{-2.5mm}
$\otimes$                      & \hspace{-2.5mm}
$\mathrm{I}_{\frac{5}{2}}]$    & \multicolumn{2}{c}{$21 $}                            
                               &
                               & \multicolumn{2}{c}{\hphantom{$-$}$ 1 $\hphantom{$-$}}
                               &
                               & \multicolumn{2}{c}{\hphantom{$-$}$ 1 $\hphantom{$-$}}
                               &
                               & \multicolumn{2}{c}{$ 0 $}                            
                               &
                               & $-3 $\hphantom{$-$}
\\
$\{\mathrm{I}_{\frac{5}{2}}$   & \hspace{-2.5mm}
$\otimes$                      & \hspace{-2.5mm}
$\mathrm{I}_{\frac{5}{2}}\}$   & \multicolumn{2}{c}{$15 $}                            
                               &
                               & \multicolumn{2}{c}{\hphantom{$-$}$ 0 $\hphantom{$-$}}
                               &
                               & \multicolumn{2}{c}{\hphantom{$-$}$ 0 $\hphantom{$-$}}
                               &
                               & \multicolumn{2}{c}{$ 0 $}                            
                               &
                               & \hphantom{$-$}$ 3 $\hphantom{$-$}
\\
$\mathrm{I}_{\frac{5}{2}}$     & \hspace{-2.5mm}
$\otimes$                      & \hspace{-2.5mm}
$\mathrm{G}_{\frac{3}{2}}$     & \multicolumn{2}{c}{$24 $}                            
                               &
                               & \multicolumn{2}{c}{$-1 $\hphantom{$-$}}              
                               &
                               & \multicolumn{2}{c}{$-1 $\hphantom{$-$}}              
                               &
                               & \multicolumn{2}{c}{$ 0 $}                            
                               &
                               & \hphantom{$-$}$ 0 $\hphantom{$-$}
\\
$[\mathrm{G}_{\frac{3}{2}}$    & \hspace{-2.5mm}
$\otimes$                      & \hspace{-2.5mm}
$\mathrm{G}_{\frac{3}{2}}]$    & \multicolumn{2}{c}{$10 $}                            
                               &
                               & \multicolumn{2}{c}{\hphantom{$-$}$ 0 $\hphantom{$-$}}
                               &
                               & \multicolumn{2}{c}{\hphantom{$-$}$ 0 $\hphantom{$-$}}
                               &
                               & \multicolumn{2}{c}{$ 1 $}                            
                               &
                               & $-2 $\hphantom{$-$}
\\
$\{\mathrm{G}_{\frac{3}{2}}$   & \hspace{-2.5mm}
$\otimes$                      & \hspace{-2.5mm}
$\mathrm{G}_{\frac{3}{2}}\}$   & \multicolumn{2}{c}{$ 6 $}
                               &
                               & \multicolumn{2}{c}{\hphantom{$-$}$ 1 $\hphantom{$-$}}
                               &
                               & \multicolumn{2}{c}{\hphantom{$-$}$ 1 $\hphantom{$-$}}
                               &
                               & \multicolumn{2}{c}{$ 0 $}                            
                               &
                               & \hphantom{$-$}$ 2 $\hphantom{$-$}
\\
$[\mathrm{E}_{\frac{1}{2}}$    & \hspace{-2.5mm}
$\otimes$                      & \hspace{-2.5mm}
$\mathrm{E}_{\frac{1}{2}}]$    & \multicolumn{2}{c}{$ 3 $}
                               &
                               & \multicolumn{2}{c}{$ \frac{1+\sqrt{5}}{2} $}
                               &
                               & \multicolumn{2}{c}{$ \frac{1-\sqrt{5}}{2} $}
                               &
                               & \multicolumn{2}{c}{$ 0 $}
                               &
                               & $-1 $\hphantom{$-$}
\\
$\{\mathrm{E}_{\frac{1}{2}}$   & \hspace{-2.5mm}
$\otimes$                      & \hspace{-2.5mm}
$\mathrm{E}_{\frac{1}{2}}\}$   & \multicolumn{2}{c}{$ 1 $}
                               &
                               & \multicolumn{2}{c}{$ 1 $}
                               &
                               & \multicolumn{2}{c}{$ 1 $}
                               &
                               & \multicolumn{2}{c}{$ 1 $}
                               &
                               & $ 1 $
\\
$\mathrm{E}_{\frac{1}{2}}$     & \hspace{-2.5mm}
$\otimes$                      & \hspace{-2.5mm}
$\mathrm{E}_{\frac{7}{2}}$     & \multicolumn{2}{c}{$ 4 $}
                               &
                               & \multicolumn{2}{c}{$-1 $\hphantom{$-$}}
                               &
                               & \multicolumn{2}{c}{$-1 $\hphantom{$-$}}
                               &
                               & \multicolumn{2}{c}{$ 1 $}
                               &
                               & $ 0 $
\\
$\mathrm{E}_{\frac{1}{2}}$     & \hspace{-2.5mm}
$\otimes$                      & \hspace{-2.5mm}
$\mathrm{G}_{\frac{3}{2}}$     & \multicolumn{2}{c}{$ 8 $}
                               &
                               & \multicolumn{2}{c}{$ \frac{1+\sqrt{5}}{2} $}
                               &
                               & \multicolumn{2}{c}{$ \frac{1-\sqrt{5}}{2} $}
                               &
                               & \multicolumn{2}{c}{$-1 $\hphantom{$-$}}
                               &
                               & $ 0 $
\\
$\mathrm{E}_{\frac{1}{2}}$     & \hspace{-2.5mm}
$\otimes$                      & \hspace{-2.5mm}
$\mathrm{I}_{\frac{5}{2}}$     & \multicolumn{2}{c}{$12 $}
                               &
                               & \multicolumn{2}{c}{$-\frac{1+\sqrt{5}}{2} $\hphantom{$-$}}
                               &
                               & \multicolumn{2}{c}{$-\frac{1-\sqrt{5}}{2} $\hphantom{$-$}}
                               &
                               & \multicolumn{2}{c}{$ 0 $}
                               &
                               & $ 0 $
\\
$[\mathrm{E}_{\frac{7}{2}}$    & \hspace{-2.5mm}
$\otimes$                      & \hspace{-2.5mm}
$\mathrm{E}_{\frac{7}{2}}]$    & \multicolumn{2}{c}{$ 3 $}
                               &
                               & \multicolumn{2}{c}{$ \frac{1-\sqrt{5}}{2} $}
                               &
                               & \multicolumn{2}{c}{$ \frac{1+\sqrt{5}}{2} $}
                               &
                               & \multicolumn{2}{c}{$ 0 $}
                               &
                               & $-1 $\hphantom{$-$}
\\
$\{\mathrm{E}_{\frac{7}{2}}$   & \hspace{-2.5mm}
$\otimes$                      & \hspace{-2.5mm}
$\mathrm{E}_{\frac{7}{2}}\}$   & \multicolumn{2}{c}{$ 1 $}
                               &
                               & \multicolumn{2}{c}{$ 1 $}
                               &
                               & \multicolumn{2}{c}{$ 1 $}
                               &
                               & \multicolumn{2}{c}{$ 1 $}
                               &
                               & $ 1 $
\\
$\mathrm{E}_{\frac{7}{2}}$     & \hspace{-2.5mm}
$\otimes$                      & \hspace{-2.5mm}
$\mathrm{G}_{\frac{3}{2}}$     & \multicolumn{2}{c}{$ 8 $}
                               &
                               & \multicolumn{2}{c}{$ \frac{1-\sqrt{5}}{2} $}
                               &
                               & \multicolumn{2}{c}{$ \frac{1+\sqrt{5}}{2} $}
                               &
                               & \multicolumn{2}{c}{$-1 $\hphantom{$-$}}
                               &
                               & $ 0 $
\\
$\mathrm{E}_{\frac{7}{2}}$     & \hspace{-2.5mm}
$\otimes$                      & \hspace{-2.5mm}
$\mathrm{I}_{\frac{5}{2}}$     & \multicolumn{2}{c}{$12 $}
                               &
                               & \multicolumn{2}{c}{$-\frac{1-\sqrt{5}}{2} $\hphantom{$-$}}
                               &
                               & \multicolumn{2}{c}{$-\frac{1+\sqrt{5}}{2} $\hphantom{$-$}}
                               &
                               & \multicolumn{2}{c}{$ 0 $}
                               &
                               & $ 0 $
\\[1.5mm]
\hline\hline
\end{tabular}
\label{AT:characterDPRItilde}
\centering
\end{table}
\vspace{-10mm}
\begin{table}[htb]
\caption{Direct-product representations made of double-valued irreducible representations of
         the double group $\widetilde{\mathbf{T}}$ and their characters.}
\vspace{2mm}
\begin{tabular}{rclcccccccccccccc}
\hline\hline
\raisebox{2mm}[0mm][0mm]{$\widetilde{\varXi}_{i}$} & \hspace{-2.5mm}
\raisebox{2mm}[0mm][0mm]{$\otimes$}                & \hspace{-2.5mm}
\raisebox{2mm}[0mm][0mm]{$\widetilde{\varXi}_{j}$}
                         & \raisebox{2mm}[0mm][0mm]{$\{\overline{E}\}$}\hspace{-0.5mm}
                         & \hspace{-0.5mm}\raisebox{2mm}[0mm][0mm]{$\{\underline{E}\}$}
                         &
                         & \rule{0mm}{8.2mm}
                           \shortstack{
                           \vphantom{{\large S}}
                           $\{3\overline{{C}_{2}},$ \\
                           $\,\,\,\,\,\,\,3\underline{{C}_{2}}\,\}$
                           }
                         &
                         & \raisebox{2mm}[0mm][0mm]{$\{4\overline{C_{3}}\}$}\hspace{-0.5mm}
                         & \hspace{-0.5mm}\raisebox{2mm}[0mm][0mm]{$\{4\underline{C_{3}}\}$}
                         &
                         & \raisebox{2mm}[0mm][0mm]{$\{4\overline{C_{3}^{2}}\}$}\hspace{-0.5mm}
                         & \hspace{-0.5mm}\raisebox{2mm}[0mm][0mm]{$\{4\underline{C_{3}^{2}}\}$}
\\[0.6mm]
\hline
$[\mathrm{G}_{\frac{3}{2}}^{(2)}$  & \hspace{-2.5mm}
$\otimes$                          & \hspace{-2.5mm}
$\mathrm{G}_{\frac{3}{2}}^{(2)}]$  & \multicolumn{2}{c}{$ 3 $}                       &
                                   & $-1 $\hphantom{$-$}                             &
                                   & \multicolumn{2}{c}{$ 0 $}                       &
                                   & \multicolumn{2}{c}{$ 0 $}
\\
$\{\mathrm{G}_{\frac{3}{2}}^{(2)}$ & \hspace{-2.5mm}
$\otimes$                          & \hspace{-2.5mm}
$\mathrm{G}_{\frac{3}{2}}^{(2)}\}$ & \multicolumn{2}{c}{$ 1 $}                       &
                                   & $ 1 $                                           &
                                   & \multicolumn{2}{c}{${e}^{-{i}\frac{2}{3}\pi} $} &
                                   & \multicolumn{2}{c}{${e}^{-{i}\frac{4}{3}\pi} $}
\\
$\mathrm{G}_{\frac{3}{2}}^{(2)}$   & \hspace{-2.5mm}
$\otimes$                          & \hspace{-2.5mm}
$\mathrm{E}_{\frac{1}{2}}$         & \multicolumn{2}{c}{$ 4 $}                       &
                                   & \hphantom{$-$}$0$\hphantom{$-$}                 &
                                   & \multicolumn{2}{c}{${e}^{-{i}\frac{4}{3}\pi} $} &
                                   & \multicolumn{2}{c}{${e}^{-{i}\frac{2}{3}\pi} $}
\\
$[\mathrm{E}_{\frac{1}{2}}$        & \hspace{-2.5mm}
$\otimes$                          & \hspace{-2.5mm}
$\mathrm{E}_{\frac{1}{2}}]$        & \multicolumn{2}{c}{$ 3 $} &
                                   & $-1 $\hphantom{$-$}       &
                                   & \multicolumn{2}{c}{$ 0 $} &
                                   & \multicolumn{2}{c}{$ 0 $}
\\
$\{\mathrm{E}_{\frac{1}{2}}$       & \hspace{-2.5mm}
$\otimes$                          & \hspace{-2.5mm}
$\mathrm{E}_{\frac{1}{2}}\}$       & \multicolumn{2}{c}{$ 1 $} &
                                   & \hphantom{$-$}$1$\hphantom{$-$}                 &
                                   & \multicolumn{2}{c}{$ 1 $} &
                                   & \multicolumn{2}{c}{$ 1 $}
\\
$\mathrm{G}_{\frac{3}{2}}^{(1)}$   & \hspace{-2.5mm}
$\otimes$                          & \hspace{-2.5mm}
$\mathrm{G}_{\frac{3}{2}}^{(2)}$   & \multicolumn{2}{c}{$ 4 $}                       &
                                   & \hphantom{$-$}$0$\hphantom{$-$}                 &
                                   & \multicolumn{2}{c}{$ 1 $}                       &
                                   & \multicolumn{2}{c}{$ 1 $}
\\
$\mathrm{G}_{\frac{3}{2}}^{(1)}$   & \hspace{-2.5mm}
$\otimes$                          & \hspace{-2.5mm}
$\mathrm{E}_{\frac{1}{2}}$         & \multicolumn{2}{c}{$ 4 $}                       &
                                   & \hphantom{$-$}$0$\hphantom{$-$}                 &
                                   & \multicolumn{2}{c}{${e}^{-{i}\frac{2}{3}\pi} $} &
                                   & \multicolumn{2}{c}{${e}^{-{i}\frac{4}{3}\pi} $}
\\
$[\mathrm{G}_{\frac{3}{2}}^{(1)}$  & \hspace{-2.5mm}
$\otimes$                          & \hspace{-2.5mm}
$\mathrm{G}_{\frac{3}{2}}^{(1)}]$  & \multicolumn{2}{c}{$ 3 $}                       &
                                   & $-1 $\hphantom{$-$}                             &
                                   & \multicolumn{2}{c}{$ 0 $}                       &
                                   & \multicolumn{2}{c}{$ 0 $}
\\
$\{\mathrm{G}_{\frac{3}{2}}^{(1)}$ & \hspace{-2.5mm}
$\otimes$                          & \hspace{-2.5mm}
$\mathrm{G}_{\frac{3}{2}}^{(1)}\}$ & \multicolumn{2}{c}{$ 1 $}                       &
                                   & \hphantom{$-$}$ 1 $\hphantom{$-$}               &
                                   & \multicolumn{2}{c}{${e}^{-{i}\frac{4}{3}\pi} $} &
                                   & \multicolumn{2}{c}{${e}^{-{i}\frac{2}{3}\pi} $}
\\[2.5mm]
\hline\hline
\end{tabular}
\label{AT:characterDPRTtilde}
\centering
\end{table}
\vspace{-10mm}
\begin{table}[htb]
\centering
\caption{Direct-product representations made of double-valued irreducible representations of
         the double group $\widetilde{\mathbf{O}}$ and their characters.}
\vspace{2mm}
\begin{tabular}{rclcccccccccccccc}
\hline\hline
\raisebox{2.2mm}[0mm][0mm]{$\widetilde{\varXi}_{i}$}      & \hspace{-2.5mm}
\raisebox{2.2mm}[0mm][0mm]{$\otimes$}                     & \hspace{-2.5mm}
\raisebox{2.2mm}[0mm][0mm]{$\widetilde{\varXi}_{j}$}
                             & \raisebox{2.2mm}[0mm][0mm]{$\{\overline{E}\}$}\hspace{-0.5mm}
                             & \hspace{-0.5mm}\raisebox{2.2mm}[0mm][0mm]{$\{\underline{E}\}$}
                             &
                             & \raisebox{2.2mm}[0mm][0mm]{$\{6\overline{C_{4}}\}$}\hspace{-0.5mm}
                             & \hspace{-0.5mm}\raisebox{2.2mm}[0mm][0mm]{$\{6\underline{C_{4}}\}$}
                             &
                             & \rule{0mm}{8.5mm}
                               \shortstack{
                                $\{3\overline{{C}_{2}},$ \\
                                $\,\,\,\,\,\,\,3\underline{{C}_{2}}\,\}$}
                             &
                             & \shortstack{
                                $\{6\overline{{C}_{2}'},$ \\
                                $\,\,\,\,\,\,\,6\underline{{C}_{2}'}\,\}$}
                             &
                             & \raisebox{2.2mm}[0mm][0mm]{$\{8\overline{C_{3}}\}$}\hspace{-0.5mm}
                             & \hspace{-0.5mm}\raisebox{2.2mm}[0mm][0mm]{$\{8\underline{C_{3}}\}$}
\\[0.7mm]
\hline
$[\mathrm{E}_{\frac{1}{2}}$  & \hspace{-2.5mm}
$\otimes$                    & \hspace{-2.5mm}
$\mathrm{E}_{\frac{1}{2}}]$  & \multicolumn{2}{c}{$ 3 $}
                             &
                             & \multicolumn{2}{c}{\hphantom{$-$}$ 1 $\hphantom{$-$}}
                             &
                             & $-1 $\hphantom{$-$}
                             &
                             & $-1 $\hphantom{$-$}
                             &
                             & \multicolumn{2}{c}{$ 0 $}
\\
$\{\mathrm{E}_{\frac{1}{2}}$ & \hspace{-2.5mm}
$\otimes$                    & \hspace{-2.5mm}
$\mathrm{E}_{\frac{1}{2}}\}$ & \multicolumn{2}{c}{$ 1 $}
                             &
                             & \multicolumn{2}{c}{\hphantom{$-$}$ 1 $\hphantom{$-$}}
                             &
                             & $ 1 $
                             &
                             & $ 1 $
                             &
                             & \multicolumn{2}{c}{$ 1 $}
\\
$\mathrm{E}_{\frac{1}{2}}$   & \hspace{-2.5mm}
$\otimes$                    & \hspace{-2.5mm}
$\mathrm{E}_{\frac{5}{2}}$   & \multicolumn{2}{c}{$ 4 $}
                             &
                             & \multicolumn{2}{c}{$-2 $\hphantom{$-$}}
                             &
                             & $ 0 $
                             &
                             & $ 0 $
                             &
                             & \multicolumn{2}{c}{$ 1 $}
\\
$[\mathrm{E}_{\frac{5}{2}}$  & \hspace{-2.5mm}
$\otimes$                    & \hspace{-2.5mm}
$\mathrm{E}_{\frac{5}{2}}]$  & \multicolumn{2}{c}{$ 3 $}
                             &
                             & \multicolumn{2}{c}{\hphantom{$-$}$ 1 $\hphantom{$-$}}
                             &
                             & $-1 $\hphantom{$-$}
                             &
                             & $-1 $\hphantom{$-$}
                             &
                             & \multicolumn{2}{c}{$ 0 $}
\\
$\{\mathrm{E}_{\frac{5}{2}}$ & \hspace{-2.5mm}
$\otimes$                    & \hspace{-2.5mm}
$\mathrm{E}_{\frac{5}{2}}\}$ & \multicolumn{2}{c}{$ 1 $}
                             &
                             & \multicolumn{2}{c}{\hphantom{$-$}$ 1 $\hphantom{$-$}}
                             &
                             & $ 1 $
                             &
                             & $ 1 $
                             &
                             & \multicolumn{2}{c}{$ 1 $}
\\
$\mathrm{G}_{\frac{3}{2}}$   & \hspace{-2.5mm}
$\otimes$                    & \hspace{-2.5mm}
$\mathrm{E}_{\frac{1}{2}}$   & \multicolumn{2}{c}{$ 8 $}
                             &
                             & \multicolumn{2}{c}{\hphantom{$-$}$ 0 $\hphantom{$-$}}
                             &
                             & $ 0 $
                             &
                             & $ 0 $
                             &
                             & \multicolumn{2}{c}{$-1 $\hphantom{$-$}}
\\
$\mathrm{G}_{\frac{3}{2}}$   & \hspace{-2.5mm}
$\otimes$                    & \hspace{-2.5mm}
$\mathrm{E}_{\frac{5}{2}}$   & \multicolumn{2}{c}{$ 8 $}
                             &
                             & \multicolumn{2}{c}{\hphantom{$-$}$ 0 $\hphantom{$-$}}
                             &
                             & $ 0 $
                             &
                             & $ 0 $
                             &
                             & \multicolumn{2}{c}{$-1 $\hphantom{$-$}}
\\
$[\mathrm{G}_{\frac{3}{2}}$  & \hspace{-2.5mm}
$\otimes$                    & \hspace{-2.5mm}
$\mathrm{G}_{\frac{3}{2}}]$  & \multicolumn{2}{c}{$10 $}
                             &
                             & \multicolumn{2}{c}{\hphantom{$-$}$ 0 $\hphantom{$-$}}
                             &
                             & $-2 $\hphantom{$-$}
                             &
                             & $-2 $\hphantom{$-$}
                             &
                             & \multicolumn{2}{c}{$ 1 $}
\\
$\{\mathrm{G}_{\frac{3}{2}}$ & \hspace{-2.5mm}
$\otimes$                    & \hspace{-2.5mm}
$\mathrm{G}_{\frac{3}{2}}\}$ & \multicolumn{2}{c}{$ 6 $}
                             &
                             & \multicolumn{2}{c}{\hphantom{$-$}$ 0 $\hphantom{$-$}}
                             &
                             & $ 2 $
                             &
                             & $ 2 $
                             &
                             & \multicolumn{2}{c}{$ 0 $}
\\[1.5mm]
\hline\hline
\end{tabular}
\centering
\label{AT:characterDPROtilde}
\end{table}

\begin{table}[htb]
\caption{Direct-product representations made of double-valued irreducible representations of
         the double group $\widetilde{\mathbf{O}_{\mathrm{h}}}$ and their characters.}
\vspace{2mm}
\begin{tabular}{rclcccccccccccccccccccccccccccccccccc}
\hline\hline
\raisebox{2.1mm}[0mm][0mm]{$\widetilde{\varXi}_{i}$} & \hspace{-1.5mm}
\raisebox{2.1mm}[0mm][0mm]{$\otimes$}\hspace{-1.5mm} & \rule{0mm}{8.6mm}
\raisebox{2.1mm}[0mm][0mm]{$\widetilde{\varXi}_{j}$}
                         & \raisebox{2.1mm}[0mm][0mm]{$\{\overline{E}\}$}\hspace{-0.5mm}
                         & \hspace{-0.5mm}\raisebox{2.1mm}[0mm][0mm]{$\{\underline{E}\}$}
                         &
                         & \shortstack{
                           $\{6\overline{{C}_{4}},$ \\
                           $\,\,\,\,\,6\underline{{C}_{4}}\,\}$}
                         &
                         & \shortstack{
                           $\{3\overline{C_{2}},$ \\
                           $\,\,\,\,\,3\underline{C_{2}}\,\}$}
                         &
                         & \shortstack{
                           $\{6\overline{C_{2}'},$ \\
                           $\,\,\,\,\,6\underline{C_{2}'}\,\}$}
                         &
                         & \raisebox{2.1mm}[0mm][0mm]{$\{8\overline{C_{3}}\}$}\hspace{-0.5mm}
                         & \hspace{-0.5mm}\raisebox{2.1mm}[0mm][0mm]{$\{8\underline{C_{3}}\}$}
                         &
                         & \shortstack{
                           $\{\overline{I},$ \\
                           $\,\,\,\,\,\underline{I}\,\}$}
                         &
                         & \shortstack{
                           $\{6\overline{IC_{4}},$ \\
                           $\,\,\,\,\,6\underline{IC_{4}}\,\}$}
                         &
                         & \shortstack{
                           $\{3\overline{IC_{2}},$ \\
                           $\,\,\,\,\,3\underline{IC_{2}}\,\}$}
                         &
                         & \shortstack{
                           $\{6\overline{IC_{2}'},$ \\
                           $\,\,\,\,\,6\underline{IC_{2}'}\,\}$}
                         &
                         & \raisebox{2.1mm}[0mm][0mm]{$\{8\overline{IC_{3}}\}$}\hspace{-0.5mm}
                         & \hspace{-0.5mm}\raisebox{2.1mm}[0mm][0mm]{$\{8\underline{IC_{3}}\}$}                 
\\[0.7mm]\hline
$[\mathrm{G}_{\frac{1}{2}+\frac{5}{2}}$   &
\hspace{-2mm}$\otimes$\hspace{-2mm}       &
$\mathrm{G}_{\frac{1}{2}+\frac{5}{2}}]$   & \multicolumn{2}{c}{$10 $}
                                          &
                                          & $ 0 $                                  
                                          &
                                          & $-2 $\hphantom{$-$}                                  
                                          &
                                          & $-2 $\hphantom{$-$}                                  
                                          &
                                          & \multicolumn{2}{c}{$ 1 $}
                                          &
                                          & $-2 $\hphantom{$-$}                                  
                                          &
                                          & $ 0 $                                  
                                          &
                                          & $ 2 $                    
                                          &
                                          & $-2 $\hphantom{$-$}                                  
                                          &
                                          & \multicolumn{2}{c}{$ 1 $}
\\
$\{\mathrm{G}_{\frac{1}{2}+\frac{5}{2}}$  &
\hspace{-2mm}$\otimes$\hspace{-2mm}       &
$\mathrm{G}_{\frac{1}{2}+\frac{5}{2}}\}$  & \multicolumn{2}{c}{$ 6 $}              
                                          &
                                          & $ 0 $                                  
                                          &
                                          & $ 2 $                                  
                                          &
                                          & $ 2 $                                  
                                          &
                                          & \multicolumn{2}{c}{$ 3 $}
                                          &
                                          & $ 2 $                                  
                                          &
                                          & $ 0 $                                  
                                          &
                                          & $-2 $\hphantom{$-$}                    
                                          &
                                          & $ 2 $                                  
                                          &
                                          & \multicolumn{2}{c}{$-1 $\hphantom{$-$}}
\\
$\mathrm{G}_{\frac{3}{2}}^{\mathrm{g}}$   &
\hspace{-2mm}$\otimes$\hspace{-2mm}       &
$\mathrm{G}_{\frac{1}{2}+\frac{5}{2}}$    & \multicolumn{2}{c}{$16 $}              
                                          &
                                          & $ 0 $                                  
                                          &
                                          & $ 0 $                                  
                                          &
                                          & $ 0 $                                  
                                          &
                                          & \multicolumn{2}{c}{$-2 $\hphantom{$-$}}
                                          &
                                          & $ 0 $                                  
                                          &
                                          & $ 0 $                                  
                                          &
                                          & \hphantom{$-$}$ 0 $\hphantom{$-$}      
                                          &
                                          & $ 0 $                                  
                                          &
                                          & \multicolumn{2}{c}{\hphantom{$-$}$ 0 $\hphantom{$-$}}
\\
$[\mathrm{G}_{\frac{3}{2}}^{\mathrm{g}}$  &
\hspace{-2mm}$\otimes$\hspace{-2mm}       &
$\mathrm{G}_{\frac{3}{2}}^{\mathrm{g}}]$  & \multicolumn{2}{c}{$10 $}              
                                          &
                                          & $ 0 $                                  
                                          &
                                          & $-2 $\hphantom{$-$}
                                          &
                                          & $-2 $\hphantom{$-$}
                                          &
                                          & \multicolumn{2}{c}{$ 1 $}              
                                          &
                                          & $-2 $\hphantom{$-$}                    
                                          &
                                          & $ 0 $                                  
                                          &
                                          & $ 2 $                    
                                          &
                                          & $-2 $\hphantom{$-$}                                  
                                          &
                                          & \multicolumn{2}{c}{\hphantom{$-$}$ 1 $\hphantom{$-$}}
\\
$\{\mathrm{G}_{\frac{3}{2}}^{\mathrm{g}}$ &
\hspace{-2mm}$\otimes$\hspace{-2mm}       &
$\mathrm{G}_{\frac{3}{2}}^{\mathrm{g}}\}$ & \multicolumn{2}{c}{$ 6 $}              
                                          &
                                          & $ 0 $                                  
                                          &
                                          & $ 2 $                                  
                                          &
                                          & $ 2 $                                  
                                          &
                                          & \multicolumn{2}{c}{$ 0 $}              
                                          &
                                          & $ 2 $                                  
                                          &
                                          & $ 0 $                                  
                                          &
                                          & $-2 $\hphantom{$-$}                    
                                          &
                                          & $ 2 $                                  
                                          &
                                          & \multicolumn{2}{c}{\hphantom{$-$}$ 2 $\hphantom{$-$}}
\\
$\mathrm{G}_{\frac{3}{2}}^{\mathrm{u}}$   &
\hspace{-2mm}$\otimes$\hspace{-2mm}       &
$\mathrm{G}_{\frac{1}{2}+\frac{5}{2}}$    & \multicolumn{2}{c}{$16 $}              
                                          &
                                          & $ 0 $                                  
                                          &
                                          & $ 0 $                                  
                                          &
                                          & $ 0 $                                  
                                          &
                                          & \multicolumn{2}{c}{$-2 $\hphantom{$-$}}
                                          &
                                          & $ 0 $                                  
                                          &
                                          & $ 0 $                                  
                                          &
                                          & \hphantom{$-$}$ 0 $\hphantom{$-$}      
                                          &
                                          & $ 0 $                                  
                                          &
                                          & \multicolumn{2}{c}{\hphantom{$-$}$ 0 $\hphantom{$-$}}
\\
$\mathrm{G}_{\frac{3}{2}}^{\mathrm{g}}$   &
\hspace{-2mm}$\otimes$\hspace{-2mm}       &
$\mathrm{G}_{\frac{3}{2}}^{\mathrm{u}}$   & \multicolumn{2}{c}{$16 $}              
                                          &
                                          & $ 0 $                                  
                                          &
                                          & $ 0 $                                  
                                          &
                                          & $ 0 $                                  
                                          &
                                          & \multicolumn{2}{c}{$ 1 $}              
                                          &
                                          & $ 0 $                                  
                                          &
                                          & $ 0 $                                  
                                          &
                                          & \hphantom{$-$}$ 0 $\hphantom{$-$}      
                                          &
                                          & $ 0 $                                  
                                          &
                                          & \multicolumn{2}{c}{$-3 $\hphantom{$-$}}
\\
$[\mathrm{G}_{\frac{3}{2}}^{\mathrm{u}}$  &
\hspace{-2mm}$\otimes$\hspace{-2mm}       &
$\mathrm{G}_{\frac{3}{2}}^{\mathrm{u}}]$  & \multicolumn{2}{c}{$10 $}              
                                          &
                                          & $ 0 $                                  
                                          &
                                          & $-2 $\hphantom{$-$}
                                          &
                                          & $-2 $\hphantom{$-$}
                                          &
                                          & \multicolumn{2}{c}{$ 1 $}              
                                          &
                                          & $-2 $\hphantom{$-$}                    
                                          &
                                          & $ 0 $                                  
                                          &
                                          & $ 2 $                    
                                          &
                                          & $-2 $\hphantom{$-$}                                  
                                          &
                                          & \multicolumn{2}{c}{\hphantom{$-$}$ 1 $\hphantom{$-$}}
\\
$\{\mathrm{G}_{\frac{3}{2}}^{\mathrm{u}}$ &
\hspace{-2mm}$\otimes$\hspace{-2mm}       &
$\mathrm{G}_{\frac{3}{2}}^{\mathrm{u}}\}$ & \multicolumn{2}{c}{$ 6 $}              
                                          &
                                          & $ 0 $                                  
                                          &
                                          & $ 2 $                                  
                                          &
                                          & $ 2 $                                  
                                          &
                                          & \multicolumn{2}{c}{$ 0 $}              
                                          &
                                          & $ 2 $                                  
                                          &
                                          & $ 0 $                                  
                                          &
                                          & $-2 $\hphantom{$-$}                    
                                          &
                                          & $ 2 $                                  
                                          &
                                          & \multicolumn{2}{c}{\hphantom{$-$}$ 2 $\hphantom{$-$}}
\\[2mm]
\hline\hline
\end{tabular}
\centering
\label{AT:characterDPROhtilde}
\end{table}

\begin{table}[tb]
\centering
\caption{Direct-product representations made of double-valued irreducible representations
         $\widetilde{\varXi}_i\otimes\widetilde{\varXi}_j$
         $(i,j=n_{\mathcal{C}}^{\mathbf{P}}+1,\cdots,n_{\mathcal{C}}^{\widetilde{\mathbf{P}}})$
         and their decompositions into single-valued irreducible representations
         $\widetilde{\varXi}_k$ $(k=1,\cdots,n_{\mathcal{C}}^{\mathbf{P}})$,
         which are doubly or singly underlined when they are relevant to
         inelastic (Raman) or elastic (Rayleigh) light scatterings,
         for various double groups $\widetilde{\mathbf{P}}$.
         Note that $\widetilde{\varXi}_k$ of $\widetilde{\mathbf{P}}$ is nothing but
         $\varXi_k$ of $\mathbf{P}$.}
\begin{tabular}{crcll}
\hline \hline
$\widetilde{\mathbf{P}}$               & \multicolumn{3}{c}
                                         {$\widetilde{\varXi}_i\otimes\widetilde{\varXi}_j$}
                                       & $\bigoplus_k\widetilde{\varXi}_k
                                         =\bigoplus_k\varXi_k$ \hfill \\
\hline
\raisebox{-23.5mm}[0mm][0mm]{
$\widetilde{\mathbf{I}}$}              & $\mathrm{I}_{\frac{5}{2}}$
                                       & \hspace{-2.5mm} $\otimes$ \hspace{-2.5mm}
                                       & $\mathrm{I}_{\frac{5}{2}}$
                                       & $\{\underline{\mathrm{A}}\}\!\oplus\!
                                          2[\mathrm{T}_{1}]\!\oplus\!
                                          2[\mathrm{T}_{2}]\!\oplus\!
                                           [\mathrm{G}]\!\oplus\!
                                           \{\mathrm{G}\}\!\oplus\!
                                           [\underline{\underline{\mathrm{H}}}]\!\oplus\!
                                          2\{\underline{\underline{\mathrm{H}}}\}$   \\
                                       & $\mathrm{I}_{\frac{5}{2}}$
                                       & \hspace{-2.5mm} $\otimes$ \hspace{-2.5mm}
                                       & $\mathrm{G}_{\frac{3}{2}}$
                                       & $\mathrm{T}_{1}\!\oplus\!
                                          \mathrm{T}_{2}\!\oplus\!
                                         2\mathrm{G}\!\oplus\!
                                         2\underline{\underline{\mathrm{H}}}$   \\
                                       & $\mathrm{G}_{\frac{3}{2}}$
                                       & \hspace{-2.5mm} $\otimes$ \hspace{-2.5mm} 
                                       & $\mathrm{G}_{\frac{3}{2}}$
                                       & $\{\underline{\mathrm{A}}\}\!\oplus\!
                                           [\mathrm{T}_{1}]\!\oplus\!
                                           [\mathrm{T}_{2}]\!\oplus\!
                                           [\mathrm{G}]\!\oplus\!
                                          \{\underline{\underline{\mathrm{H}}}\}$   \\
                                       & $\mathrm{E}_{\frac{1}{2}}$
                                       & \hspace{-2.5mm} $\otimes$ \hspace{-2.5mm} 
                                       & $\mathrm{E}_{\frac{1}{2}}$
                                       & $\{\underline{\mathrm{A}}\}\!\oplus\!
                                           [\mathrm{T}_{1}]$   \\
                                       & $\mathrm{E}_{\frac{1}{2}}$
                                       & \hspace{-2.5mm} $\otimes$ \hspace{-2.5mm} 
                                       & $\mathrm{E}_{\frac{7}{2}}$
                                       & $\mathrm{G}$   \\
                                       & $\mathrm{E}_{\frac{1}{2}}$
                                       & \hspace{-2.5mm} $\otimes$ \hspace{-2.5mm} 
                                       & $\mathrm{G}_{\frac{3}{2}}$
                                       & $\mathrm{T}_{1}\!\oplus\!
                                          \underline{\underline{\mathrm{H}}}$   \\
                                       & $\mathrm{E}_{\frac{1}{2}}$
                                       & \hspace{-2.5mm} $\otimes$ \hspace{-2.5mm} 
                                       & $\mathrm{I}_{\frac{5}{2}}$
                                       & $\mathrm{T}_{2}\!\oplus\!
                                          \mathrm{G}\!\oplus\!
                                          \underline{\underline{\mathrm{H}}}$   \\
                                       & $\mathrm{E}_{\frac{7}{2}}$
                                       & \hspace{-2.5mm} $\otimes$ \hspace{-2.5mm} 
                                       & $\mathrm{E}_{\frac{7}{2}}$
                                       & $\{\underline{\mathrm{A}}\}\!\oplus\!
                                           [\mathrm{T}_{2}]$   \\
                                       & $\mathrm{E}_{\frac{7}{2}}$
                                       & \hspace{-2.5mm} $\otimes$ \hspace{-2.5mm} 
                                       & $\mathrm{G}_{\frac{3}{2}}$
                                       & $\mathrm{T}_{2}\!\oplus\!
                                          \underline{\underline{\mathrm{H}}}$   \\
                                       & $\mathrm{E}_{\frac{7}{2}}$
                                       & \hspace{-2.5mm} $\otimes$ \hspace{-2.5mm} 
                                       & $\mathrm{I}_{\frac{5}{2}}$
                                       & $\mathrm{T}_{1}\!\oplus\!
                                          \mathrm{G}\!\oplus\!
                                          \underline{\underline{\mathrm{H}}}$
\\[1.5mm]
\hline
\raisebox{-14.5mm}[0mm][0mm]{
$\widetilde{\mathbf{T}}$}              
                                       & $\mathrm{G}_{\frac{3}{2}}^{(2)}$
                                       & \hspace{-2.5mm} $\otimes$ \hspace{-2.5mm}
                                       & $\mathrm{G}_{\frac{3}{2}}^{(2)}$
                                       & $\{\underline{\underline{\mathrm{E}^{(1)}}}\}\!\oplus\!
                                          [\underline{\underline{\mathrm{T}}}]$  \\
                                       & $\mathrm{G}_{\frac{3}{2}}^{(2)}$
                                       & \hspace{-2.5mm} $\otimes$ \hspace{-2.5mm}
                                       & $\mathrm{E}_{\frac{1}{2}}$
                                       & $\underline{\underline{\mathrm{E}^{(2)}}}\!\oplus\!
                                          \underline{\underline{\mathrm{T}}}$   \\
                                       & $\mathrm{E}_{\frac{1}{2}}$
                                       & \hspace{-2.5mm} $\otimes$ \hspace{-2.5mm}
                                       & $\mathrm{E}_{\frac{1}{2}}$
                                       & $\{\underline{\mathrm{A}}\}\!\oplus\!
                                          [\underline{\underline{\mathrm{T}}}]$   \\
                                       & $\mathrm{G}_{\frac{3}{2}}^{(1)}$
                                       & \hspace{-2.5mm} $\otimes$ \hspace{-2.5mm}
                                       & $\mathrm{G}_{\frac{3}{2}}^{(2)}$
                                       & $\underline{\mathrm{A}}\!\oplus\!
                                          \underline{\underline{\mathrm{T}}}$   \\
                                       & $\mathrm{G}_{\frac{3}{2}}^{(1)}$
                                       & \hspace{-2.5mm} $\otimes$ \hspace{-2.5mm}
                                       & $\mathrm{E}_{\frac{1}{2}}$
                                       & $\underline{\underline{\mathrm{E}^{(1)}}}\!\oplus\!
                                          \underline{\underline{\mathrm{T}}}$   \\
                                       & $\mathrm{G}_{\frac{3}{2}}^{(1)}$
                                       & \hspace{-2.5mm} $\otimes$ \hspace{-2.5mm}
                                       & $\mathrm{G}_{\frac{3}{2}}^{(1)}$
                                       & $\{\underline{\underline{\mathrm{E}^{(2)}}}\}\!\oplus\!
                                          [\underline{\underline{\mathrm{T}}}]$   \\[2mm]
\hline
\raisebox{-13mm}[0mm][0mm]{
$\widetilde{\mathbf{O}}$}              & $\mathrm{E}_{\frac{1}{2}}$
                                       & \hspace{-2.5mm} $\otimes$ \hspace{-2.5mm}
                                       & $\mathrm{E}_{\frac{1}{2}}$
                                       & $\{\underline{\mathrm{A_{1}}}\}\!\oplus\!
                                           [\mathrm{T}_{1}]$   \\
                                       & $\mathrm{E}_{\frac{1}{2}}$
                                       & \hspace{-2.5mm} $\otimes$ \hspace{-2.5mm}
                                       & $\mathrm{E}_{\frac{5}{2}}$
                                       & $\mathrm{A_{2}}\!\oplus\!
                                          \underline{\underline{\mathrm{T}_{2}}}$   \\
                                       & $\mathrm{E}_{\frac{5}{2}}$
                                       & \hspace{-2.5mm} $\otimes$ \hspace{-2.5mm}
                                       & $\mathrm{E}_{\frac{5}{2}}$
                                       & $\{\underline{\mathrm{A_{1}}}\}\!\oplus\!
                                           [\mathrm{T}_{1}]$   \\
                                       & $\mathrm{G}_{\frac{3}{2}}$
                                       & \hspace{-2.5mm} $\otimes$ \hspace{-2.5mm}
                                       & $\mathrm{E}_{\frac{1}{2}}$
                                       & $\underline{\underline{\mathrm{E}}}\!\oplus\!
                                          \mathrm{T_{1}}\!\oplus\!
                                          \underline{\underline{\mathrm{T}_{2}}}$\\
                                       & $\mathrm{G}_{\frac{3}{2}}$
                                       & \hspace{-2.5mm} $\otimes$ \hspace{-2.5mm}
                                       & $\mathrm{E}_{\frac{5}{2}}$
                                       & $\underline{\underline{\mathrm{E}}}\!\oplus\!
                                          \mathrm{T_{1}}\!\oplus\!
                                          \underline{\underline{\mathrm{T}_{2}}}$   \\
                                       & $\mathrm{G}_{\frac{3}{2}}$
                                       & \hspace{-2.5mm} $\otimes$ \hspace{-2.5mm}
                                       & $\mathrm{G}_{\frac{3}{2}}$
                                       & $\{\underline{\mathrm{A_{1}}}\}\!\oplus\!
                                          [\mathrm{A_{2}}]\!\oplus\!
                                          \{\underline{\underline{\mathrm{E}}}\}\!\oplus\!
                                         2[\mathrm{T_{1}}]\!\oplus\!
                                          [\underline{\underline{\mathrm{T}_{2}}}]\!\oplus\!
                                          \{\underline{\underline{\mathrm{T}_{2}}}\}$   \\[1.5mm]
\hline
\raisebox{-14mm}[0mm][0mm]{
$\widetilde{\mathbf{O}_{\mathrm{h}}}$} & $\mathrm{G}_{\frac{1}{2}+\frac{5}{2}}$
                                       & \hspace{-2.5mm} $\otimes$ \hspace{-2.5mm}
                                       & $\mathrm{G}_{\frac{1}{2}+\frac{5}{2}}$
                                       & $\{\underline{\mathrm{A_{1g}}}\}\!\oplus\!
                                          \{\mathrm{A_{1u}}\}\!\oplus\!
                                           [\mathrm{A_{2g}}]\!\oplus\!
                                          \{\mathrm{A_{2u}}\}\!\oplus\!
                                           [\mathrm{T_{1g}}]\!\oplus\!
                                           [\mathrm{T_{1u}}]\!\oplus\!
                                          \{\underline{\underline{\mathrm{T_{2g}}}}\}\!\oplus\!
                                           [\mathrm{T_{2u}}]$   \\
                                       & $\mathrm{G}_{\frac{3}{2}}^{\mathrm{g}}$
                                       & \hspace{-2.5mm} $\otimes$ \hspace{-2.5mm}
                                       & $\mathrm{G}_{\frac{1}{2}+\frac{5}{2}}$
                                       & $\underline{\underline{\mathrm{E_{g}}}}\!\oplus\!
                                          \mathrm{E_{u}}\!\oplus\!
                                          \mathrm{T_{1g}}\!\oplus\!
                                          \mathrm{T_{1u}}\!\oplus\!
                                          \underline{\underline{\mathrm{T_{2g}}}}\!\oplus\!
                                          \mathrm{T_{2u}}$   \\
                                       & $\mathrm{G}_{\frac{3}{2}}^{\mathrm{g}}$
                                       & \hspace{-2.5mm} $\otimes$ \hspace{-2.5mm}
                                       & $\mathrm{G}_{\frac{3}{2}}^{\mathrm{g}}$
                                       & $\{\underline{\mathrm{A_{1g}}}\}\!\oplus\!
                                           [\mathrm{A_{2g}}]\!\oplus\!
                                          \{\mathrm{E_{u}}\}\!\oplus\!
                                           [\mathrm{T_{1g}}]\!\oplus\!
                                           [\mathrm{T_{1u}}]\!\oplus\!
                                          \{\underline{\underline{\mathrm{T_{2g}}}}\}\!\oplus\!
                                           [\mathrm{T_{2u}}]$   \\
                                       & $\mathrm{G}_{\frac{3}{2}}^{\mathrm{u}}$
                                       & \hspace{-2.5mm} $\otimes$ \hspace{-2.5mm}
                                       & $\mathrm{G}_{\frac{1}{2}+\frac{5}{2}}$
                                       & $\underline{\underline{\mathrm{E_{g}}}}\!\oplus\!
                                          \mathrm{E_{u}}\!\oplus\!
                                          \mathrm{T_{1g}}\!\oplus\!
                                          \mathrm{T_{1u}}\!\oplus\!
                                          \underline{\underline{\mathrm{T_{2g}}}}\!\oplus\!
                                          \mathrm{T_{2u}}$   \\
                                       & $\mathrm{G}_{\frac{3}{2}}^{\mathrm{g}}$
                                       & \hspace{-2.5mm} $\otimes$ \hspace{-2.5mm}
                                       & $\mathrm{G}_{\frac{3}{2}}^{\mathrm{u}}$
                                       & $\mathrm{A_{1u}}\!\oplus\!
                                          \mathrm{A_{2u}}\!\oplus\!
                                          \underline{\underline{\mathrm{E_{g}}}}\!\oplus\!
                                          \mathrm{T_{1g}}\!\oplus\!
                                          \mathrm{T_{1u}}\!\oplus\!
                                          \underline{\underline{\mathrm{T_{2g}}}}\!\oplus\!
                                          \mathrm{T_{2u}}\!\!\!$   \\
                                       & $\mathrm{G}_{\frac{3}{2}}^{\mathrm{u}}$
                                       & \hspace{-2.5mm} $\otimes$ \hspace{-2.5mm}
                                       & $\mathrm{G}_{\frac{3}{2}}^{\mathrm{u}}$
                                       & $\{\underline{\mathrm{A_{1g}}}\}\!\oplus\!
                                           [\mathrm{A_{2g}}]\!\oplus\!
                                          \{\mathrm{E_{u}}\}\!\oplus\!
                                           [\mathrm{T_{1g}}]\!\oplus\!
                                           [\mathrm{T_{1u}}]\!\oplus\!
                                          \{\underline{\underline{\mathrm{T_{2g}}}}\}\!\oplus\!
                                           [\mathrm{T_{2u}}]$   \\[2mm]
\hline\hline
\end{tabular}
\label{AT:DPRintoIrrep}
\end{table}
\clearpage

\section{POLARIZATION DEPENDENCES OF RAMAN SPECTRA}
\label{AS:RamanPD}

    The ground-state Raman scattering intensity of a Kitaev gauged lattice within the LF scheme
\cite{F514,S1068,S365,K187201} reads
\begin{align}
   &
   I(\omega)
  =\frac{1}{2\pi\hbar L}
   \int_{-\infty}^\infty
   \langle 0|
    e^{\frac{{i}\mathscr{H}t}{\hbar}}
    \mathscr{R}^\dagger
    e^{-\frac{{i}\mathscr{H}t}{\hbar}}
    \mathscr{R}
   |0\rangle
   e^{i\omega t}dt
  =\frac{1}{2\pi\hbar L}
   \int_{-\infty}^\infty
   \langle 0|
    e^{\frac{{i}\mathscr{H}t}{\hbar}}
    \mathscr{R}
    e^{-\frac{{i}\mathscr{H}t}{\hbar}}
    \mathscr{R}
   |0\rangle
   e^{i\omega t}dt;
   \nonumber \\
   &
   \mathscr{R}
  \equiv
   \sum_{\mu=x,y,z}\sum_{\nu=x,y,z}
   e_{\mathrm{in}}^{\mu}
   e_{\mathrm{sc}}^{\nu}
   \mathcal{R}^{\mu\nu}
  \equiv
   {}^{\textrm{t}}\!\bm{e}_{\mathrm{in}}
   \mathcal{R}
   \bm{e}_{\mathrm{sc}},\ 
   \mathcal{R}^{\mu\nu}
  \equiv
  -J
   \sum_{\lambda=x,y,z}\sum_{<m,n>_{\lambda}}
   d_{mn}^{\mu}
   d_{mn}^{\nu}
   \sigma_{m}^{\lambda}\sigma_{n}^{\lambda},
   \label{AE:I(w)LF}
\end{align}
where
$\bm{e}_{\mathrm{in}}
\equiv{}^{\mathrm{t}}\!\left[e_{\mathrm{in}}^x\,e_{\mathrm{in}}^y\,e_{\mathrm{in}}^z\right]$ and
$\bm{e}_{\mathrm{sc}}
\equiv{}^{\mathrm{t}}\!\left[e_{\mathrm{sc}}^x\,e_{\mathrm{sc}}^y\,e_{\mathrm{sc}}^z\right]$ are
the unit column vectors indicating the polarizations of incoming and outgoing photons,
respectively, while
$\mathcal{R}\equiv\left[\mathcal{R}^{\mu\nu}\right]$ is
the matrix representation of the Raman operator in Cartesian coordinates.
The matrix elements $\mathcal{R}^{\mu\nu}$ are expressed in terms of Majorana fermions and spinons,
\begin{align}
   \!\!
   \mathcal{R}^{\mu\nu}
   &
 =iJ
   \sum_{\lambda=x,y,z}\sum_{<m,n>_{\lambda}}
   d_{mn}^{\mu}
   d_{mn}^{\nu}
   \hat{u}_{<m,n>_{\lambda}}
   c_{m}c_{n}
   \nonumber \\
   &
 =iJ
   \sum_{\lambda=x,y,z}\sum_{<m,n>_{\lambda}}
   \sum_{k=1}^{L/2}\sum_{k'=1}^{L/2}
   d_{mn}^{\mu}
   d_{mn}^{\nu}
   \hat{u}_{<m,n>_{\lambda}}
   \nonumber \\
   &\ \times
   \bigl[
    (\psi_{m,2k-1}+i\psi_{m,2k})\alpha_k^\dagger
   +(\psi_{m,2k-1}-i\psi_{m,2k})\alpha_k
   \bigr]
   \bigl[
    (\psi_{n,2k'-1}+i\psi_{n,2k'})\alpha_{k'}^\dagger
   +(\psi_{n,2k'-1}-i\psi_{n,2k'})\alpha_{k'}
   \bigr].
   \label{AE:Rspinon}
\end{align}
\noindent
The LF vertex can be decomposed in terms of single-valued irreducible representations of
the double group $\widetilde{\mathbf{P}}$ of the background gauged lattice,
i.e. irreducible representations of the corresponding point symmetry group $\mathbf{P}$
\cite{D175,C172406,K024414},
\begin{align}
   \mathscr{R}
  ={\sum_i}'
   \sum_{\mu=1}^{d_{\widetilde{\varXi}_i}^{\widetilde{\mathbf{P}}}}
   E_{\widetilde{\varXi}_i:\mu}^{\widetilde{\mathbf{P}}}
   \mathcal{R}_{\widetilde{\varXi}_i:\mu}^{\widetilde{\mathbf{P}}}
   \color{black}
  ={\sum_{i}}'
   \sum_{\mu=1}^{d_{\varXi_{i}}^{\mathbf{P}}}
   E_{\varXi_{i}:\mu}^{\mathbf{P}}\mathcal{R}_{\varXi_{i}:\mu}^{\mathbf{P}},
   \label{AE:Rirrep}
\end{align}
where
$E_{\widetilde{\varXi}_{i}:\mu}^{\widetilde{\mathbf{P}}}$ ($E_{\varXi_{i}:\mu}^{\mathbf{P}}$)
is the $\mu$th polarization-vector basis function for the $\widetilde{\varXi}_{i}$ ($\varXi_{i}$)
irreducible representation of $\widetilde{\mathbf{P}}$ ($\mathbf{P}$),
$\mathcal{R}_{\widetilde{\varXi}_{i}:\mu}^{\widetilde{\mathbf{P}}}$
($\mathcal{R}_{\varXi_{i}:\mu}^{\mathbf{P}}$) is the symmetry-definite LF vertex accompanying it,
and $\sum_{i}'$ runs over the \textit{LF-active real} irreducible representations.
Within the LF formulation, the nonvanishing vertices and corresponding basis functions read
\begin{alignat}{3}
   &
   E_{\mathrm{A}_{1}:1}^{\mathbf{C}_{6\mathrm{v}}}
  =\frac{e_{\mathrm{in}}^{x}e_{\mathrm{sc}}^{x}
   +e_{\mathrm{in}}^{y}e_{\mathrm{sc}}^{y}}{\sqrt{2}},\ 
   &&
   E_{\mathrm{E}_{2}:1}^{\mathbf{C}_{6\mathrm{v}}}
  =\frac{e_{\mathrm{in}}^{x}e_{\mathrm{sc}}^{x}
   -e_{\mathrm{in}}^{y}e_{\mathrm{sc}}^{y}}{\sqrt{2}},\ 
   &&
   E_{\mathrm{E_{2}}:2}^{\mathbf{C}_{6\mathrm{v}}}
  =\frac{e_{\mathrm{in}}^{x}e_{\mathrm{sc}}^{y}
    +e_{\mathrm{in}}^{y}e_{\mathrm{sc}}^{x}}{\sqrt{2}},
   \nonumber \\
   &
   \mathcal{R}_{\mathrm{A_{1}}:1}^{\mathbf{C}_{6\mathrm{v}}}
  =\frac{\mathcal{R}^{xx}+\mathcal{R}^{yy}}{\sqrt{2}},
   &&
   \mathcal{R}_{\mathrm{E}_{2}:1}^{\mathbf{C}_{6\mathrm{v}}}
  =\frac{\mathcal{R}^{xx}-\mathcal{R}^{yy}}{\sqrt{2}},
   &&
   \mathcal{R}_{\mathrm{E}_{2}:2}^{\mathbf{C}_{6\mathrm{v}}}
  =\frac{\mathcal{R}^{xy}+\mathcal{R}^{yx}}{\sqrt{2}}
   \label{AE:LFV(C6v)}
\end{alignat}
for the two-dimensional $\widetilde{\mathbf{C}_{6\mathrm{v}}}$ gauged honeycomb and
\begin{alignat}{2}
   &
   E_{\mathrm{A}:1}^{\mathbf{I}}
  =E_{\mathrm{A}:1}^{\mathbf{T}}
  =E_{\mathrm{A_{1g}}:1}^{\mathbf{O}_{\mathrm{h}}}
  =\frac{e_{\mathrm{in}}^{x}e_{\mathrm{sc}}^{x}
   +e_{\mathrm{in}}^{y}e_{\mathrm{sc}}^{y}
   +e_{\mathrm{in}}^{z}e_{\mathrm{sc}}^{z}}{\sqrt{3}},
   &\quad&
   \mathcal{R}_{\mathrm{A}:1}^{\mathbf{I}}
  =\mathcal{R}_{\mathrm{A}:1}^{\mathbf{T}}
  =\mathcal{R}_{\mathrm{A_{1g}}:1}^{\mathbf{O}_{\mathrm{h}}}
  =\frac{\mathcal{R}^{xx}+\mathcal{R}^{yy}+\mathcal{R}^{zz}}{\sqrt{3}},
   \nonumber \\
   &
   E_{\mathrm{H}:1}^{\mathbf{I}}
  =E_{\mathrm{E}:1}^{\mathbf{T}}
  =E_{\mathrm{E}_{\mathrm{g}}:1}^{\mathbf{O}_{\mathrm{h}}}
  =\frac{2e_{\mathrm{in}}^{z}e_{\mathrm{sc}}^{z}
    -e_{\mathrm{in}}^{x}e_{\mathrm{sc}}^{x}
    -e_{\mathrm{in}}^{y}e_{\mathrm{sc}}^{y}}{\sqrt{6}},
   &&
   \mathcal{R}_{\mathrm{H}:1}^{\mathbf{I}}
  =\mathcal{R}_{\mathrm{E}:1}^{\mathbf{T}}
  =\mathcal{R}_{\mathrm{E_{g}}:1}^{\mathbf{O}_{\mathrm{h}}}
  =\frac{2\mathcal{R}^{zz}-\mathcal{R}^{xx}-\mathcal{R}^{yy}}{\sqrt{6}},
   \nonumber \\
   &
   E_{\mathrm{H}:2}^{\mathbf{I}}
  =E_{\mathrm{E}:2}^{\mathbf{T}}
  =E_{\mathrm{E}_{\mathrm{g}}:2}^{\mathbf{O}_{\mathrm{h}}}
  =\frac{e_{\mathrm{in}}^{x}e_{\mathrm{sc}}^{x}
   -e_{\mathrm{in}}^{y}e_{\mathrm{sc}}^{y}}{\sqrt{2}},
   &&
   \mathcal{R}_{\mathrm{H}:2}^{\mathbf{I}}
  =\mathcal{R}_{\mathrm{E}:2}^{\mathbf{T}}
  =\mathcal{R}_{\mathrm{E_{g}}:2}^{\mathbf{O}_{\mathrm{h}}}
  =\frac{\mathcal{R}^{xx}-\mathcal{R}^{yy}}{\sqrt{2}},
   \nonumber \\
   &
   E_{\mathrm{H}:3}^{\mathbf{I}}
  =E_{\mathrm{T}:1}^{\mathbf{T}}
  =E_{\mathrm{T}_{2\mathrm{g}}:1}^{\mathbf{O}_{\mathrm{h}}}
  =\frac{e_{\mathrm{in}}^{x}e_{\mathrm{sc}}^{y}
    +e_{\mathrm{in}}^{y}e_{\mathrm{sc}}^{x}}{\sqrt{2}},
   &&
   \mathcal{R}_{\mathrm{H}:3}^{\mathbf{I}}
  =\mathcal{R}_{\mathrm{T}:1}^{\mathbf{T}}
  =\mathcal{R}_{\mathrm{T_{2g}}:1}^{\mathbf{O}_{\mathrm{h}}}
  =\frac{\mathcal{R}^{xy}+\mathcal{R}^{yx}}{\sqrt{2}},
   \nonumber \\
   &
   E_{\mathrm{H}:4}^{\mathbf{I}}
  =E_{\mathrm{T}:2}^{\mathbf{T}}
  =E_{\mathrm{T}_{2\mathrm{g}}:2}^{\mathbf{O}_{\mathrm{h}}}
  =\frac{e_{\mathrm{in}}^{y}e_{\mathrm{sc}}^{z}
   +e_{\mathrm{in}}^{z}e_{\mathrm{sc}}^{y}}{\sqrt{2}},
   &&
   \mathcal{R}_{\mathrm{H}:4}^{\mathbf{I}}
  =\mathcal{R}_{\mathrm{T}:2}^{\mathbf{T}}
  =\mathcal{R}_{\mathrm{T_{2g}}:2}^{\mathbf{O}_{\mathrm{h}}}
  =\frac{\mathcal{R}^{yz}+\mathcal{R}^{zy}}{\sqrt{2}},
   \nonumber \\
   &
   E_{\mathrm{H}:5}^{\mathbf{I}}
  =E_{\mathrm{T}:3}^{\mathbf{T}}
  =E_{\mathrm{T}_{2\mathrm{g}}:3}^{\mathbf{O}_{\mathrm{h}}}
  =\frac{e_{\mathrm{in}}^{z}e_{\mathrm{sc}}^{x}
    +e_{\mathrm{in}}^{x}e_{\mathrm{sc}}^{z}}{\sqrt{2}},
   &&
   \mathcal{R}_{\mathrm{H}:5}^{\mathbf{I}}
  =\mathcal{R}_{\mathrm{T}:3}^{\mathbf{T}}
  =\mathcal{R}_{\mathrm{T_{2g}}:3}^{\mathbf{O}_{\mathrm{h}}}
  =\frac{\mathcal{R}^{zx}+\mathcal{R}^{xz}}{\sqrt{2}}
   \label{AE:E&R(I&T&Oh)}
\end{alignat}
for the $\widetilde{\mathbf{I}}$, $\widetilde{\mathbf{T}}$, and
$\widetilde{\mathbf{O}_{\mathrm{h}}}$ gauged polyhedra,
where $\mathcal{R}_{\mathrm{A_{1}}:1}^{\mathbf{C}_{6\mathrm{v}}}$ and
$\mathcal{R}_{\mathrm{A}:1}^{\mathbf{I}}
=\mathcal{R}_{\mathrm{A}:1}^{\mathbf{T}}
=\mathcal{R}_{\mathrm{A_{1g}}:1}^{\mathbf{O}_{\mathrm{h}}}$,
belonging to the identity representations in two and three dimensions, respectively,
all commute with the corresponding Hamiltonians to contribute merely to elastic (Rayleigh)
scattering.

   Decomposing the Raman operator into irreducible representations (\ref{AE:Rirrep}) and taking
account of their orthogonality (\ref{E:RiRjOrthogonality}), we write the Raman scattering
intensity as
\begin{align}
   &
   I(\omega)
  ={\sum_{i}}'{\sum_{j}}'
   \sum_{\mu=1}^{d_{\varXi_{i}}^{\mathbf{P}}}
   \sum_{\nu=1}^{d_{\varXi_{j}}^{\mathbf{P}}}
   E_{\varXi_{i}:\mu}^{\mathbf{P}}E_{\varXi_{j}:\nu}^{\mathbf{P}}
   \int_{-\infty}^\infty
   \frac{dt\,e^{i\omega t}}{2\pi\hbar L}
   \langle 0|
   e^{\frac{{i}\mathscr{H}t}{\hbar}}
   \mathcal{R}_{\varXi_i:\mu}^{\mathbf{P}}
   e^{-\frac{{i}\mathscr{H}t}{\hbar}}
   \mathcal{R}_{\varXi_j:\nu}^{\mathbf{P}}
   |0\rangle
   \nonumber \\
   &\qquad
  ={\sum_{i}}'
   \sum_{\mu=1}^{d_{\varXi_{i}}^{\mathbf{P}}}
   \left(E_{\varXi_{i}:\mu}^{\mathbf{P}}\right)^2
   \int_{-\infty}^\infty
   \frac{dt\,e^{i\omega t}}{2\pi\hbar L}
   \langle 0|
   e^{\frac{{i}\mathscr{H}t}{\hbar}}
   \mathcal{R}_{\varXi_i:\mu}^{\mathbf{P}}
   e^{-\frac{{i}\mathscr{H}t}{\hbar}}
   \mathcal{R}_{\varXi_i:\mu}^{\mathbf{P}}
   |0\rangle
  \equiv
   {\sum_{i}}'
   \sum_{\mu=1}^{d_{\varXi_{i}}^{\mathbf{P}}}
   \left(E_{\varXi_{i}:\mu}^{\mathbf{P}}\right)^{2}
   I_{\varXi_{i}:\mu}^{\mathbf{P}}(\omega).
   \label{AE:I(w)irrep}
\end{align}
We write the Raman vertices in Cartesian coordinates (\ref{AE:E&R(I&T&Oh)}) and then in terms of
spinon operators (\ref{AE:Rspinon}).
Having in mind that $\alpha_k|0\rangle=0$ and discarding Rayleigh terms, we can express
$I(\omega)$ by Fermi's golden rule,
\begin{align}
   &
   I(\omega)
  ={\sum_{i}}'\sum_{\mu=1}^{d_{\varXi_{i}}^{\mathbf{P}}}
   \left(E_{\varXi_{i}:\mu}^{\mathbf{P}}\right)^{2}
   \int_{-\infty}^\infty
   \frac{dt\,e^{i\omega t}}{2\pi\hbar L}
   \sum_{q=0}^{2^{\frac{L}{2}+1}-1}\sum_{\kappa=0}^{2^{\frac{L}{2}-1}-1}
   \langle 0|
   e^{\frac{{i}\mathscr{H}t}{\hbar}}
   \mathcal{R}_{\varXi_i:\mu}^{\mathbf{P}}
   e^{-\frac{{i}\mathscr{H}t}{\hbar}}
   |\{n_k\}\rangle_\kappa\otimes|\{W_p\}\rangle_q
   {}_q\langle\{W_p\}|\otimes{}_\kappa\langle\{n_k\}|
   \mathcal{R}_{\varXi_i:\mu}^{\mathbf{P}}
   |0\rangle
   \nonumber \\
   &\qquad
  =\frac{1}{L}
   {\sum_{i}}'\sum_{\mu=1}^{d_{\varXi_{i}}^{\mathbf{P}}}
   \left(E_{\varXi_{i}:\mu}^{\mathbf{P}}\right)^{2}
   \frac{1}{2\pi\hbar}
   \int_{-\infty}^\infty
   e^{i\left(\omega-\frac{\epsilon_k}{\hbar}-\frac{\epsilon_{k'}}{\hbar}\right)t}dt
   \sum_{q=0}^{2^{\frac{L}{2}+1}-1}
   {}_0\langle\{W_p\}|\{W_p\}\rangle_q{}_q\langle\{W_p\}|\{W_p\}\rangle_0
   \nonumber \\
   &\qquad\quad
   \times
   \sum_{1=k<k'=\frac{L}{2}}
   {}_0\langle\{n_k\}|
   \left.
    \mathcal{R}_{\varXi_i:\mu}^{\mathbf{P}}
   \right|_{\{u_{<m,n>_{\lambda}}\}_{0(r)}}
   \alpha_{k'}^\dagger\alpha_k^\dagger
   |\{n_k\}\rangle_0
   {}_0\langle\{n_k\}|
   \alpha_k\alpha_{k'}
   \left.
    \mathcal{R}_{\varXi_i:\mu}^{\mathbf{P}}
   \right|_{\{u_{<m,n>_{\lambda}}\}_{0(r)}}
   |\{n_k\}\rangle_0
   \nonumber \\
   &\qquad
  =\frac{1}{L}
   {\sum_i}'\sum_{\mu=1}^{d_{\varXi_{i}}^{\mathbf{P}}}
   \left(E_{\varXi_{i}:\mu}^{\mathbf{P}}\right)^2
   \sum_{1=k<k'=\frac{L}{2}}
   \left|
     \langle 0|\alpha_{k}\alpha_{k'}
       \mathcal{R}_{\varXi_i:\mu}^{\mathbf{P}}
    |0\rangle
   \right|^{2}
   \delta(\hbar\omega-\varepsilon_{k}-\varepsilon_{k'}),
   \label{AE:I(w)LFirrep}
\end{align}
where 
$\left.
  \mathcal{R}_{\varXi_i:\mu}^{\mathbf{P}}
 \right|_{\{u_{<m,n>_{\lambda}}\}_{0(r)}}$
are the \textit{gauge-ground} LF vertices.

   The spectral degeneracy within each multidimensional irreducible representation \cite{P094439}
is the case with Kitaev spin balls as well.
Considering the QSL ground state (\ref{E:|0>}) is invariant under every symmetry operation
$P\in\mathbf{P}$, the Raman response with
$\mathcal{P}\bm{e}_{\mathrm{in}}\equiv\tilde{\bm{e}}_{\mathrm{in}}$ and
$\mathcal{P}\bm{e}_{\mathrm{sc}}\equiv\tilde{\bm{e}}_{\mathrm{sc}}$,
which we shall denote by $\tilde{I}(\omega)$, should remain the same as
$I(\omega)$ with $\bm{e}_{\mathrm{in}}$ and $\bm{e}_{\mathrm{sc}}$, where we denote
the matrix representation in Cartesian coordinates for a point symmetry operation $P$ by
$\mathcal{P}$.
With Eq. (\ref{AE:I(w)LF}) in mind, a point symmetry operation of the Raman operator reads
\begin{align}
   {}^{\mathrm{t}}\!\tilde{\bm{e}}_{\mathrm{in}}
   \mathcal{R}
   \tilde{\bm{e}}_{\mathrm{sc}}
  \equiv\widetilde{\mathscr{R}}
  =\sum_{\mu,\nu=x,y,z}
   \sum_{\mu',\nu'=x,y,z}
   e_{\mathrm{in}}^\mu
   {}^{\mathrm{t}}\!\mathcal{P}^{\mu\mu'}
   \mathcal{R}^{\mu'\nu'}
   \mathcal{P}^{\nu'\nu}
   e_{\mathrm{sc}}^\nu
  \equiv
   \sum_{\mu,\nu=x,y,z}
   e_{\mathrm{in}}^\mu
   \widetilde{\mathcal{R}}^{\mu\nu}(P)
   e_{\mathrm{sc}}^\nu
  \equiv
   {}^{\mathrm{t}}\!\bm{e}_{\mathrm{in}}
   \widetilde{\mathcal{R}}(P)
   \bm{e}_{\mathrm{sc}},
   \label{AE:PmathscrR}
\end{align}
and therefore, we have an intensity
\begin{align}
   &
   \tilde{I}(\omega)
  =\int_{-\infty}^\infty
   \frac{dt\,e^{i\omega t}}{2\pi\hbar L}
   \langle 0|
   e^{\frac{{i}\mathscr{H}t}{\hbar}}
   \widetilde{\mathscr{R}}
   e^{-\frac{{i}\mathscr{H}t}{\hbar}}
   \widetilde{\mathscr{R}}
   |0\rangle
  ={\sum_{i}}'
   \sum_{\mu=1}^{d_{\varXi_{i}}^{\mathbf{P}}}
   \left(E_{\varXi_{i}:\mu}^{\mathbf{P}}\right)^2
   \int_{-\infty}^\infty
   \frac{dt\,e^{i\omega t}}{2\pi\hbar L}
   \langle 0|
   e^{\frac{{i}\mathscr{H}t}{\hbar}}
   \widetilde{\mathcal{R}}_{\varXi_i:\mu}^{\mathbf{P}}(P)
   e^{-\frac{{i}\mathscr{H}t}{\hbar}}
   \widetilde{\mathcal{R}}_{\varXi_i:\mu}^{\mathbf{P}}(P)
   |0\rangle
   \nonumber \\
   &\qquad
  \equiv
   {\sum_{i}}'
   \sum_{\mu=1}^{d_{\varXi_{i}}^{\mathbf{P}}}
   \left(E_{\varXi_{i}:\mu}^{\mathbf{P}}\right)^{2}
   \tilde{I}_{\varXi_{i}:\mu}^{\mathbf{P}}(\omega)
  ={\sum_{i}}'
   \sum_{\mu=1}^{d_{\varXi_{i}}^{\mathbf{P}}}
   \left(E_{\varXi_{i}:\mu}^{\mathbf{P}}\right)^{2}
   I_{\varXi_{i}:\mu}^{\mathbf{P}}(\omega).
   \label{AE:tildeI(w)}
\end{align}
Arbitrary polarization vectors $\bm{e}_{\mathrm{in}}$ and $\bm{e}_{\mathrm{sc}}$ yield
arbitrary coefficients $\left(E_{\varXi_{i}:\mu}^{\mathbf{P}}\right)^{2}$ and therefore
demand that
$
   \tilde{I}_{\varXi_{i}:\mu}^{\mathbf{P}}(\omega)
  =I_{\varXi_{i}:\mu}^{\mathbf{P}}(\omega)
$
for every Raman-active mode $\varXi_{i}:\mu$.
It is instructive to review the Raman-active $\mathrm{E}_2$ symmetry species of
the $\mathbf{C}_{6\mathrm{v}}$ honeycomb lattice \cite{P094439} on the $xy$ plane.
The threefold rotation about the $z$ axis of the polarization vectors reads converting
the Raman operator into
\begin{align}
   C_{3(z)}\mathcal{R}
  \equiv
   \widetilde{\mathcal{R}}(C_{3(z)})
  =\left[
    \begin{array}{cc}
     \widetilde{\mathcal{R}}^{xx}(C_{3(z)}) & \widetilde{\mathcal{R}}^{xy}(C_{3(z)}) \\
     \widetilde{\mathcal{R}}^{yx}(C_{3(z)}) & \widetilde{\mathcal{R}}^{zz}(C_{3(z)}) \\
    \end{array}
   \right]
  =\left[
    \begin{array}{cc}
    -\frac{1}{2} &  \frac{\sqrt{3}}{2} \\
    -\frac{\sqrt{3}}{2} & -\frac{1}{2} \\
    \end{array}
   \right]
   \left[
    \begin{array}{cc}
     \mathcal{R}^{xx} & \mathcal{R}^{xy} \\
     \mathcal{R}^{yx} & \mathcal{R}^{zz} \\
    \end{array}
   \right]
   \left[
    \begin{array}{cc}
    -\frac{1}{2} & -\frac{\sqrt{3}}{2} \\
     \frac{\sqrt{3}}{2} & -\frac{1}{2} \\
    \end{array}
   \right].
   \label{AE:C3R}
\end{align}
Then the Raman vertices of $\mathrm{E}_2$ symmetry species behave as
\begin{align}
   C_{3(z)}\mathcal{R}_{\mathrm{E}_{2}:1}^{\mathbf{C}_{6\mathrm{v}}}
  \equiv
   \widetilde{\mathcal{R}}_{\mathrm{E}_{2}:1}^{\mathbf{C}_{6\mathrm{v}}}(C_{3(z)})
  =\frac{\widetilde{\mathcal{R}}^{xx}(C_{3(z)})-\widetilde{\mathcal{R}}^{yy}(C_{3(z)})}
   {\sqrt{2}}
 =-\frac{1}{2}
   \mathcal{R}_{\mathrm{E}_{2}:1}^{\mathbf{C}_{6\mathrm{v}}}
  -\frac{\sqrt{3}}{2}
   \mathcal{R}_{\mathrm{E}_{2}:2}^{\mathbf{C}_{6\mathrm{v}}}.
   \label{AE:C3RE2:1}
\end{align}
The Raman response of the Kitaev honeycomb QSL remains unchanged against the symmetry operation
$C_{3(z)}\in\mathbf{C}_{6\mathrm{v}}$,
\begin{align}
   I_{\mathrm{E}_{2}:1}^{\mathbf{C}_{6\mathrm{v}}}(\omega)
  =C_{3(z)}
   I_{\mathrm{E}_{2}:1}^{\mathbf{C}_{6\mathrm{v}}}(\omega)
  =\int_{-\infty}^\infty
   \frac{dt\,e^{i\omega t}}{2\pi\hbar L}
   \langle 0|
   e^{\frac{{i}\mathscr{H}t}{\hbar}}
   \widetilde{\mathcal{R}}_{\mathrm{E}_{2}:1}^{\mathbf{C}_{6\mathrm{v}}}(C_{3(z)})
   e^{-\frac{{i}\mathscr{H}t}{\hbar}}
   \widetilde{\mathcal{R}}_{\mathrm{E}_{2}:1}^{\mathbf{C}_{6\mathrm{v}}}(C_{3(z)})
   |0\rangle
  =\frac{1}{4}I_{\mathrm{E}_{2}:1}^{\mathbf{C}_{6\mathrm{v}}}(\omega)
  +\frac{3}{4}I_{\mathrm{E}_{2}:2}^{\mathbf{C}_{6\mathrm{v}}}(\omega),
  \label{AE:C3RI(w)E2:1}
\end{align}
and therefore, we find that
$I_{\mathrm{E}_{2}:1}^{\mathbf{C}_{6\mathrm{v}}}(\omega)
=I_{\mathrm{E}_{2}:2}^{\mathbf{C}_{6\mathrm{v}}}(\omega)$.
Next we consider rotating the $\mathbf{T}$ and $\mathbf{O}_{\mathrm{h}}$ polyhedra by
$\frac{2\pi}{3}$ about the $[111]$ axis, which reads converting the Raman operator into
\begin{align}
   C_{3(111)}\mathcal{R}
  \equiv
   \left[
    \begin{array}{ccc}
     \widetilde{\mathcal{R}}^{xx}(C_{3(111)}) &
     \widetilde{\mathcal{R}}^{xy}(C_{3(111)}) &
     \widetilde{\mathcal{R}}^{xz}(C_{3(111)}) \\
     \widetilde{\mathcal{R}}^{yx}(C_{3(111)}) &
     \widetilde{\mathcal{R}}^{yy}(C_{3(111)}) &
     \widetilde{\mathcal{R}}^{yz}(C_{3(111)}) \\
     \widetilde{\mathcal{R}}^{zx}(C_{3(111)}) &
     \widetilde{\mathcal{R}}^{zy}(C_{3(111)}) &
     \widetilde{\mathcal{R}}^{zz}(C_{3(111)}) \\
    \end{array}
   \right]
  =\left[
    \begin{array}{ccc}
     0 & 1 & 0 \\
     0 & 0 & 1 \\
     1 & 0 & 0 \\
    \end{array}
   \right]
   \left[
    \begin{array}{ccc}
     \mathcal{R}^{xx} & \mathcal{R}^{xy} & \mathcal{R}^{xz} \\ 
     \mathcal{R}^{yx} & \mathcal{R}^{yy} & \mathcal{R}^{yz} \\ 
     \mathcal{R}^{zx} & \mathcal{R}^{zy} & \mathcal{R}^{zz} \\ 
    \end{array}
   \right]
   \left[
    \begin{array}{ccc}
     0 & 0 & 1 \\
     1 & 0 & 0 \\
     0 & 1 & 0 \\
    \end{array}
   \right].
   \label{AE:C3(111)R}
\end{align}
They each have the two Raman-active symmetry species
$\mathrm{E}/\mathrm{E}_{\mathrm{g}}$ and
$\mathrm{T}/\mathrm{T}_{2\mathrm{g}}$
and the corresponding Raman vertices behave under the threefold rotation as
\begin{align}
   &
   C_{3(111)}
   \mathcal{R}_{\mathrm{E}/\mathrm{E}_{\mathrm{g}}:1}^{\mathbf{T}/\mathbf{O}_{\mathrm{h}}}
  \equiv
   \widetilde{\mathcal{R}}_{\mathrm{E}/\mathrm{E}_{\mathrm{g}}:1}
                          ^{\mathbf{T}/\mathbf{O}_{\mathrm{h}}}(C_{3(111)})
  =\frac{2\widetilde{\mathcal{R}}^{zz}(C_{3(111)})
         -\widetilde{\mathcal{R}}^{xx}(C_{3(111)})
         -\widetilde{\mathcal{R}}^{yy}(C_{3(111)})}
   {\sqrt{6}}
 =-\frac{1}{2}
   \mathcal{R}_{\mathrm{E}/\mathrm{E}_{\mathrm{g}}:1}^{\mathbf{T}/\mathbf{O}_{\mathrm{h}}}
  +\frac{\sqrt{3}}{2}
   \mathcal{R}_{\mathrm{E}/\mathrm{E}_{\mathrm{g}}:2}^{\mathbf{T}/\mathbf{O}_{\mathrm{h}}}.
   \allowdisplaybreaks
   \nonumber \\
   &
   C_{3(111)}
   \mathcal{R}_{\mathrm{T}/\mathrm{T}_{2\mathrm{g}}:1}^{\mathbf{T}/\mathbf{O}_{\mathrm{h}}}
  \equiv
   \widetilde{\mathcal{R}}_{\mathrm{T}/\mathrm{T}_{2\mathrm{g}}:1}
                          ^{\mathbf{T}/\mathbf{O}_{\mathrm{h}}}(C_{3(111)})
  =\frac{\widetilde{\mathcal{R}}^{xy}(C_{3(111)})
        +\widetilde{\mathcal{R}}^{yx}(C_{3(111)})}
   {\sqrt{2}}
  =\mathcal{R}_{\mathrm{T}/\mathrm{T}_{2\mathrm{g}}:2}^{\mathbf{T}/\mathbf{O}_{\mathrm{h}}},
   \allowdisplaybreaks
   \nonumber \\
   &
   C_{3(111)}
   \mathcal{R}_{\mathrm{T}/\mathrm{T}_{2\mathrm{g}}:2}^{\mathbf{T}/\mathbf{O}_{\mathrm{h}}}
  \equiv
   \widetilde{\mathcal{R}}_{\mathrm{T}/\mathrm{T}_{2\mathrm{g}}:2}
                          ^{\mathbf{T}/\mathbf{O}_{\mathrm{h}}}(C_{3(111)})
  =\frac{\widetilde{\mathcal{R}}^{yz}(C_{3(111)})
        +\widetilde{\mathcal{R}}^{zy}(C_{3(111)})}
   {\sqrt{2}}
  =\mathcal{R}_{\mathrm{T}/\mathrm{T}_{2\mathrm{g}}:3}^{\mathbf{T}/\mathbf{O}_{\mathrm{h}}},
   \allowdisplaybreaks
   \nonumber \\
   &
   C_{3(111)}
   \mathcal{R}_{\mathrm{T}/\mathrm{T}_{2\mathrm{g}}:3}^{\mathbf{T}/\mathbf{O}_{\mathrm{h}}}
  \equiv
   \widetilde{\mathcal{R}}_{\mathrm{T}/\mathrm{T}_{2\mathrm{g}}:3}
                          ^{\mathbf{T}/\mathbf{O}_{\mathrm{h}}}(C_{3(111)})
  =\frac{\widetilde{\mathcal{R}}^{zx}(C_{3(111)})
        +\widetilde{\mathcal{R}}^{xz}(C_{3(111)})}
   {\sqrt{2}}
  =\mathcal{R}_{\mathrm{T}/\mathrm{T}_{2\mathrm{g}}:1}^{\mathbf{T}/\mathbf{O}_{\mathrm{h}}}.
  \label{AE:C3(111)RE&O}
\end{align}
The Raman responses of these Kitaev polyhedral QSLs are invariant under their common symmetry
operation $C_{3(111)}$,
\begin{align}
   &
   I_{\mathrm{E}/\mathrm{E}_{\mathrm{g}}:1}^{\mathbf{T}/\mathbf{O}_{\mathrm{h}}}(\omega)
  =C_{3(111)}
   I_{\mathrm{E}/\mathrm{E}_{\mathrm{g}}:1}^{\mathbf{T}/\mathbf{O}_{\mathrm{h}}}(\omega)
  =\frac{1}{4}I_{\mathrm{E}/\mathrm{E}_{\mathrm{g}}:1}
               ^{\mathbf{T}/\mathbf{O}_{\mathrm{h}}}(\omega)
  +\frac{3}{4}I_{\mathrm{E}/\mathrm{E}_{\mathrm{g}}:2}
               ^{\mathbf{T}/\mathbf{O}_{\mathrm{h}}}(\omega),
   \allowdisplaybreaks
   \nonumber \\
   &
   I_{\mathrm{T}/\mathrm{T}_{2\mathrm{g}}:1}^{\mathbf{T}/\mathbf{O}_{\mathrm{h}}}(\omega)
  =C_{3(111)}
   I_{\mathrm{T}/\mathrm{T}_{2\mathrm{g}}:1}^{\mathbf{T}/\mathbf{O}_{\mathrm{h}}}(\omega)
  =I_{\mathrm{T}/\mathrm{T}_{2\mathrm{g}}:2}^{\mathbf{T}/\mathbf{O}_{\mathrm{h}}}(\omega),
   \allowdisplaybreaks
   \nonumber \\
   &
   I_{\mathrm{T}/\mathrm{T}_{2\mathrm{g}}:2}^{\mathbf{T}/\mathbf{O}_{\mathrm{h}}}(\omega)
  =C_{3(111)}
   I_{\mathrm{T}/\mathrm{T}_{2\mathrm{g}}:2}^{\mathbf{T}/\mathbf{O}_{\mathrm{h}}}(\omega)
  =I_{\mathrm{T}/\mathrm{T}_{2\mathrm{g}}:3}^{\mathbf{T}/\mathbf{O}_{\mathrm{h}}}(\omega),
   \allowdisplaybreaks
   \nonumber \\
   &
   I_{\mathrm{T}/\mathrm{T}_{2\mathrm{g}}:3}^{\mathbf{T}/\mathbf{O}_{\mathrm{h}}}(\omega)
  =C_{3(111)}
   I_{\mathrm{T}/\mathrm{T}_{2\mathrm{g}}:3}^{\mathbf{T}/\mathbf{O}_{\mathrm{h}}}(\omega)
  =I_{\mathrm{T}/\mathrm{T}_{2\mathrm{g}}:1}^{\mathbf{T}/\mathbf{O}_{\mathrm{h}}}(\omega),
  \label{AE:C3(111)I(w)E&O}
\end{align}
and therefore, we find that
$I_{\mathrm{E}/\mathrm{E}_{\mathrm{g}}:1}^{\mathbf{T}/\mathbf{O}_{\mathrm{h}}}(\omega)
=I_{\mathrm{E}/\mathrm{E}_{\mathrm{g}}:2}^{\mathbf{T}/\mathbf{O}_{\mathrm{h}}}(\omega)$ and
$I_{\mathrm{T}/\mathrm{T}_{2\mathrm{g}}:1}^{\mathbf{T}/\mathbf{O}_{\mathrm{h}}}(\omega)
=I_{\mathrm{T}/\mathrm{T}_{2\mathrm{g}}:2}^{\mathbf{T}/\mathbf{O}_{\mathrm{h}}}(\omega)
=I_{\mathrm{T}/\mathrm{T}_{2\mathrm{g}}:3}^{\mathbf{T}/\mathbf{O}_{\mathrm{h}}}(\omega)$.
For the Raman-active $\mathrm{H}$ symmetry species of the Kitaev dodecahedral QSL as well,
we can similarly find the spectral degeneracy
$I_{\mathrm{H}:1}^{\mathbf{I}}(\omega)
=I_{\mathrm{H}:2}^{\mathbf{I}}(\omega)
=I_{\mathrm{H}:3}^{\mathbf{I}}(\omega)
=I_{\mathrm{H}:4}^{\mathbf{I}}(\omega)
=I_{\mathrm{H}:5}^{\mathbf{I}}(\omega)$.

   Now that Eq. (\ref{AE:I(w)irrep}) reduces to
\begin{align}
   I(\omega)
  ={\sum_i}'
   \sum_{\mu=1}^{d_{\varXi_{i}}^{\mathbf{P}}}
   \left(E_{\varXi_{i}:\mu}^{\mathbf{P}}\right)^2
   I_{\varXi_{i}:\mu}^{\mathbf{P}}(\omega)
  ={\sum_i}'
   I_{\varXi_{i}:1}^{\mathbf{P}}(\omega)
   \sum_{\mu=1}^{d_{\varXi_{i}}^{\mathbf{P}}}
   \left(E_{\varXi_{i}:\mu}^{\mathbf{P}}\right)^2,
   \label{AE:I(w)irrep1stMode}
\end{align}
how many Raman-active modes are possible in the lattice geometry is most decisive of
whether and how the scattering intensity depends on the light polarization.
In Eq. (\ref{AE:I(w)irrep1stMode}), we have
\begin{align}
   \sum_{\mu=1}^{2}
   \left(E_{\mathrm{E}_{2}:\mu}^{\mathbf{C}_{6\mathrm{v}}}\right)^2
  =\smash{\frac{1}{2}}\sin^2\vartheta_{\mathrm{in}}\sin^2\vartheta_{\mathrm{sc}}
  \label{AE:E^2C6vE2}
\end{align}
for the two-dimensional honeycomb lattice,
\begin{align}
   \sum_{\mu=1}^{5}
   \left(E_{\mathrm{H}:\mu}^{\mathbf{I}}\right)^{2}
   &
  =\frac{1}{6}
   \left[
    2\cos\vartheta_{\mathrm{in}}\cos\vartheta_{\mathrm{sc}}
   -\sin\vartheta_{\mathrm{in}}\sin\vartheta_{\mathrm{sc}}
    \cos(\varphi_{\mathrm{in}}-\varphi_{\mathrm{sc}})
   \right]^2
   \allowdisplaybreaks
   \nonumber \\
   &
  +\frac{1}{2}
   [\sin\vartheta_{\mathrm{in}}\sin\vartheta_{\mathrm{sc}}
    \cos(\varphi_{\mathrm{in}}+\varphi_{\mathrm{sc}})]^2
   \allowdisplaybreaks
   \nonumber \\
   &
  +\frac{1}{2}
   \left[
    \sin\vartheta_{\mathrm{in}}\sin\vartheta_{\mathrm{sc}}
    \sin(\varphi_{\mathrm{in}}+\varphi_{\mathrm{sc}})
   \right]^2
   \allowdisplaybreaks
   \nonumber \\
   &
  +\frac{1}{2}
   (\sin\vartheta_{\mathrm{in}}\sin\varphi_{\mathrm{in}}\cos\vartheta_{\mathrm{sc}}
   +\cos\vartheta_{\mathrm{in}}\sin\vartheta_{\mathrm{sc}}\sin\varphi_{\mathrm{sc}})^2
   \allowdisplaybreaks
   \nonumber \\
   &
  +\frac{1}{2}
   (\cos\vartheta_{\mathrm{in}}\sin\vartheta_{\mathrm{sc}}\cos\varphi_{\mathrm{sc}}
   +\sin\vartheta_{\mathrm{in}}\cos\varphi_{\mathrm{in}}\cos\vartheta_{\mathrm{sc}})^2
   \label{AE:E^2IH}
\end{align}
for the dodecahedral lattice, and
\begin{align}
   \sum_{\mu=1}^{2}
   \left(E_{\mathrm{E}:\mu}^{\mathbf{T}}\right)^2
  =\sum_{\mu=1}^2
   \left(E_{\mathrm{E}_{\mathrm{g}}:\mu}^{\mathbf{O}_{\mathrm{h}}}\right)^2
   &
  =\frac{1}{6}
   \left[
    2\cos\vartheta_{\mathrm{in}}\cos\vartheta_{\mathrm{sc}}
   -\sin\vartheta_{\mathrm{in}}\sin\vartheta_{\mathrm{sc}}
    \cos(\varphi_{\mathrm{in}}-\varphi_{\mathrm{sc}})
   \right]^2
   \allowdisplaybreaks
   \nonumber \\
   &
  +\frac{1}{2}
   \left[
    \sin\vartheta_{\mathrm{in}}\sin\vartheta_{\mathrm{sc}}
    \cos(\varphi_{\mathrm{in}}+\varphi_{\mathrm{sc}})
   \right]^2,
   \allowdisplaybreaks
   \nonumber \\
   \sum_{\mu=1}^{3}
   \left(E_{\mathrm{T}:\mu}^{\mathbf{T}}\right)^2
  =\sum_{\mu=1}^{3}
   \left(E_{\mathrm{T}_{2\mathrm{g}}:\mu}^{\mathbf{O}_{\mathrm{h}}}\right)^2
   \allowdisplaybreaks
   &
  =\frac{1}{2}
   \left[
    \sin\vartheta_{\mathrm{in}}\sin\vartheta_{\mathrm{sc}}
    \sin(\varphi_{\mathrm{in}}+\varphi_{\mathrm{sc}})
   \right]^2 
   \nonumber \\
   &
  +\frac{1}{2}
   (\sin\vartheta_{\mathrm{in}}\sin\varphi_{\mathrm{in}}\cos\vartheta_{\mathrm{sc}}
   +\cos\vartheta_{\mathrm{in}}\sin\vartheta_{\mathrm{sc}}\sin\varphi_{\mathrm{sc}})^2
   \allowdisplaybreaks
   \nonumber \\
   &
  +\frac{1}{2}
   (\cos\vartheta_{\mathrm{in}}\sin\vartheta_{\mathrm{sc}}\cos\varphi_{\mathrm{sc}}
   +\sin\vartheta_{\mathrm{in}}\cos\varphi_{\mathrm{in}}\cos\vartheta_{\mathrm{sc}})^2
   \label{AE:E^2T&O_E&O}
\end{align}
for the truncated tetrahedral and octahedral lattices.
For the honeycomb lattice, we take interest only in the polarization vectors parallel to the plain,
\begin{align}
   \left.
   \sum_{\mu=1}^2
   \left(E_{\mathrm{E}_2:\mu}^{\mathbf{C}_{6\mathrm{v}}}\right)^2
   \right|_{\vartheta_{\mathrm{in}}=\vartheta_{\mathrm{sc}}=\frac{\pi}{2}}
   &
  =\smash{\frac{1}{2}},
  \label{AE:E^2C6vE2xyplain}
\end{align}
and find no polarization dependence of the Raman response within the LF scheme.
For the dodecahedral lattice, even if we restrict the polarization vectors to the $xy$ plain,
the Raman response still exhibits weak polarization dependence even within the LF scheme,
\begin{align}
   \left.
   \sum_{\mu=1}^{5}
   \left(E_{\mathrm{H}:\mu}^{\mathbf{I}}\right)^{2}
   \right|_{\vartheta_{\mathrm{in}}=\vartheta_{\mathrm{sc}}=\frac{\pi}{2}}
  =\frac{1}{6}
   \cos^{2}(\varphi_{\mathrm{in}}-\varphi_{\mathrm{sc}})
  +\frac{1}{2},
  \label{AE:E^2IHxyplain}
\end{align}
i.e., the spectra peak exactly the same but weigh differently according to the light polarization.
For the truncated tetrahedral and octahedral lattices, even if we consider the Raman scattering
within the LF scheme and restrict the polarization vectors to the $xy$ plain,
we have two Raman-active symmetry species to find strong polarization dependence of the spectra,
\begin{align}
   &
   \left.
   \sum_{\mu=1}^{2}
   \left(E_{\mathrm{E}:\mu}^{\mathbf{T}}\right)^{2}
   \right|_{\vartheta_{\mathrm{in}}=\vartheta_{\mathrm{sc}}=\frac{\pi}{2}}
  =\left.
   \sum_{\mu=1}^{2}
   \left(E_{\mathrm{E_{g}}:\mu}^{\mathbf{O}_{\mathrm{h}}}\right)^{2}
   \right|_{\vartheta_{\mathrm{in}}=\vartheta_{\mathrm{sc}}=\frac{\pi}{2}}
  =\frac{1}{6}
   \cos^{2}(\varphi_{\mathrm{in}}-\varphi_{\mathrm{sc}})
  +\frac{1}{2}
   \cos^{2}(\varphi_{\mathrm{in}}+\varphi_{\mathrm{sc}}),
   \allowdisplaybreaks
   \nonumber \\
   &
   \left.
   \sum_{\mu=1}^{3}
   \left(E_{\mathrm{T}:\mu}^{\mathbf{T}}\right)^{2}
   \right|_{\vartheta_{\mathrm{in}}=\vartheta_{\mathrm{sc}}=\frac{\pi}{2}}
  =\left.
   \sum_{\mu=1}^{3}
   \left(E_{\mathrm{T_{2g}}:\mu}^{\mathbf{O}_{\mathrm{h}}}\right)^{2}
   \right|_{\vartheta_{\mathrm{in}}=\vartheta_{\mathrm{sc}}=\frac{\pi}{2}}
  =\frac{1}{2}\sin^{2}(\varphi_{\mathrm{in}}+\varphi_{\mathrm{sc}}),
  \label{AE:E^2T&O_E&Oxyplain}
\end{align}
i.e., spectra peak and weigh differently according to the light polarization.
Note in this context that we do not have any accidental degeneracy, i.e.,
neither $I_{\mathrm{E}:1}^{\mathbf{T}}(\omega)$ equals $I_{\mathrm{T}:1}^{\mathbf{T}}(\omega)$
nor $I_{\mathrm{E}_{\mathrm{g}}:1}^{\mathbf{O}_{\mathrm{h}}}(\omega)$ equals
$I_{\mathrm{T}_{2\mathrm{g}}:1}^{\mathbf{O}_{\mathrm{h}}}(\omega)$.
\end{appendix}
\end{widetext}

\end{document}